\documentclass[a4paper,iop,twocolumn,numberedappendix]{emulateapj} 
\usepackage[colorlinks=true,linkcolor=blue,anchorcolor=blue,citecolor=blue,urlcolor=blue,draft=false]{hyperref}\usepackage{color}
\usepackage{graphicx}
\usepackage{amsmath} 
\usepackage{amssymb}

\usepackage{placeins}
\usepackage{times}

\usepackage{xcolor}

\shorttitle{AMiBA 13-Element Expansion}
\shortauthors{Lin et~al.}

\usepackage{graphicx}\usepackage{epstopdf}

\usepackage{natbib}

\begin{document}

\title{AMiBA: Cluster Sunyaev-Zel'dovich Effect Observations with the Expanded 13-Element Array}

\author{
Kai-Yang Lin\altaffilmark{1}, 
Hiroaki Nishioka\altaffilmark{1},
Fu-Cheng Wang\altaffilmark{2},
Chih-Wei Locutus Huang\altaffilmark{1},
Yu-Wei Liao\altaffilmark{1},
Jiun-Huei Proty Wu\altaffilmark{2},
Patrick M. Koch\altaffilmark{1},
Keiichi Umetsu\altaffilmark{1},
Ming-Tang Chen\altaffilmark{1},
Shun-Hsiang Chan\altaffilmark{2},
Shu-Hao Chang\altaffilmark{1},
Wen-Hsuan Lucky Chang\altaffilmark{2},
Tai-An Cheng\altaffilmark{2},
Hoang Ngoc Duy\altaffilmark{1},
Szu-Yuan Fu\altaffilmark{2},
Chih-Chiang Han\altaffilmark{1},
Solomon Ho\altaffilmark{3},
Ming-Feng Ho\altaffilmark{2},
Paul T.P. Ho\altaffilmark{1,4},
Yau-De Huang\altaffilmark{1},
Homin Jiang\altaffilmark{1},
Derek Y. Kubo\altaffilmark{3},
Chao-Te Li\altaffilmark{1},
Yu-Chiung Lin\altaffilmark{2},
Guo-Chin Liu\altaffilmark{5},
Pierre Martin-Cocher\altaffilmark{1},
Sandor M. Molnar\altaffilmark{1},
Emmanuel Nunez\altaffilmark{3},
Peter Oshiro\altaffilmark{3},
Shang-Ping Pai\altaffilmark{2},
Philippe Raffin\altaffilmark{3},
Anthony Ridenour\altaffilmark{3},
Chia-You Shih\altaffilmark{3},
Sara Stoebner\altaffilmark{3},
Giap-Siong Teo\altaffilmark{2},
Jia-Long Johnny Yeh\altaffilmark{2},
Joshua Williams\altaffilmark{3},
and Mark Birkinshaw\altaffilmark{6} 
}

\altaffiltext{1}{Academia Sinica Institute of Astronomy and Astrophysics, P.O. Box 23-141, Taipei, Taiwan 106; kylin@asiaa.sinica.edu.tw}
\altaffiltext{2}{Physics Department, National Taiwan University, Taipei, Taiwan 106; jhpw@phys.ntu.edu.tw}
\altaffiltext{3}{ASIAA Hilo Office, 645 N. A'ohoku Place, University Park, Hilo, Hawaii 96720, U.S.A.}
\altaffiltext{4}{East Asian Observatory, 660 N. A’ohoku Place, University Park, Hilo, Hawaii 96720, U.S.A.}
\altaffiltext{5}{Department of Physics, Tamkang University, 251-37 Tamsui, New Taipei City, Taiwan}
\altaffiltext{6}{HH Wills Physics Laboratory, University of Bristol, Tyndall Avenue, Bristol BS8 1TL, UK}


\begin{abstract}
 The Yuan-Tseh Lee Array for Microwave Background Anisotropy (AMiBA) is a co-planar interferometer array operating at a wavelength of 3\,mm to measure the Sunyaev-Zel'dovich effect (SZE) of galaxy clusters at arcminute scales. 
 The first phase of operation -- with a compact 7-element array with 0.6\,m antennas (AMiBA-7) -- observed six clusters at angular scales from $5\arcmin$ to $23\arcmin$. 
 Here, we describe the expansion of AMiBA to a 13-element array with 1.2\,m antennas (AMiBA-13), its subsequent commissioning, and cluster SZE observing program. 
 The most noticeable changes compared to AMiBA-7 are (1) array re-configuration with baselines ranging from 1.4\,m to 4.8\,m, allowing us to sample structures between $2\arcmin$ and $10\arcmin$, (2) thirteen new lightweight carbon-fiber-reinforced plastic (CFRP) 1.2\,m reflectors, and (3) additional correlators and six new receivers. 
 Since the reflectors are co-mounted on and distributed over the entire six-meter CFRP platform, a refined hexapod pointing error model and phase error correction scheme have been developed for AMiBA-13. 
 These effects -- entirely negligible for the earlier central close-packed AMiBA-7 configuration -- can lead to additional geometrical delays during observations. 
 Our correction scheme recovers at least $80 \pm 5\%$ of point source fluxes. We, therefore, apply an upward correcting factor of 1.25 to our visibilities to correct for phase decoherence, and a $\pm5\%$ systematic uncertainty is added in quadrature with our statistical errors. 
 We demonstrate the absence of further systematics with a noise level consistent with zero in stacked $uv$-visibilities. 
 From the AMiBA-13 SZE observing program, we present here maps of a subset of twelve clusters with signal-to-noise ratios above five. 
 We demonstrate combining AMiBA-7 with AMiBA-13 observations on Abell 1689, by jointly fitting their data to a generalized Navarro--Frenk--White (gNFW) model. Our cylindrically-integrated Compton-$y$ values for five radii are consistent with results from the Berkeley-Illinois-Maryland Array (BIMA), Owens Valley Radio Observatory (OVRO), Sunyaev-Zel'dovich Array (SZA), and the {\em Planck} Observatory.
 We also report the first targeted SZE detection towards the optically selected cluster RCS J1447+0828, and we demonstrate the ability of AMiBA SZE data to serve as a proxy for the total cluster mass.
Finally, we show that our AMiBA-SZE derived cluster masses are
 consistent with recent lensing mass measurements in the literature.
\end{abstract}


\section{Introduction}

The Yuan-Tseh Lee Array for Microwave Background Anisotropy (AMiBA)\footnote{http://amiba.asiaa.sinica.edu.tw}  is a platform-mounted interferometer operating at a wavelength of 3\,mm to study arcminute-scale fluctuations in the cosmic microwave background (CMB) radiation \citep{ho2009}. 
While the primary anisotropies in the CMB are measured to high accuracy over the whole sky, and the cosmological parameters are tightly constrained by the Wilkinson Microwave Anisotropy Probe \citep[WMAP,][]{hinshaw2013} and the {\em Planck} mission \citep{planckcollaboration2015xiii}, the arcminute-scale fluctuations resulting from secondary perturbations along the line of sight are less resolved.
One of the most prominent perturbations comes from galaxy clusters, which are the largest bound objects in the framework of cosmological hierarchical structure formation. Hot electrons that reside in the deep gravitational cluster potential scatter off and transfer energy to the cold CMB photons.  
This Sunyaev-Zel'dovich effect \citep[SZE,][see also \citealp{birkinshaw1999, carlstrom2002}]{Sunyaev1970, Sunyaev1972} is directly related to the density and temperature of the hot cluster gas, which traces the underlying dark matter distribution, and is complementary to information derived from X-ray, gravitational lensing, and kinematic observations of the galaxy cluster. The AMiBA observing wavelength of 3\,mm was chosen to minimize the combined contamination from both radio sources and dusty galaxies.

The SZE is nearly redshift-independent and is, thus, suitable to search for high-redshift galaxy clusters. 
To date, several extensive blind SZE surveys with catalogs of hundreds of clusters have been conducted by the Atacama Cosmology Telescope \citep[ACT,][]{hasselfield2013}, the South Pole Telescope
\citep[SPT,][]{bleem2015}, and the {\em Planck} mission \citep{planckcollaboration2015xxvii}. 
These surveys generally have arcminute scale resolutions and are, in some cases, able to resolve the pressure profile of the hot cluster gas. Compared to X-ray-selected cluster samples, SZE-selected samples tend to have shallower cores, hinting at a population at dynamically younger states that may have been under-represented in X-ray surveys \citep{planckcollaboration2011ix}.

The AMiBA is sited within the Mauna Loa Observatory at an altitude of $3,400$~m on the Big Island of Hawaii.  The telescope consists of a novel hexapod mount \citep{koch2009} with a carbon-fiber-reinforced polymer (CFRP) platform \citep{raffin2004, raffin2006, koch2008, huang2011}. Dual linear polarization heterodyne receivers \citep{chen2009}, powered by high electron-mobility transistor (HEMT) low-noise amplifiers (LNAs) and monolithic microwave integrated circuit (MMIC) mixers, are co-mounted on the steerable platform. A wideband analog correlator system \citep{li2010} correlates, integrates, and records the signal on the platform. 

The interferometer was built and operated in two phases. The first phase was comprised of seven close-packed $0.6$\,m antennas  \citep[hereafter AMiBA-7,][]{ho2009}. Scientific observations were conducted during 2007 - 2008, and six galaxy clusters in the redshift range of 0.09 to 0.32 were mapped with an angular resolution of 6\arcmin\citep{wu2009}.  
We carefully examined noise properties \citep{nishioka2009}, system performance \citep{lin2009}, and contamination by CMB and foreground sources \citep{liu2010} in our science data.
\citet{huang2010} derived the cylindrically-integrated Compton-$y$ parameter $Y_{2500}$ of the small sample and found consistent scaling relations with X-ray-derived temperature $T_e$, mass $M$ and luminosity $L_x$ (all within $r_{2500})$. 
\citet{liao2010} further tested recovering temperature $T_e$, gas mass $M_{gas}$, and total mass $M_{tot}$ of the cluster from AMiBA-7 data using different cluster gas models and found the results to be also consistent with values in the literature. Four of the six clusters also had Subaru weak-lensing observations, and \citet{umetsu2009} derived gas fraction profiles from the SZE and lensing mass data.

The second phase expanded the array to thirteen 1.2\,m antennas (hereafter
AMiBA-13) with a synthesized beam of 2\farcm5, enhancing the ability to
detect clusters at higher redshifts. \citet{molnar2010} tested the ability of 
AMiBA-13 to constrain the temperature distribution for non-isothermal
$\beta$-model mock observations of hydrodynamic simulations and
concluded that the scale radius of the temperature distribution can be
constrained to about 50\% accuracy. Scientific observations using AMiBA-13
started in mid-2011 and ended in late 2014. The targets observed with
AMiBA-13 include (a) the six clusters observed with AMiBA-7, (b)
high-mass clusters selected from the Cluster Lensing And Supernova survey with Hubble \citep[CLASH,][]{postman2012} sample, and (c) a small sample drawn from the Red-sequence Cluster Survey 2 \citep[RCS2,][]{gilbank2011}.

We will describe changes made to the instrument in Section~\ref{sec:change} and demonstrate the performance and systematics of the array in Section~\ref{sec:performance}. In Section~\ref{sec:observation}, we will detail our observing strategy, calibration, and data flagging. 
Section~\ref{sec:discussion} discusses radio source contamination and  interpretation of our cluster SZE data.
Our conclusions are summarized in Section~\ref{sec:conclusion}.
We adopt a flat $\Lambda$ cold dark matter ($\Lambda$CDM) cosmology with
$H_0=67.1$\,km\,s$^{-1}$\,Mpc$^{-1}$, $\Omega_\mathrm{m}=0.3175$, and
$T_\mathrm{CMB}=2.725$\,K \citep{PlanckCollaboration2014xvi}.

\section{Changes Compared to AMiBA-7}\label{sec:change}

To complete the 13-element array, six additional receivers were built with design and component specifications identical to the first seven receivers. All of the new receivers, except one, have noise temperatures around $55-75$~K which is comparable to the old receivers \citep{chen2009}, while the one exception shows a higher noise temperature at $85$~K. Additional correlators and intermediate frequency (IF) distribution networks were also built following the 7-element design \citep{li2010}. The new correlators are housed in the same enclosures on the platform that were previously only partially populated by the 7-element correlators. Table~\ref{characteristics} summarizes the changes between AMiBA-7 and AMiBA-13. The system performance is discussed in Section~\ref{sec:performance}.

\begin{figure*}[!htb]
\begin{center}
 \includegraphics[width=0.45\textwidth,angle=0,clip]{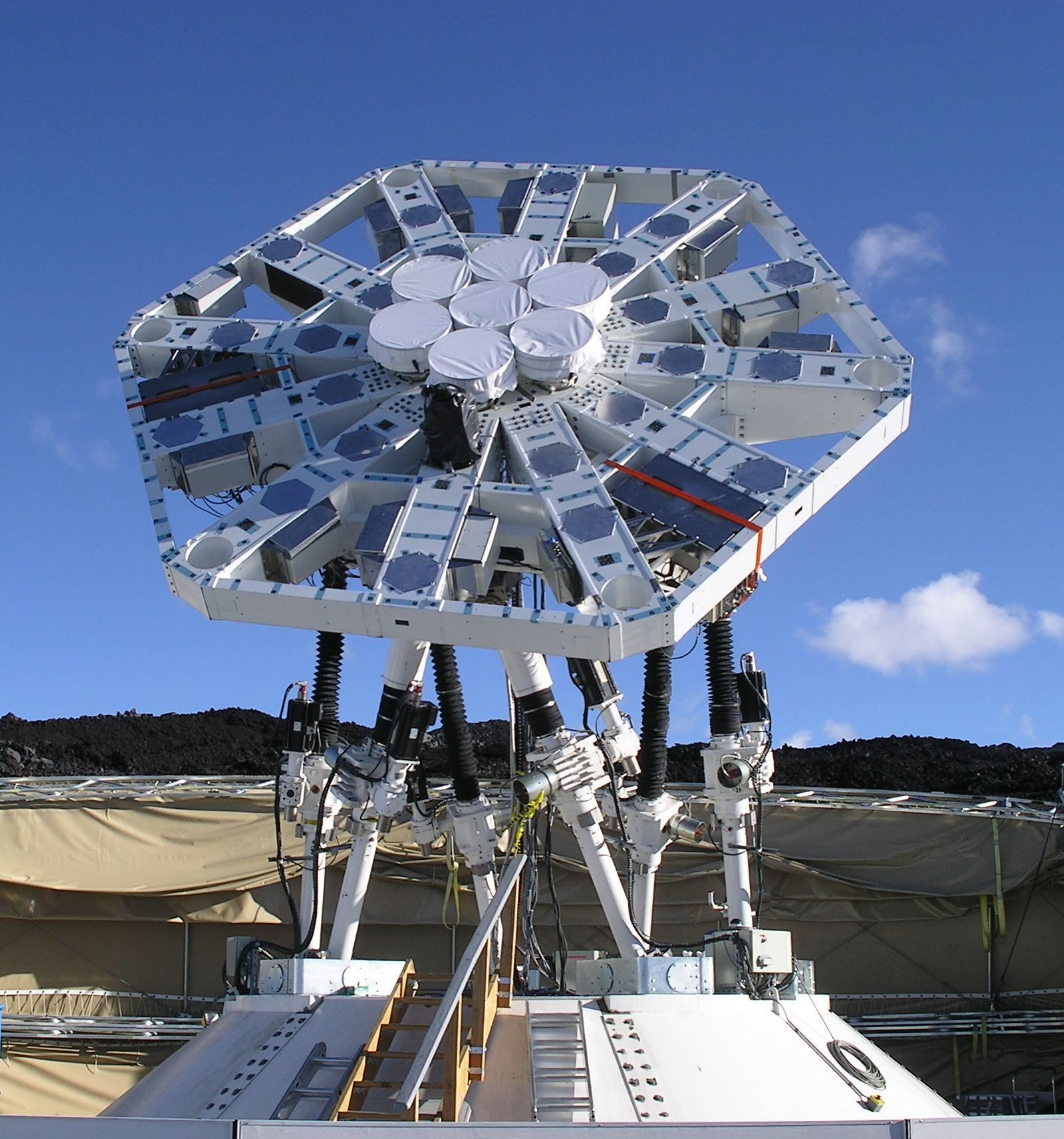}
 \includegraphics[width=0.45\textwidth,angle=0,clip]{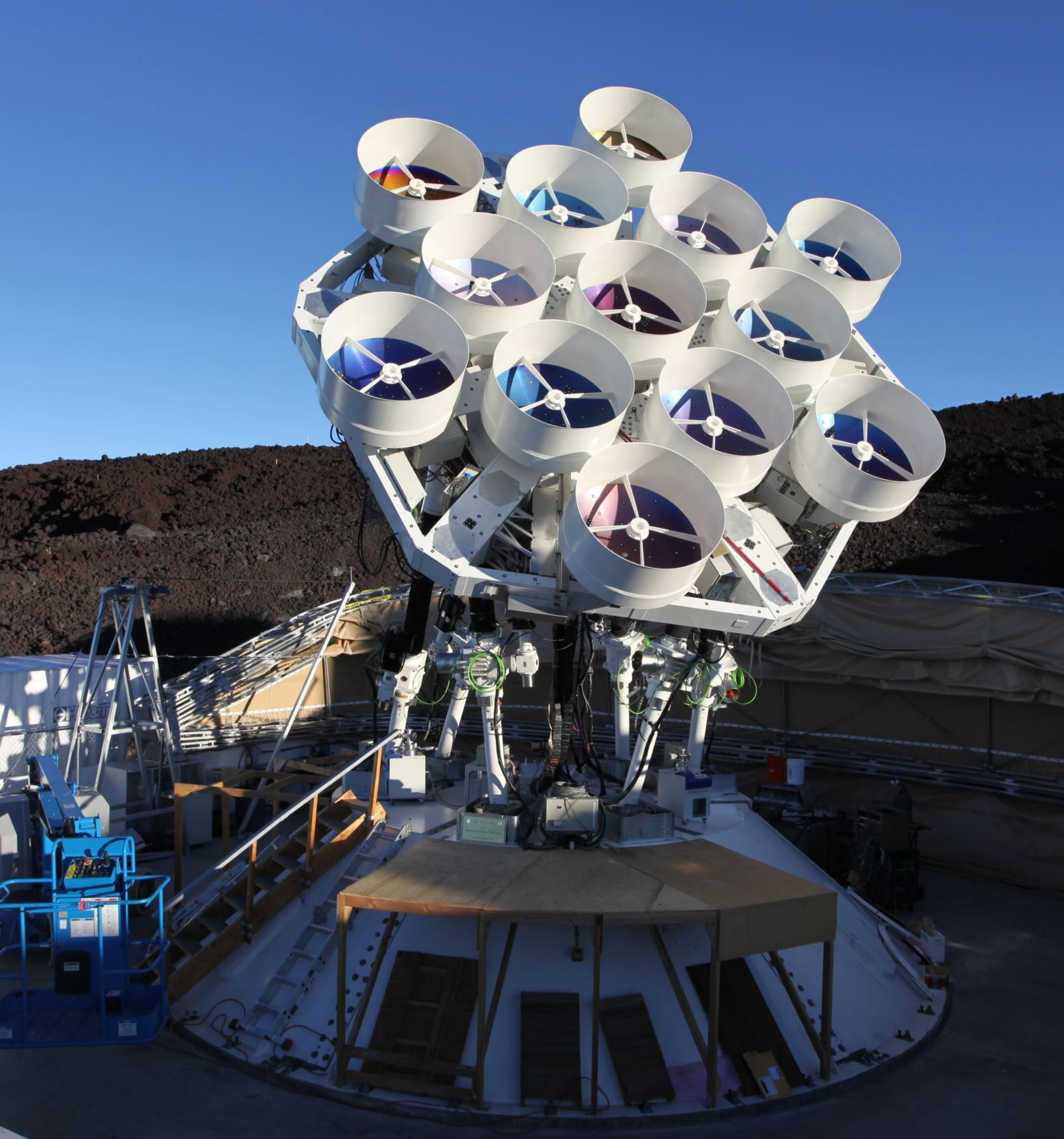}
\end{center}
 \caption{Pictures of the AMiBA-7 (left) and AMiBA-13 (right). The AMiBA-7 had 0.6\,m antennas in a closely-packed configuration with shortest spacing 0.6\,m. The AMiBA-13 has 1.2\,m antennas with a shortest spacing of 1.4\,m.
 Visible in the background is the retractable shelter that is closed when the telescope is not in operation.}
 \label{amiba_photos}
\end{figure*}

\begin{deluxetable}{lcc}
\tablecaption{Comparison of AMiBA-7 and AMiBA-13 \label{characteristics}}
\tablecolumns{3}
\tablehead{\colhead{} & \colhead{AMiBA-7} & \colhead{AMiBA-13} \\}

\startdata
Number of Antennas & 7 & 13 \\ 
Number of Baselines & 21 & 78 \\
Polarizations & XX \& YY & XX \& YY \\
Antenna Diameter (m) & 0.6 & 1.2 \\
Baseline Range (m) & $0.6 - 1.2$ & $1.4 - 4.8$ \\
Primary Beam (FWHM) & $22$\arcmin & $11$\arcmin\\
Synthesized Beam (FWHM) & $6$\arcmin & $2\farcm5$ \\
Elevation Limit (deg) & 30 & 40\tablenotemark{a} \\
Point Source Sensitivity (mJy/$\sqrt{\mathrm{hr}}$) & $64$ & $8$\\
Extended Source Sensitivity ($\mu$K/$\sqrt{\mathrm{hr}}$) & $238$ & $174$
\enddata

\tablenotetext{a}{The CFRP platform was repaired prior to the AMiBA-13 observations. After the repair, we limited the operating range of the hexapod as a safeguard.}
\end{deluxetable}

\subsection{Array Configuration}
\begin{figure*}
 \centering
 \includegraphics[width=2.1in]{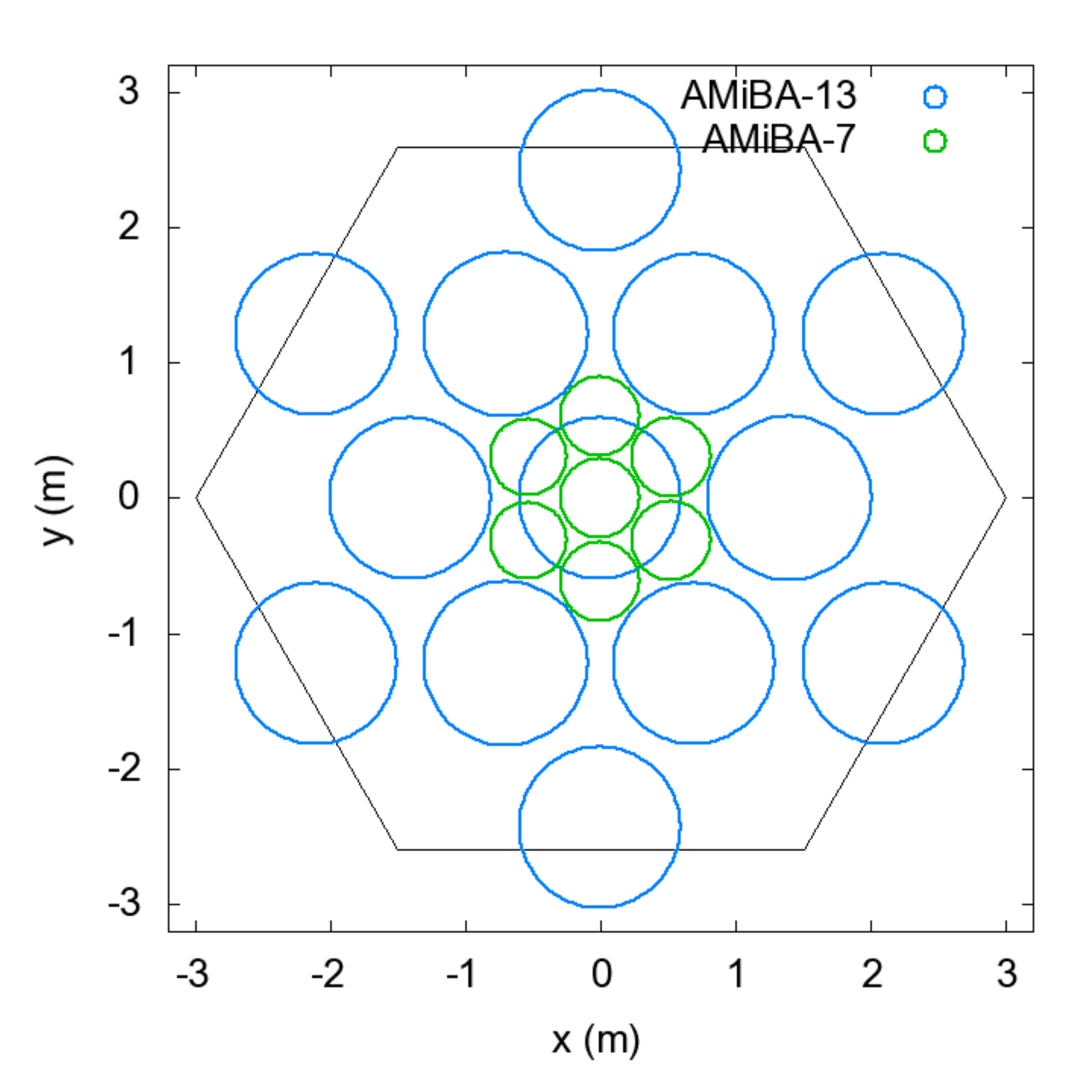}
 \includegraphics[width=2.4in]{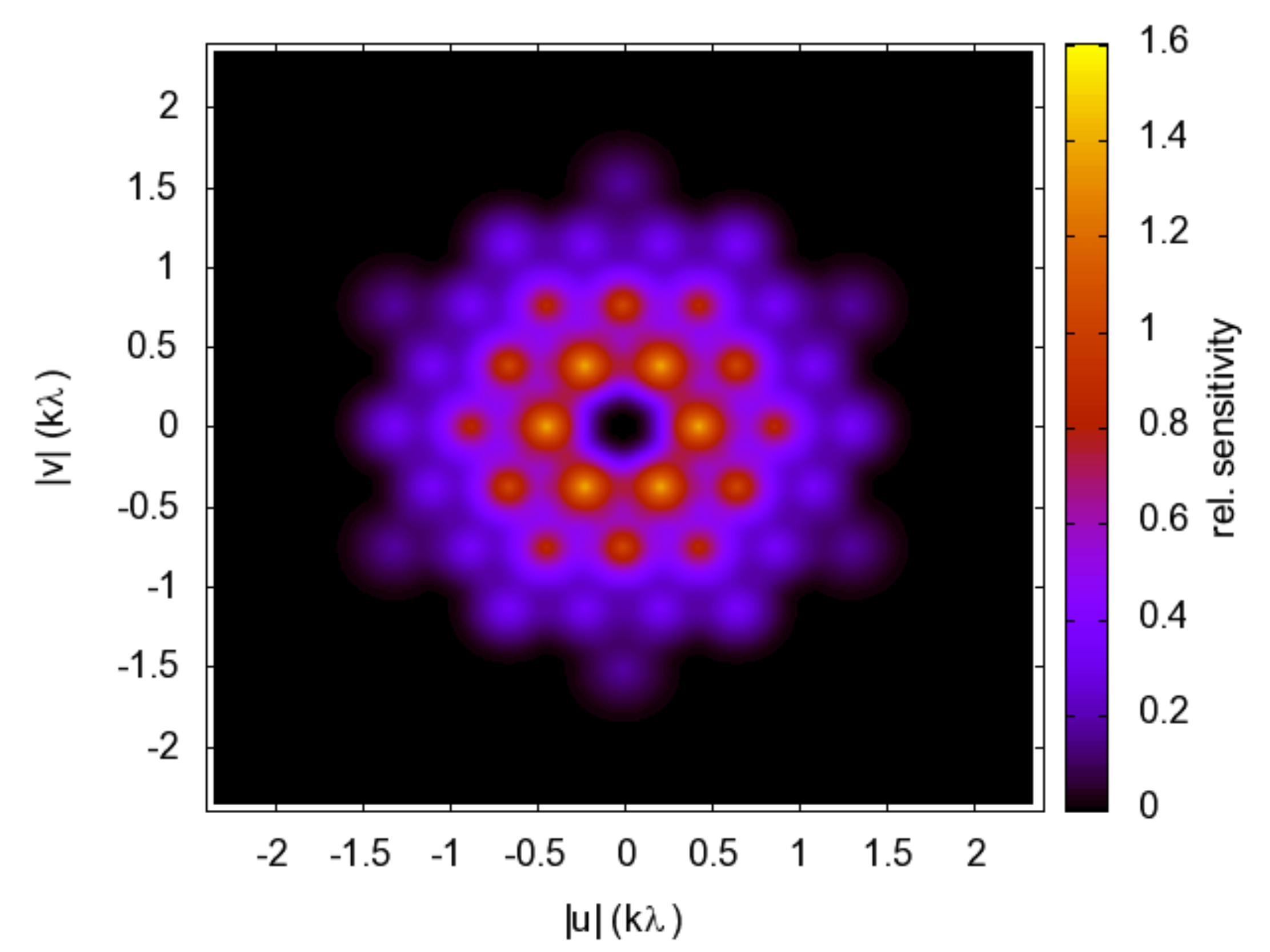}
 \includegraphics[width=2.4in]{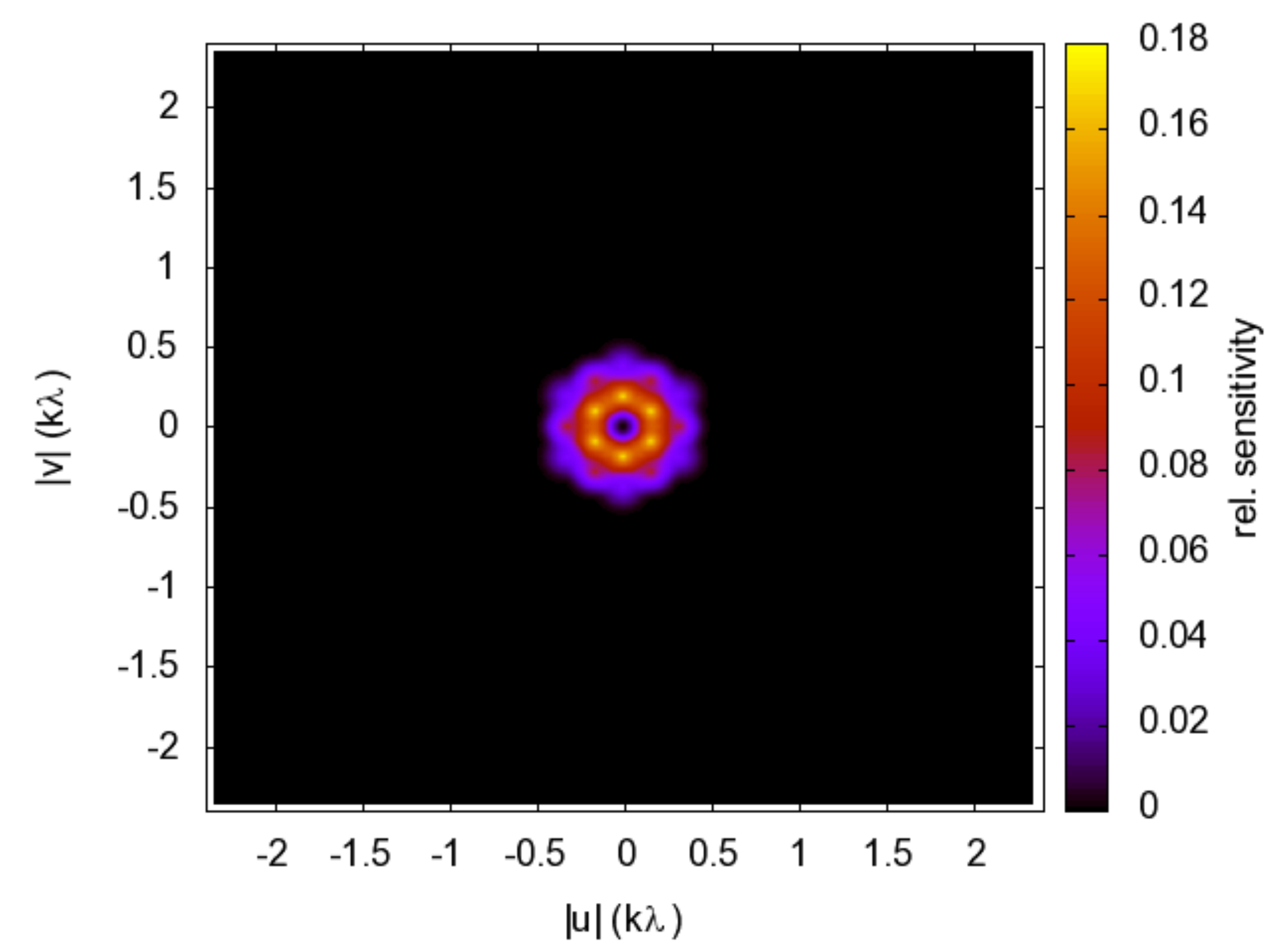}
 \caption{Left panel: array configuration in platform coordinates. The larger circles represent the dish sizes and locations of the 13-element array. For comparison, the smaller circles in the center of the platform indicate where the AMiBA-7 antennas were. Large hexagon identifies the edge of the platform. Middle and right panel: instantaneous $uv$-coverage of AMiBA-13 and AMiBA-7, respectively, with their relative sensitivities in color scales.}
 \label{array-config}
\end{figure*}

Figure~\ref{amiba_photos} shows the AMiBA-7 and AMiBA-13. Equipped now with larger antennas, six of the original receivers were relocated further out on the platform. Similar to the 7-element array, the 13-element array has a hexagonally close-packed configuration. Two choices of shortest baseline lengths are available for the $1.2$\,m diameter reflectors, namely $1.2$\,m and $1.4$\,m. We chose the configuration with the $1.4$\,m separations, which has about a 10 times lower cross-talk between neighboring dishes \citep[a measured $-135$~dB on the 1.4\,m versus an estimated $-125$~dB on the 1.2\,m baseline;][]{koch2011}. Figure~\ref{array-config} shows the array configuration in the platform coordinate system and the corresponding instantaneous $uv$-coverage assuming a single frequency of $94$~GHz. 

\begin{figure}
\begin{center}
\plotone{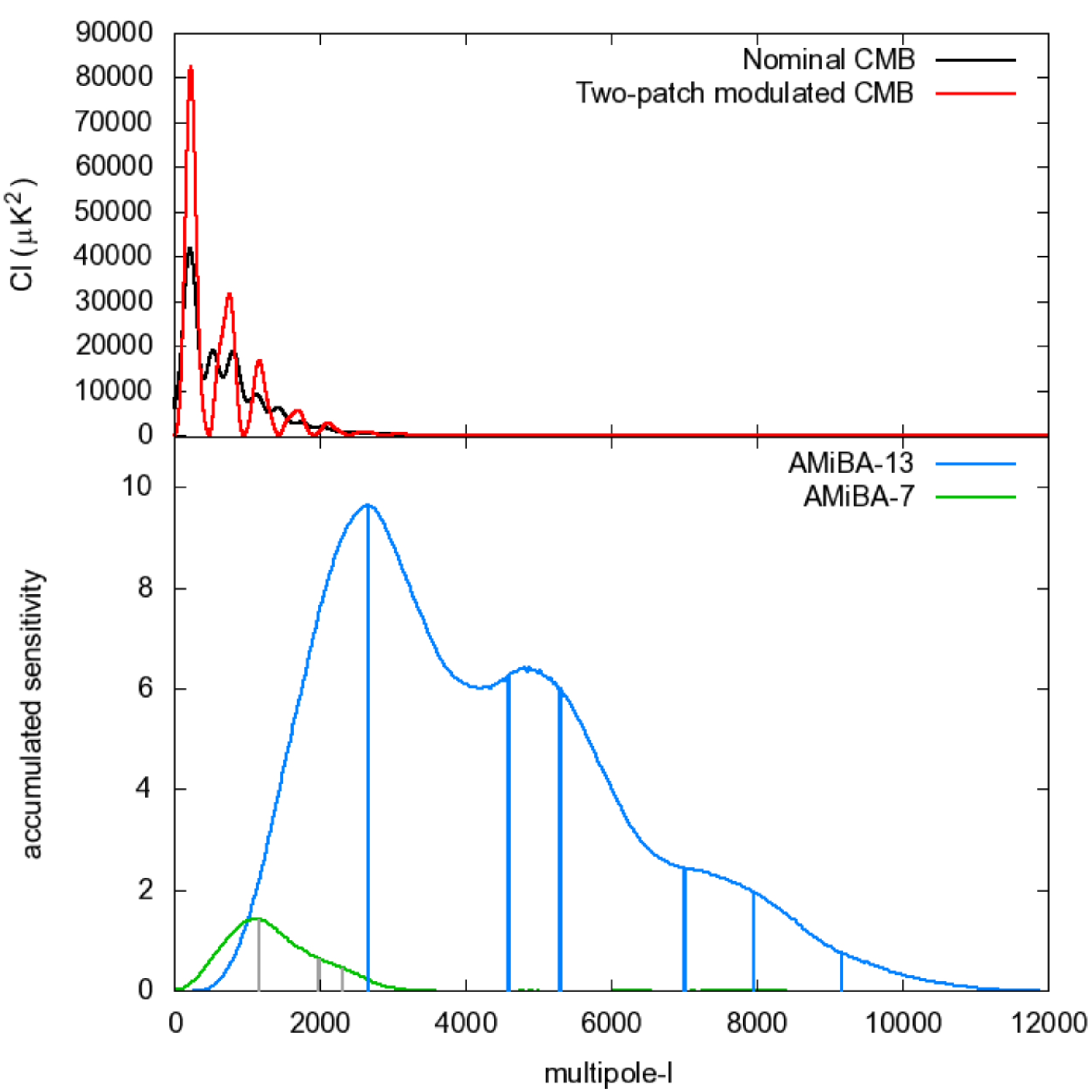}
\caption{Top panel: CMB primary anisotropy spectrum (black) and modulated spectrum after a two-patch subtraction with a separation of 45\arcmin (red). Bottom panel: accumulated sensitivity in $uv$-space, summed in each annulus of multipole-$l$, for AMiBA-13 and AMiBA-7, respectively. The vertical lines denote the centers of the baselines. The conversion from baseline length to multipole-$l$ assumes a single frequency of $94$~GHz.}
\label{uv-sensitivity}
\end{center}
\end{figure}

Compared to the close-packed configuration, the 1.4\,m separation between dishes also helps to suppress the primary CMB leakage, in favor of cleaner cluster SZE observations. Given the angular power spectrum $C_l$ of the CMB, we can estimate the rms fluctuation that is picked up by a baseline following the steps outlined in \citet{liu2010} as
\begin{multline}
\langle V^2(u_b,v_b)\rangle = \int du dv \tilde{A}^2(u_b-u, v_b-v) C_l\\
\times (1 - \cos[2\pi (u\Delta x + v\Delta y)])\,,
\end{multline}
where $(u_b, v_b)$ corresponds to the center of a particular baseline. The modulating factor $(1-\cos[2\pi (u\Delta x + v\Delta y)])$ comes from subtracting the trailing patch from the target patch, with a sky separation of $(\Delta x, \Delta y)$. The 'two-patch' observation scheme is discussed in more detail in Section~\ref{sec:two-patch}.
The CMB leakage is stronger for shorter baselines. With the 1.2\,m dishes, we estimate that the $1.2$\,m baseline has an rms fluctuation of $\sim 20$~mJy from the CMB.
The $1.4$\,m baseline has about a factor of two lower level of rms fluctuation, at $\sim 11$~mJy.
By comparison, the AMiBA-7 configuration ($0.6$~m dishes separated by $0.6$~m) had fluctuations of roughly $170$~mJy at the $0.6$\,m and $24$~mJy at the longer $1.2$\,m baseline.
Figure~\ref{uv-sensitivity} shows the unmodulated CMB power spectrum and a two-patch modulated spectrum with a typical patch separation of $\Delta x = 45\arcmin$ and $\Delta y = 0$. Also shown are the footprints in $uv$-space, accumulated in each annular bin, as a function of multipole-$l$ for both AMiBA-13 and AMiBA-7 to show where the sensitivity lies.

\subsection{$1.2$\,m Reflector}
The design of AMiBA has the 1.2\,m diameter f/0.35 Cassegrain reflector \citep[Table~\ref{dish_char},][]{koch2011} mounted onto the top plate of the receiver assembly while the receiver itself is directly attached onto the CFRP platform. This avoids having the reflector directly on the platform and eliminates any additional misalignment between reflector optical axis and the receiver feed. 
A detailed Finite-Element Analysis (FEA) of the entire CFRP reflector helped to reduce the weight to a final 25\,kg from an original prototype that weighed almost 50~kg.  An equally stiff antenna made out of aluminium would be at least 35~kg. CFRP was chosen as a lightweight material in order to minimize torque and structural deformations under various load cases. Excellent structural behavior is found from the FEA for the  lightweight CFRP reflector.  
Thermal load cases introduce tilts in the optical axis of only around 1\arcsec. Strong winds of 10\,m/s lead to tilts between 0.5$\arcmin$ and $\sim 1\arcmin$ depending on pointing elevation. 
Deformation under gravity is largest, at about 1 arcmin, at the lowest operating elevation of $30^\circ$. All these tilts are within 10\% of the 11-arcmin full width at half maximum (FWHM) of the antennas, and introduce less than 3\% of loss.

Primary and secondary mirror surfaces were measured after manufacturing. Fitting for a primary paraboloid and a secondary hyperboloid shows random surface rms errors of about 30~$\mu$m and 15~$\mu$m, respectively. Following \citet{ruze1966}, these small manufacturing errors keep the surface efficiencies at 98.5\% and 99\% for primary and secondary at a frequency of 94~GHz. 
After assembly, the resulting alignment errors are between 50~$\mu$m and 100~$\mu$m which reduce the aperture efficiency by less than 1\%. 
The final antenna aperture efficiency, composed of a series of independent factors --- feed-horn illumination efficiency, secondary mirror and support leg blockage efficiency, surface roughness efficiency, feed spillover efficiency, focus error efficiency, cross-polarization efficiency, diffraction and ohmic losses --- is estimated to be about 0.6, dominated by the feed spillover efficiency of $<0.78$ \citep{koch2011}. 
Both primary and secondary mirrors are aluminum-coated in vacuum with a homogeneous aluminum layer of about 2\,$\mu$m. Immediately after the aluminum sputtering an 0.3-$\mu$m $\rm Ti O_2$ layer is added for protection against oxidation, abrasion, peeling off and accidental pointing towards the Sun.

The antenna beam pattern was measured in the far field by scanning a fixed thermally stabilized 90~GHz source. The antenna response was previously simulated including the complete feed horn-antenna system with a corrugated feed horn with a semiflare illumination angle of $14^{\circ}$ with a parabolic illumination grading with a -10.5 dB edge taper. Our measurement confirms the simulated main lobe with an $11\arcmin$ FWHM. The location of the first side lobe is confirmed at $18\arcmin$, while its level is about 2--4\,dB higher than expected, peaking around -16 to -18\,dB \citep{koch2011}.

Finally, close-packed antenna configurations can cause cross-talk problems in weak cluster SZE and CMB signals. Our estimated tolerable level of cross-talk is around -127\,dB \citep{koch2011,padin2000}. 
In order to minimize this signal, a cylindrical shielding baffle is added to the reflector, similar to the earlier 0.6\,m antennas. Effectively reduced cross-talk signals were, indeed, verified on the operating AMiBA  platform where one antenna was used as an emitter with a $\sim 10$\,dBm source while in neighbouring antennas with different baseline lengths the weak cross-talk signal was measured with a spectrum analyzer. 
On the shortest 1.4\,m baseline, cross-talk signals of $\sim -135$\,dB and $\sim -115$\,dB were measured with and without the shielding baffle, respectively.  A further reduced signal of $\sim -145$\,dB was found when the separation was increased to 2.8\,m \citep{koch2011}. For baselines longer than 2.8\,m, the cross-talk is below our detection limit of $-145$\,dB. Besides shielding the reflectors, an 
additional measure to further reduce unwanted scattered signals was taken by optimizing the shape of the secondary mirror support leg structure. A triangular roof is added on the lower side of the feed leg to terminate scattered light on the sky (Lamb 1998, ALMA Memo 195; Cheng et al. 1998, ALMA Memo 197)\footnote{Main ALMA Memo Series: http://library.nrao.edu/alma.shtml}.
As a result, cross-like features in the measured beam patterns at the locations of the feed leg are reduced to an amplitude of about 1\,dB compared to more apparent peaks around 3\,dB in the earlier 0.6\,m antennas.
Additional details of the 1.2\,m Cassegrain antenna can be found in \citet{koch2011}.

\begin{deluxetable}{lc}
\tablecaption{Characteristics of $1.2$\,m Reflector \label{dish_char}}
\tablecolumns{2}
\tablehead{\colhead{Parameters} & \colhead{Values} \\}
\startdata
Reflector Type & Cassegrain \\ 
Primary Diameter & $1.2$\,m \\
FWHM of Beam Pattern & $11$\arcmin \\
Primary Focal Ratio & $0.35$ \\
Secondary Diameter & $0.19$\,m \\
Effective Focal Ratio & $2.04$ \\ 
Final Focal Position & At vertex of primary \\
Illumination Edge Taper & $-10.5$~dB \\
Antenna Efficiency & $60$~\% \\
Height of Baffle above Secondary Edge & $0.36$\,m
\enddata

\end{deluxetable}

\section{Commissioning}\label{sec:performance}

\subsection{Delay Correction\label{sec:delay}}
After new receiver units and IF distributions are installed, the path lengths need to be adjusted so that signals from within the field of view can be adequately sampled by our lag-correlator. The path length difference, excluding the geometric delays, is referred to as the instrumental delay throughout this work. 
The lag-correlator has four mixers, each separated by $25$~ps delay, corresponding to the Nyquist sampling rate for a bandwidth of $20$~GHz \citep{li2010}. The accessible delay range is, thus, around $\pm50$~ps. 
In the case of AMiBA-13, to allow a $5$\,m baseline to observe the entire $11$\arcmin field-of-view, the instrumental delay should be controlled within $\pm22$~ps. Following the method outlined in \citet{lin2009}, instrumental delays were measured  on the platform, from feed horn to correlator, using two methods that will be described below. Cables were then inserted into the IF path in order to compensate for the delays.

In the first method, we set up a broadband noise source simultaneously emitting toward two receivers (without the $1.2$~m dishes) at a time. We then moved the noise source along the baseline and recorded the fringe as a function of the geometric delay. The difference between the center of the baseline and the position where the fringe peaked, marked the instrumental delay between the pair of receivers. Additionally, by Fourier transforming the fringe with respect to the geometric delay, we could also obtain a measure of the bandpass response function, modulated by the spectral shape of the noise source. The bandpass of the new baselines was, indeed, similar to the ones previously measured for AMiBA-7, with a comparable effective bandwidth around $7-13$~GHz.

The above mentioned method is powerful but time consuming. It was performed once on all functioning baselines to establish a reference bandpass functions. During subsequent iterations of delay-tuning, we relied on the second method, in which we scanned the array, without the dishes, across the Sun and recorded the fringes. Without the dishes, the FWHM of the feed horn is about $20^\circ$ \citep{koch2011}. The resultant fringe is a convolution of the point-source fringe with the brightness distribution of the Sun, which we assumed to be a circular top-hat function. 
Depending on the instrumental delay $\tau^{\mathrm inst}$, the measured fringe peak can appear before or after the expected one  with an angular offset $\theta$ described by
\begin{equation}
 \theta B\cos\alpha = c\tau^{\mathrm{inst}}\,,
\end{equation} 
where $B\cos\alpha$ is the projected baseline length along the scanning direction. $c$ is the speed of light.
However, for baselines that are almost perpendicular to the scan, the fringes would be too slow, $\theta$ too big, and $\tau^{\mathrm{inst}}$ thus poorly determined. 
Therefore, for each measurement we scanned the Sun in four directions, namely along the right ascension (R.A.), along the declination (decl.), and two directions in between, so that all baselines had a sufficiently high fringe rate in a few of the scans. In this way, we could probe a large delay range for each baseline.

For each polarization ($XX$ or $YY$), two coaxial cables from the same IF channel are fed to the correlator rack, with one feeding a "row" of correlators from the "front" and the other feeding a "column" from the "back". 
We simplify the measured delays as the difference of electrical lengths from the IF channels $L_k$, or
\begin{equation}
\tau^{\mathrm{inst}}_{i} = \left[\delta_{f(i)k} - \delta_{b(i)k}\right]
L_k \equiv D_{ik}L_k\,,\label{eq:measure-tau}
\end{equation}
where $\delta_{f(i)k}$ and $\delta_{b(i)k}$ are Kronecker deltas that select the IF paths corresponding to the "front" $f(i)$ and "back" $b(i)$ of the $i$-th correlator.
For AMiBA-13, there are 78 delays measured for each polarization ($i \in [1,78]$).
Since the cables connecting to the "front" are independent from the ones connected to the "back" of the correlator rack, their electrical lengths are solved independently. 
There are, thus, 24 electrical lengths to solve ($k \in [1,24]$), in which 12 "fronts" connect to antenna 1 through 12 and 12 "backs" connect to antenna 2 through 13, respectively.
We also note that there are power dividers and cables inside the correlator rack in order to further distribute the signals to the 78 correlators. The electrical lengths of these paths, while short, may not be equal. These delays are included in the measured $\tau^{\mathrm{inst}}_{i}$ but they are not explicitly represented in Equation~(\ref{eq:measure-tau}) because they cannot be adjusted. This is likely the major source of our residual delays.

We used the LAPACK routine {\sc sgesvd} to perform a singular value
decomposition (SVD) of the sparse matrix $D_{ik}$, zeroing singular
values smaller than $10^{-6}$. We then constructed the pseudo-inverse
matrix $D^{-1}_{\mathrm{SVD}}$ to find the estimated electrical lengths $\tilde{L}_k$ through
\begin{equation}
    \tilde{L}_k = \left(D^{-1}_{\mathrm{SVD}}\right)_{ki} \tau^{\mathrm{inst}}_{i}\,.
\end{equation}

Adjustments to the electrical lengths were done by installing short cables of corresponding lengths to each of the IF paths. The measurement and adjustment process was iterated several times until the residuals could not be further improved.
Because of measurement uncertainties, imperfection of the pseudo-inverse matrix reconstruction, and the additional delays mentioned above, exact solutions are not possible. Consequently, some of the correlators ended up having much larger residual delays than the others. The rms scatter of these residual delays is $22$\,ps, while the maximum of these residuals can be up to twice this amount. However, we note that even correlators with the largest residual delays are still capable of detecting a source that is not too far off the pointing center of the platform, which is also the phase center after calibration. 
Correlators that have large delays can be very inefficient when additional geometrical delays are present. In this case, they will be flagged. They amount to about 10~\% of the total number of correlators. Contributions to the geometrical delays come from  target offsets from the phase center (offset observations, extended objects, or platform pointing errors) and the platform deformation. Both effects are discussed in the following sections. The overall effect on phase error is discussed in Section~\ref{sec:phase-error}.

\subsection{Platform Deformation \label{sec:deform}}
\begin{figure}
 \plotone{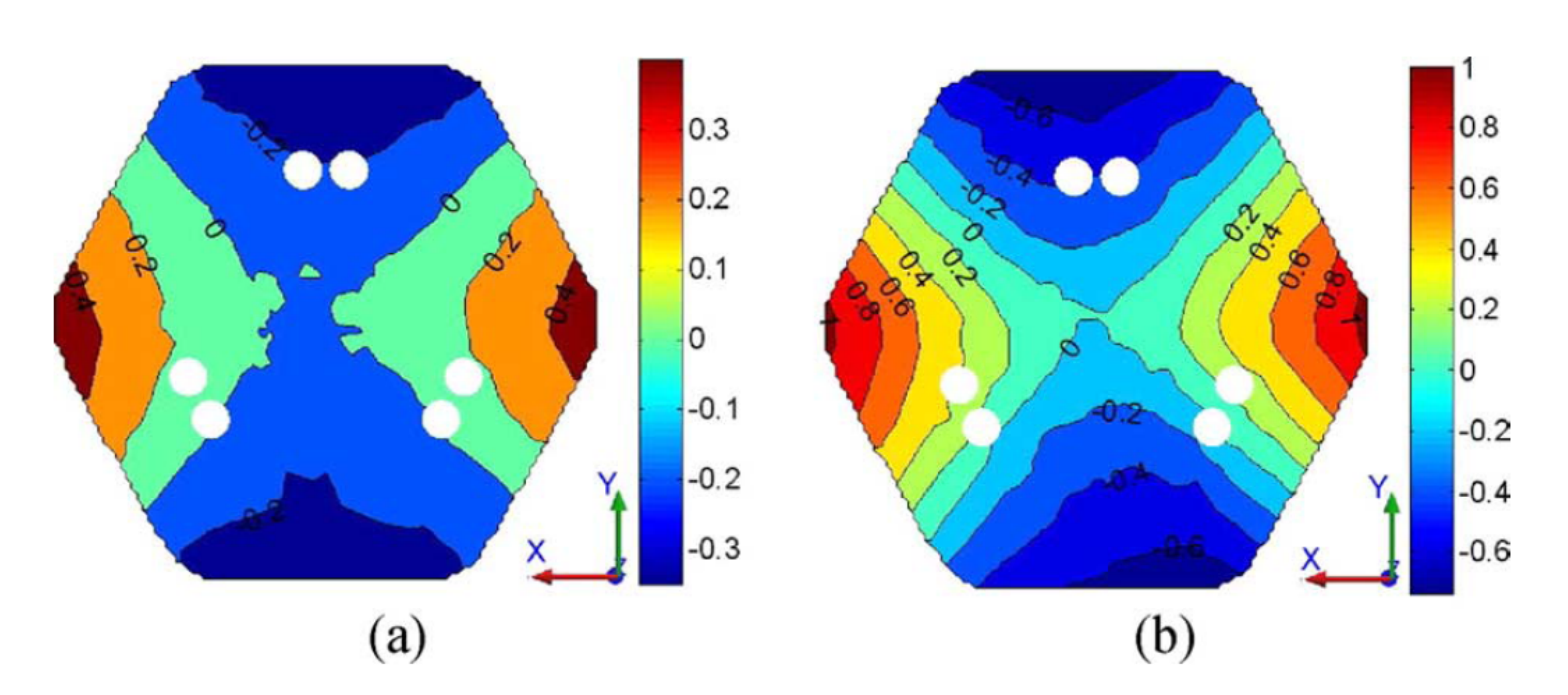}
 \caption{Photogrammetry-measured platform deformation with respect to zenith position at different elevations (a) $el=60^{\circ}$ and (b) $el=30^{\circ}$, both with $az=0^{\circ}$ and $hexpol=0^{\circ}$. Color scale is in units of mm. The three pairs of white filled circles indicate the locations of the u-joints on the under-side of the platform. \label{saddle}}
\end{figure}

AMiBA uses a CFRP platform as a lightweight solution to host the entire array and the correlator system on top of the hexapod mount \citep{raffin2004}. 
Ideally, the universal joints (u-joints) of the hexapod should be held
rigidly in two planes (one for the upper u-joints, and one for the lower
u-joints). While the lower u-joints are fixed to the supporting cone, a
steel interface ring beneath the platform is used to hold the upper
u-joints. However, the interface ring was found to be not rigid enough to entirely absorb the differential forces
from the six heavy legs, in addition to the gravitational forces from the
platform and equipment at tilted orientations. As a result, the platform
deforms as a function of its orientation, including the pointing in azimuth ($az$) and elevation
($el$), and a rotation along the pointing axis (hereafter referred to as
$hexpol$). The deformation was measured in several photogrammetry
campaigns, sampling the ($az$,$el$,$hexpol)$-parameter space with
hundreds of photos of the platform (with several hundred reflective
targets). The results show that the deformation is, indeed, repeatable within the measurement uncertainty 
of $\sim50$\,$\mu$m in rms. Figure~\ref{saddle} shows an example of the deformation pattern at two different elevations. Generally, the deformation along the pointing direction appears to be saddle-like, with a functional form $\sim A\cdot(x^2-y^2)$, where $A$ is an amplitude and $x$ and $y$ are coordinates in a platform reference frame. Across the sample of  photogrammetry-measured positions, the deformation shows characteristic properties: deformations grow with radius, reaching maximum values at the edge of the platform. Moreover, the deformation amplitude increases with lower elevation and larger platform $hexpol$ rotations. The saddle pattern rotates with a roughly constant amplitude as the hexapod changes its azimuth pointing \citep{liao2013, huang2008, koch2008}. Relative to the neutral orientation (pointing toward zenith), the maximum normal deformation at the edge of the platform increases with decreasing elevation and can reach a value as high as 1.5\,mm \citep{huang2008, koch2008}, or half of 
our observing wavelength.

In order to define a tolerance for deformation, for simplicity, we model the deformation-induced phase error with a Gaussian random distribution and an rms error of $\sigma$. The coherence efficiency is, thus, $\eta_{c} = \exp(-\sigma^2/2)$. When $\sigma=2\pi/20$ -- which corresponds to an rms deformation of a wavelength $\lambda/20$ ($\lambda=3$\,mm) or roughly 150\,$\mu$m at our frequency -- the efficiency is about $95\%$. We have set this as our nominal tolerance level for residual deformation errors.

\citet{huang2011} further investigated the platform deformation using FEA. They confirmed the insufficient stiffness to be the dominant cause of our deformation problem. However, further investigation also determined that given our constraints on load capacity of the hexapod mount and the existing shelter size and dimensions (Figure~\ref{amiba_photos}), neither a space frame platform built with steel nor a space frame built with CFRP can provide the required stiffness to keep the deformations within the 150\,$\mu$m rms error tolerance across all observing orientations. Therefore, we kept the platform and decided to remedy the problem through modelling and post-processing of the collected data.

The major effect of the platform deformation on a co-planar interferometry observation is the addition of a pointing-dependent geometric delay. Since AMiBA uses an analog delay-correlator (with four lags) to generate two spectral channels that cover $86-102$~GHz in the radio frequency (RF), the geometric delay induces a phase change at the center of each frequency band and a band-smearing effect in each 8~GHz channel. The phase change can be calculated and removed if we know how the platform deforms as a function of pointing. In order to model the band-smearing effect, on the other hand, precise knowledge of the source spectrum and the bandpass response is required for each baseline. However, \citet{lin2009} showed that even though the bandpass could be characterised to a spectral resolution of $\sim0.1$~GHz, it failed to model the band-smearing effect. Therefore, in the current work, the band-smearing effect remains a systematic factor that reduces the peak flux of a point source by up to 10~\% at the lowest elevation of $el=40^{\circ}$.

\citet{liao2013} summarize in detail the method that we use to measure and model the platform deformation. Moreover, they demonstrate its effectiveness when applied to planets and a few radio point sources. 
In short, we find changes in the large-scale deformation pattern -- measured by photogrammetry across the entire platform -- to correspond well to changes in the local tangent of the surface. 
Therefore, it is possible to use and correlate two optical telescopes (OTs; mounted on two different locations on the platform) and perform an all-sky pointing error analysis to find out the relative change between the local tangents of the two OTs.
This information is then used to solve for an all-sky deformation model. Following the trajectory of Jupiter, it was verified that geometrical delays predicted by this model match the measurements to within $\pm0.2$\,mm. 
Considering the decoherence effect due to the phase error, it was further shown that when applying this deformation correction, we are able to recover at least 95~\% of the remaining flux, after considering the band-smearing loss mentioned above, compared to a mere 75~\% recovery without the deformation correction. 

It is important to note that, although the platform deformation has been the same since the beginning of the AMiBA project, the earlier AMiBA-7 observations utilized only a small central part of the platform (Figure~\ref{amiba_photos}) where the deformation-induced geometrical delay is within 150~$\mu$m in rms and the induced loss is less than 5\%. The earlier AMiBA-7 science results are, thus, unaffected by the platform deformation problem.

\subsection{Antenna Alignment}
Another effect of the platform deformation that impacts a co-planar
array is the changing alignment between antennas. These alignments were
measured by scanning a planet (Jupiter or Saturn) along the R.A. and
decl. directions and recording their fringes. 
On top of the intrinsic fringe envelope described in Section~\ref{sec:delay}, the fringe envelope for each baseline is modulated by the combined beam attenuation of the two antennas that form its baseline. 
Note that longer baselines have faster fringes and narrower intrinsic envelopes compared to the primary beam. Therefore, any residual instrumental delay may shift the position of the intrinsic envelope and bias the measurement of the beam center. Such baselines are then flagged and not used to solve for the misalignments. 
On the other hand, baselines with shorter {\em projected} lengths along the scanning direction have slower variations in the intrinsic fringe envelopes, and the primary beam attenuation dominates the fringe envelope. 
We then fit a Gaussian to the fringe envelope to determine the offset of the combined beam center along the scanning direction. 
In some cases, baselines with too slow a fringe rate, showing no more than two fringes within the primary beam, are also discarded because their envelopes are distorted by under-sampling in delay by the lag-correlator.
Lastly, since the combined primary beam attenuation should affect both polarization in the same way, we look for and flag any inconsistency between the $XX$ and $YY$ beam center measurements that may indicate an excessive instrumental delay for one polarization or other faults in the fringe records.

The combined beam center of one baseline can be approximated by $\mathbf{C} = (\mathbf{P}_a + \mathbf{P}_b) / 2$ as in the case of Gaussian beams, where $\mathbf{P}_a$ and $\mathbf{P}_b$ denote beam centers of antenna $a$ and $b$, respectively. Our two orthogonal scans project the beam centers onto two sets of measurements that can be solved independently. Let $x$ denote the component projected along either R.A. or decl., we then have
\begin{equation}
C_{xi} = \frac{1}{2}\left[
 \delta_{a(i)k} + \delta_{b(i)k}
\right] P_{xk} \equiv M_{ik} P_{xk}\,,
\end{equation}
where the subscript $i$ runs through the subset of unmasked measurements out of the 78 baselines, and the subscript $k$ denotes the 13 independent antennas. 
Similar to what was done in Section~\ref{sec:delay} for delay measurements, we invert the sparse matrix $M$ by the SVD technique and a solution can be found for each scan.
Uncertainties in determining the fringe envelope and its centroid position are propagated through the SVD-based matrix inversion to the alignment solutions. We estimate these uncertainties to be about 0\farcm5 in rms.

\begin{figure*}
 \plotone{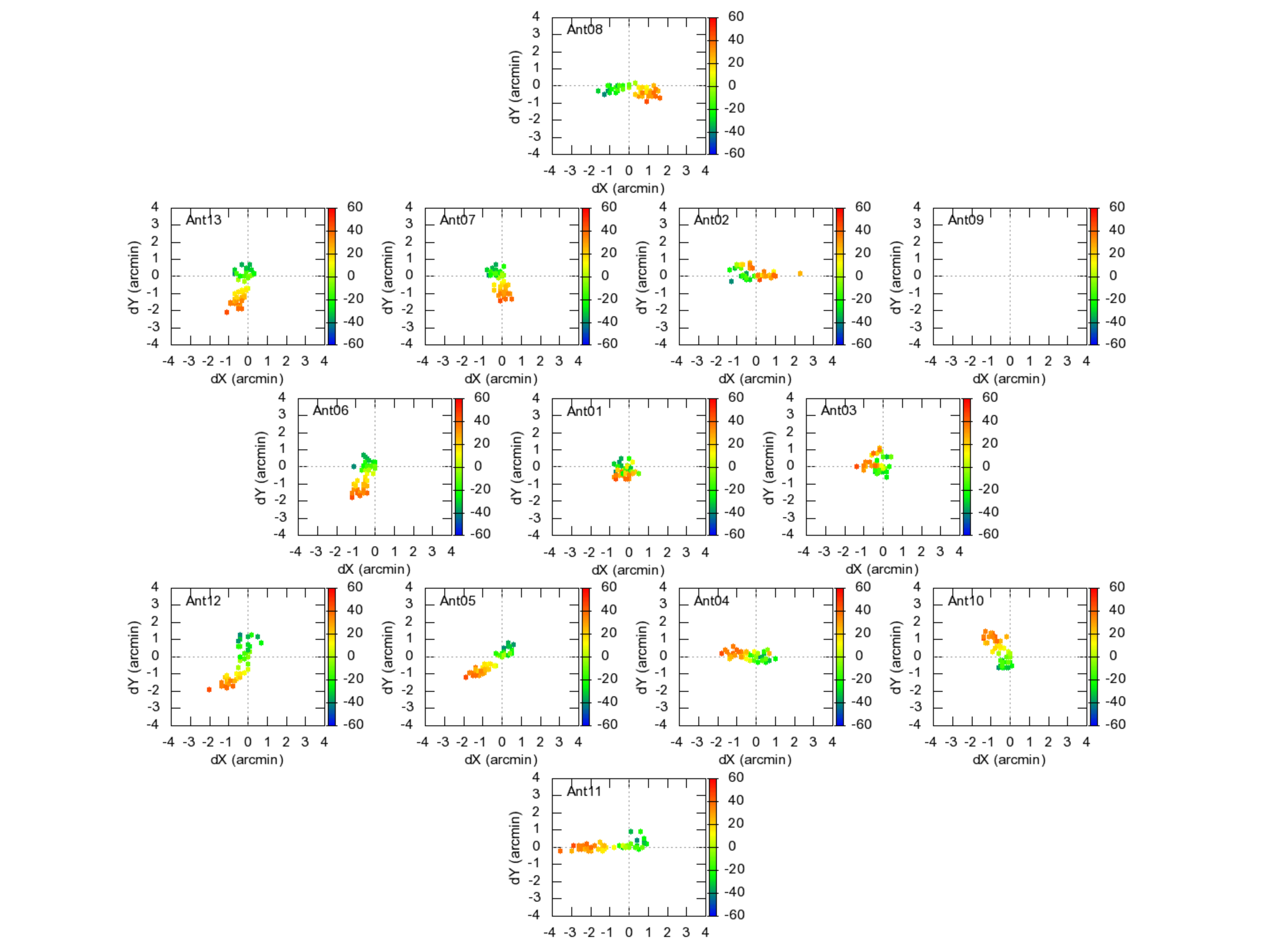}
 \caption{Misalignment of each antenna following the trajectory of Saturn as a function of hour angle of pointing (color scale, units in degree). The declination of Saturn during the observation was roughly $-4^{\circ}$. The misalignment is plotted in arcminutes projected on the platform. 
 Antenna 09 was off-line during this test. It is shown that, except Antenna 11, all other antennas have their misalignment within $\pm2$\arcmin. \label{alignment}}
\end{figure*}

Since the platform deformation changes with pointing (see Section~\ref{sec:deform}), most importantly with azimuth, the antennas sway as the deformation pattern rotates. Figure~\ref{alignment} shows the alignment solutions of repeated scans while we followed Saturn across the sky in one night. 
For this plot, only the relative change is shown, referenced to the alignment at transit of Saturn. The result shows that antennas mostly swing within $\pm 2\arcmin$, with the extremes at lower elevations.
These measured misalignments agree with the photogrammetry-measured deformation amplitudes along the z-direction (Figure~\ref{saddle}), i.e., measured maximum amplitudes of about 1.5\,mm at the outer platform lead to a tilt of about 1\farcm7 over a 3\,m platform radius. This indicates that the photogrammetry is, indeed, capturing all relevant deformation features.

\begin{figure}[!htb]
 \begin{center}
  \includegraphics[width=0.45\textwidth,angle=0,clip]{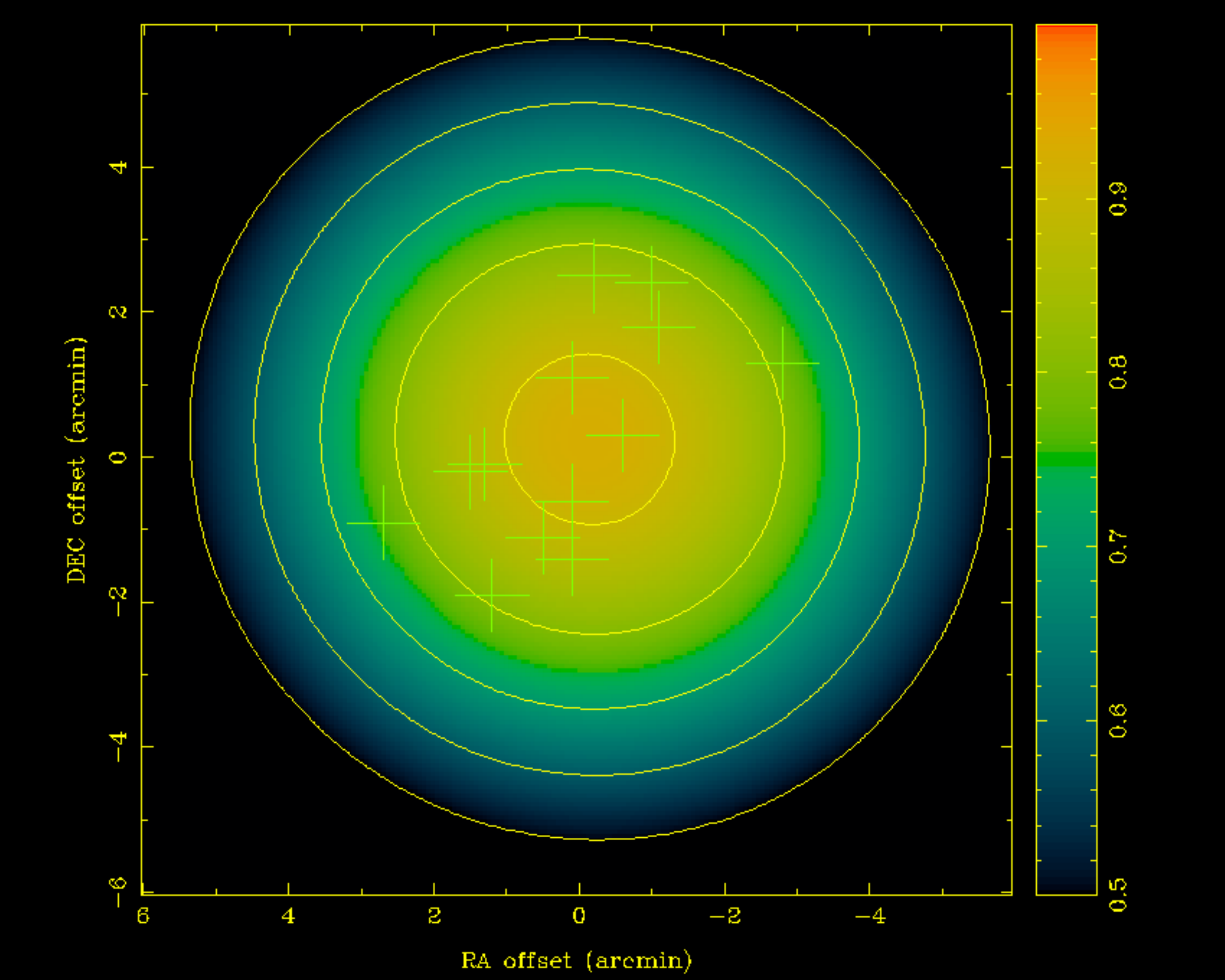}
 \end{center}
 \caption{Instance of combined alignment efficiency plot. Green crosses
 indicate misalignments of the 13 individual antennas in arcminutes. The
 color scale shows the averaged primary beam attenuation for a source at
 indicated (R.A., decl.)-offset from the pointing center. Contours denote 0.5, 0.6, 0.7, 0.8, and 0.9 times the ideal-case attenuation, assuming all antennas to be perfectly aligned. For sources without any pointing error (at the center of the plot), the sensitivity is 93\%, while the 90\% sensitivity region has a radius of about 1\arcmin. \label{align_sensitivity}}
\end{figure}

Antenna misalignments additionally lead to an efficiency loss. Figure~\ref{align_sensitivity} shows an example of this loss at lower elevation where the loss is more severe. For a source at the pointing center, the loss is about 7\%. It is less if the source is observed at higher elevation.

\subsection{All-Sky Radio Pointing}
\begin{figure}[!htb]
 \begin{center}
  \includegraphics[width=0.45\textwidth,angle=0,clip]{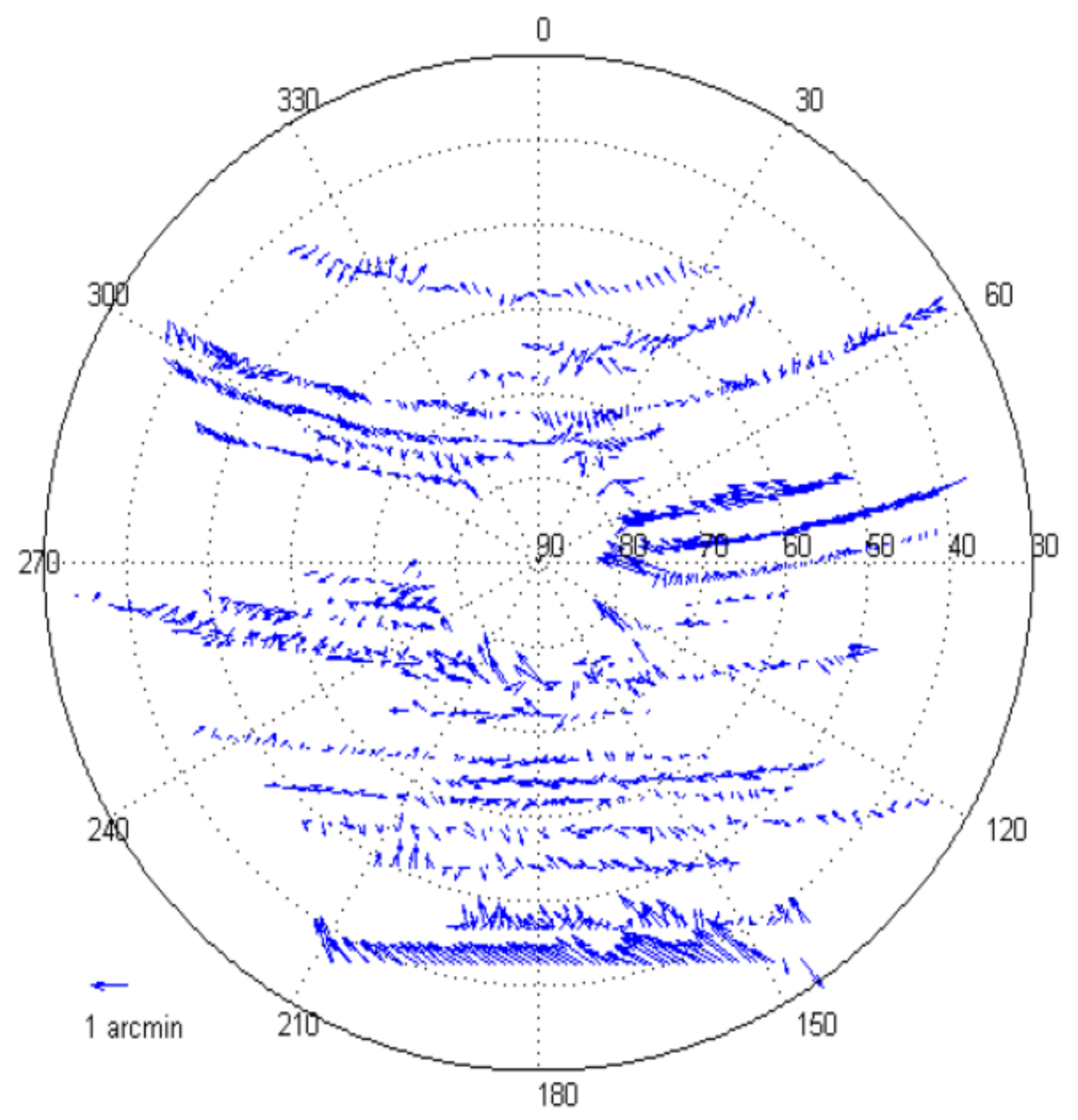}
 \end{center}
 \caption{All-sky radio pointing error verification. The chart shows the
 full azimuth range ($0^{\circ}\le az \le 360^{\circ}$) and an elevation
 range down to $30^{\circ}$.
Arrows indicate the magnitude and direction of pointing errors.
The rms pointing error for observations with $40^{\circ}<el<77^{\circ}$
 is 0\farcm4. Cluster observations are carried out in this elevation
 range. \label{rpointing}} 
\end{figure}

\citet{koch2009} describe how the pointing model of the AMiBA hexapod mount was established with an optical telescope. Further taking into account the parametric model of the platform deformation developed by \citet{liao2013}, we carefully rebuilt the pointing model by removing the tilt of the optical telescope due to the platform deformation in order to achieve a better pointing accuracy as required for the more extended AMiBA-13 array. 
We further observed a dozen radio sources, selected from the Australia Telescope Compact Array (ATCA) calibrator database\footnote{http://www.narrabri.atnf.csiro.au/calibrators/}, that have a listed 3\,mm flux density higher than 2\,Jy and that are evenly distributed in our observable decl. range in order to evaluate the residual pointing error in {\em radio} observations. 
The first round of radio pointing observations revealed a residual error pattern ranging from 0\arcmin to 2\arcmin that slowly varied with pointing. 
We then fitted a low-order polynomial function to the pattern and removed it from the pointing model. Figure~\ref{rpointing} shows the residual error distribution of the second round of radio pointing observations. It shows that some decl. ranges still have larger pointing errors ($\sim 1\arcmin$), but overall the radio pointing error is about 0\farcm4 in rms. 
Equally importantly, the pointing repeatability --- derived from night-to-night trackings over several hours of the same stars with the OTs --- is around 10\arcsec and it sets the limit of achievable pointing error for our telescope. 
If we assume the pointing error to be truly random with an rms error of 0\farcm4, for a 2\farcm5 synthesized beam, the smearing effect leads to a loss of less than 2\%. However, since for any given target the pointing error along its track is not random and is seldom symmetric, there can be systematic pointing errors after integration. To alleviate this problem, a companion pointing source within 5$^\circ$ of the target is observed every $\sim$30\,min to monitor the pointing error along the track.

\subsection{Phase Error and Flux Correction}\label{sec:phase-error}
In Section~\ref{sec:delay}, we mentioned that correlators that have large residual delays are especially sensitive to additional geometrical delays coming from pointing offset or platform deformation. These correlators are flagged during data processing. 
However, regardless of the amount of residual delay, all correlators suffer from phase errors and amplitude imbalance between the two spectral channels. The problem arises from wide-band smearing, leakage between the two channels, and leakage between real and imaginary parts in the lag-to-visibility transformation. 
The error varies with the geometric delays. If there were no pointing error nor platform deformation, the error could be calibrated out with an astronomical source. In practice, amplitude and phase errors occur because of the difference in delay between the calibrator and the target observations.
We emphasize that we have modelled and corrected for the "mean" phase shift for each spectral channel corresponding to the platform deformation. The phase error considered here, which originates from the phase slope within each spectral channel, is different.
We also note that AMiBA-7, having a much smaller range of deformation errors and shorter baselines, was much less susceptible to the variation of lag-to-visibility errors. 

Since large phase and amplitude errors are both symptoms of a large delay, it is possible to select and remove part of the data for better accuracy. 
An implication of the channel amplitude imbalance is that one of the two spectral channels has a low signal-to-noise ratio and a high noise variance after flux calibration. 
When the noise variances of two spectral channels are co-added, the resulting variance becomes substantially larger if the imbalance is stronger. By using the inverse of the co-added noise variance as weighting, we can efficiently downweight correlators that have large delays and smearing effects during the calibrator observations. 
Nevertheless, correlators that were not downweighted still have phase and amplitude errors, especially when observing a target that is further away from the calibrator.
We assess the severity of this effect by checking the point source flux-recovery ratio for a few flux standards. Table~\ref{tab:flux_error} summarizes the flux errors under various conditions. The typical flux recovery is between 75\% and 85\% unless the target is close to the track of the calibrator. 
Therefore, for visibility and flux measurements, we apply a correction factor of 1.25 and also add $\pm 5\%$ of systematic error in quadrature to the thermal noise as the final uncertainty.

\begin{deluxetable*}{lll}
\tablecaption{Flux-Recovery Ratio \label{tab:flux_error}}
\tablecolumns{3}
\tablehead{\colhead{Factors} & \colhead{Recovery Ratio} & \colhead{Remark} \\}
\startdata
Band-smearing (no phase error) & $>90\%$\tablenotemark{a} & \\ 
Deformation (without phase correction) & $>65\%$\tablenotemark{a} & Saturn, Jupiter (relative) \tablenotemark{b}\\
Residual deformation (with phase correction) & $>85\%$\tablenotemark{a} & Saturn, Jupiter (relative) \tablenotemark{b}\\
Cross-calibration (with phase correction) & $75 \sim 85\%$ & Uranus\tablenotemark{c}, 3C286\tablenotemark{d} (absolute)\\

\enddata

\tablenotetext{a}{Minimum recovery occurs at the elevation limit of 40$^\circ$.}
\tablenotetext{b}{Using only one ``two-patch'' near transit to calibrate the entire track of the planet (Saturn or Jupiter).}
\tablenotetext{c}{Flux of Uranus is calculated by assuming a disk
 brightness temperature of 120\,K, from the mean W-band results of
 WMAP-7 observations \citep{weiland2011}.}
\tablenotetext{d}{Absolute flux of 3C286 is taken to be $0.91\pm0.02$~Jy \citep{agudo2012}. }
\end{deluxetable*}

\section{Cluster SZE Observations}\label{sec:observation}

\subsection{Cluster Targets}
\begin{figure*}[!htb] 
 \begin{center}
  \includegraphics[width=0.9\textwidth,angle=0,clip]{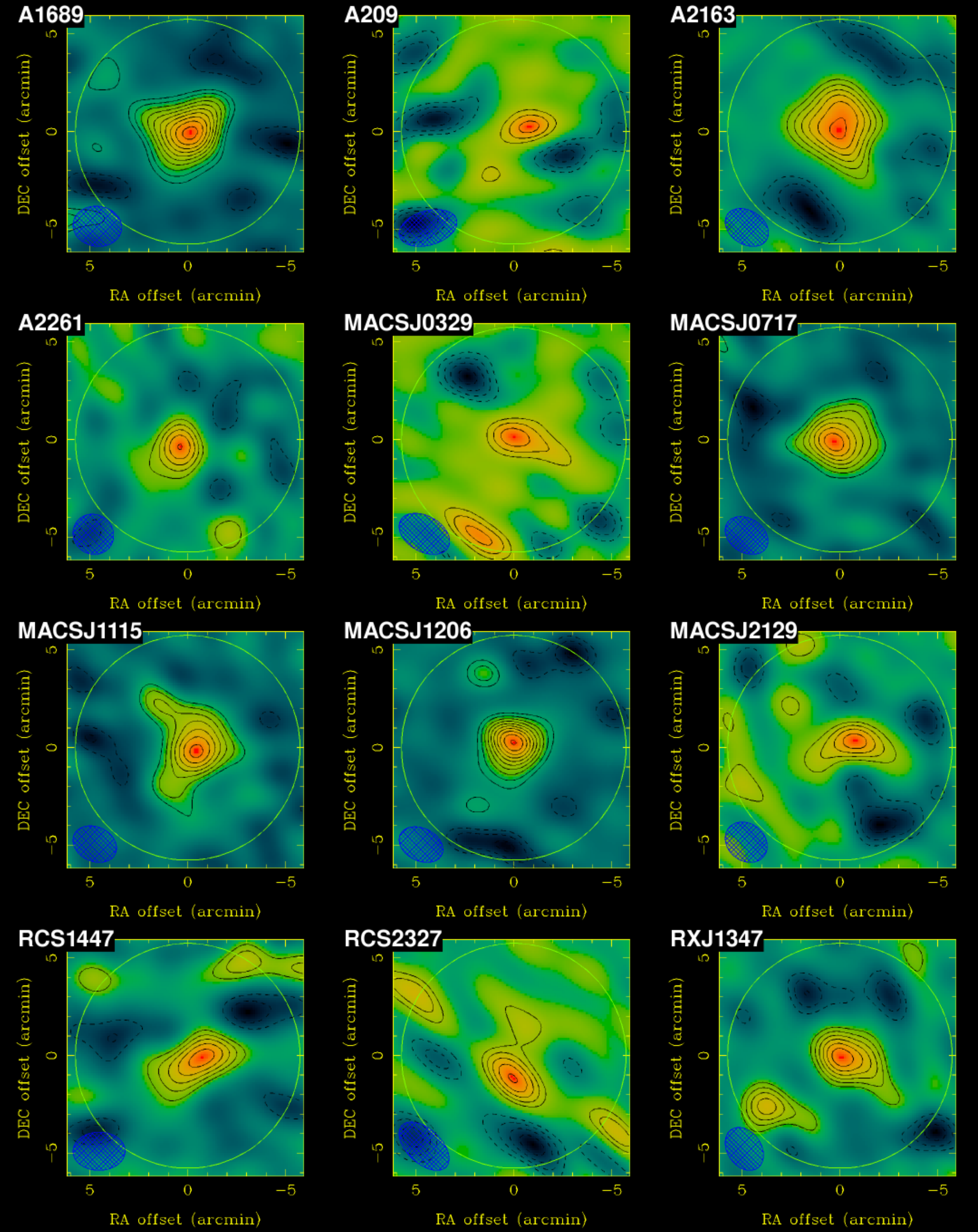}
 \end{center}
 \caption{Cleaned AMiBA-13 images of the twelve cluster targets listed
 in Table~\ref{coord_table}. The contour levels are shown in units of
 $\sigma$, starting from the $2\sigma$ detection significance level.
For each cluster, a solid circle indicates the FWHM of the
 primary beam (11\arcmin). The synthesized beam is displayed with a blue-shaded ellipse in the bottom-left corner.}
 \label{thumbnail}
\end{figure*}

Cluster targets for our AMiBA-13 observations are drawn from three different samples emphasizing different aspects of cluster studies. 
The first sample consists of the six clusters observed by AMiBA-7. Here, a combined AMiBA-7 and AMiBA-13 analysis with an improved $uv$-coverage can place tighter constraints on the cluster gas pressure profiles.
The second set of twenty clusters is selected from the CLASH sample, which has exquisite strong-lensing \citep{zitrin2015}, weak-lensing \citep{umetsu2014, umetsu2016, merten2015}, and X-ray \citep{donahue2014} data, as well as 2\,mm SZE data \citep[Bolocam,][]{sayers2013b} with angular scales similar to AMiBA-13, and additional 3\,mm SZE data with $\sim 10\arcsec$ resolution \citep[MUSTANG, e.g.][]{mason2010, mroczkowski2012}.
AMiBA-13 is complementary to these existing SZE data, allowing for joint analyses of the physical processes that govern the hot cluster gas. 
Finally, before the observing was concluded, seven optically selected cluster candidates were chosen from the Red-sequence Cluster Survey 2 \citep[RCS2,][]{gilbank2011} catalog according to their richness indicator $B_{gc}$ and added to our observations. 
Although the sample is small, we aim at comparing their SZE signals to other X-ray-selected clusters of similar richness and redshift for signs of selection biases.  From all these observed targets, twelve clusters show robust detections above $5\sigma$. Coordinates and redshifts of these clusters are listed in Table~\ref{coord_table}. Their  integration times and detection significances are given in Table~\ref{snr_table}. Figure~\ref{thumbnail} shows our SZE images of these selected clusters. For each cluster, the $uv$ data were natually weighted and inverted to produce the dirty image. The image was then cleaned with the Miriad\footnote{http://www.atnf.csiro.au/computing/software/miriad/} task CLEAN, looking for sources within the FWHM of the primary beam. The primary beam attenuation was not corrected for. In Section~\ref{sec:pntsrc}, we will discuss the possibility of point source contamination. However, for the twelve clusters presented here, no significant point source with flux $> 1$\,mJy is expected, and we have made no correction to the data.

\begin{deluxetable*}{llllccc}
\tablecaption{Clusters Detected By AMiBA-13\label{coord_table}}
\tablecolumns{5}
\tablehead{\colhead{Cluster} & \colhead{R.A. (J2000)} & \colhead{decl. (J2000)} & \colhead{Redshift} & \colhead{Sample\,\tablenotemark{a}} & \colhead{Cool-Core\,\tablenotemark{b}} & \colhead{Disturbed\,\tablenotemark{b}}}

\startdata
Abell 1689         & 13:11:29.45 & -01:20:28.1 & 0.183 & A & \nodata & \nodata \\
Abell 2163         & 16:15:46.20 & -06:08:51.3 & 0.203 & A & \nodata & \nodata\\
Abell 209          & 01:31:52.57 & -13:36:38.8 & 0.206 & C & & \\
Abell 2261         & 17:22:27.25 & +32:07:58.6 & 0.224 & A,C & $\surd$ & \\
MACS J1115.9+0129  & 11:15:52.05 & +01:29:56.6 & 0.352 & C & $\surd$ & \\
RCS J1447+0828     & 14:47:26.89 & +08:28:17.5 & 0.38  & A,R &$\surd$ & \\
MACS J1206.2-0847  & 12:06:12.28 & -08:48:02.4 & 0.440 & C & & \\
MACS J0329.7-0211  & 03:29:41.68 & -02:11:47.7 & 0.450 & C & $\surd$ & $\surd$\\
RX J1347.5-1145    & 13:47:30.59 & -11:45:10.1 & 0.451 & C & $\surd$ & $\surd$\\
MACS J0717.5+3745  & 07:17:31.65 & +37:45:18.5 & 0.548 & C & & $\surd$\\
MACS J2129.4-0741  & 21:29:26.06 & -07:41:28.0 & 0.570 & C & $\surd$ & \\
RCS J2327-0204     & 23:27:26.16 & -02:04:01.2 & 0.700 & R & \nodata &
\nodata
\enddata

\tablenotetext{a}{A: AMiBA-7; C: CLASH; R: RCS}
\tablenotetext{b}{X-ray morphology classification taken from Table~3 of \citet{sayers2013b}. We additionally identify RCS J1447.5+0828 as a cool-core cluster on the basis of \citet{hicks2013}. Although RX J1347.5-1145 was not identified as a disturbed cluster in \citet{sayers2013b}, significant substructure and large ellipticity were found for this cluster \citep[see e.g.][]{postman2012}.}
\end{deluxetable*}

\begin{deluxetable*}{lcccccc}
\tablecaption{Integration Times and Detection Significance \label{snr_table}}
\tablecolumns{6}
\tablehead{
\colhead{Cluster} & \colhead{Obs. Time} & \colhead{Used Time} & \colhead{Eff. Time} & \colhead{Peak\tablenotemark{a}} & \colhead{Noise} & \colhead{S/N\tablenotemark{a}}\\
  & (hr) & (hr) & (hr) & (mJy/b) & (mJy/b) & }
\startdata
Abell 1689         & 27.7 & 14.8 &  7.3 & -46.1 & 4.0 & 11.5\\ 
Abell 2163         & 48.7 & 34.5 &  9.5 & -28.1 & 3.9 & 7.3\\
Abell 209          & 14.4 & 2.7  &  0.8 & -29.4 & 6.6 & 4.4\\ 
Abell 2261         & 22.4 & 11.8 &  4.4 & -26.3 & 4.3 & 6.1\\ 
MACS J1115.9+0129  & 35.7 & 22.9 &  5.8 & -25.8 & 3.1 & 8.4\\ 
RCS J1447+0828     & 9.5  & 2.6  &  0.7 & -45.8 & 6.9 & 6.6\\ 
MACS J1206.2-0847  & 25.1 & 17.6 &  5.0 & -36.2 & 3.3 & 11.1\\ 
MACS J0329.7-0211  & 33.4 & 11.6 &  4.0 & -19.8 & 4.2 & 4.8\\ 
RX J1347.5-1145    & 29.5 & 11.0 &  2.5 & -44.1 & 5.7 & 7.8\\ 
MACS J0717.5+3745  & 23.9 & 18.0 &  5.1 & -44.4 & 5.1 & 8.7\\
MACS J2129.4-0741  & 57.7 & 29.4 &  7.3 & -21.9 & 3.8 & 5.7\\
RCS J2327-0204     & 36.0 & 20.6 &  3.0 & -32.7 & 5.3 & 6.1\\
\hline
Sum                & 364.0 & 197.5 & 55.4 & & &
\enddata
\tablenotetext{a}{``Raw'' peak in the cleaned image, before applying the upward-flux correction. }
\tablecomments{"Eff. Time" is defined in Section \ref{sec:noise}, "Obs. Time" and "Used Time" are defined in Section \ref{sec:data_flagging}.}
\end{deluxetable*}

\subsection{Observing Strategy}\label{sec:two-patch}
Cluster observations follow the same ``two-patch'' procedure outlined for AMiBA-7 observations \citep{wu2009}, where we track the cluster for three minutes and then move the telescope to a trailing patch that is 3m10s later in R.A. for another three minutes. 
The on-source and the off-source patches share the same telescope trajectory. Differencing of the two patches efficiently removes the ground pickup and other low-frequency contamination in the system. 
Since AMiBA-13 has a primary beam of $11\farcm5$ FWHM, or $\sigma\sim10\arcmin$ if the beam is approximated by a Gaussian, a typical cluster is roughly 3 to 4$\sigma$ away from the pointing center of the trailing patch, and any cluster pickup is negligible. 
For targets at higher declinations ($\mathrm{decl.}\gtrsim 60^{\circ}$), the integration time and the separation between two patches are both doubled to ensure that a possible source leakage to the trailing patch does not bias our measurement of the background level.

A subtle difference with the AMiBA-7 observing procedure is how we choose to populate the $uv$-plane. The instantaneous $uv$-coverage of AMiBA is highly redundant due to its six-fold symmetry for most of the baselines. In AMiBA-7, we split a cluster observation into eight parts, each with the platform position angle ($hexpol$) rotated by $7.5^{\circ}$ with respect to the sky. Combined with the six-fold symmetry, this procedure densely sampled the azimuthal angle in the $uv$-plane. However, since typically our cluster signal-to-noise ratio per $uv$-mode is less than 1 after integration, spreading the integration in the $uv$-plane does not provide a significant advantage in cluster imaging and modelling.  For AMiBA-13, we stopped actively changing the platform position angle and chose to operate the mount in its most balanced orientation in order to minimize the platform deformation. As we track a target, the sky rotates relative to the platform, and so does the $uv$-coverage. The rotation is within $\pm15^{\circ}$ for most of 
our cluster observations and has a high concentration within $\pm5^{\circ}$. The resulting synthesized beam is, thus, less circular as compared to the AMiBA-7 beam. This is especially the case when some antennas are offline during an observation.

\subsection{Calibration}\label{sec:calibration}

Similar to AMiBA-7, the flux and gain calibration of AMiBA-13 is done by observing Jupiter and Saturn for at least one hour each night. All calibration observations are also done in the ``two-patch'' observing mode with four minutes of integration per patch and a 4m10s separation in R.A. A one-hour observation provides seven sets of two-patch data that are used to gauge the performance of each baseline, including phase scatter over time and phase consistency between the two spectral channels. 

\begin{figure}[!htb] 
 \begin{center}
  \includegraphics[width=0.45\textwidth,angle=0,clip]{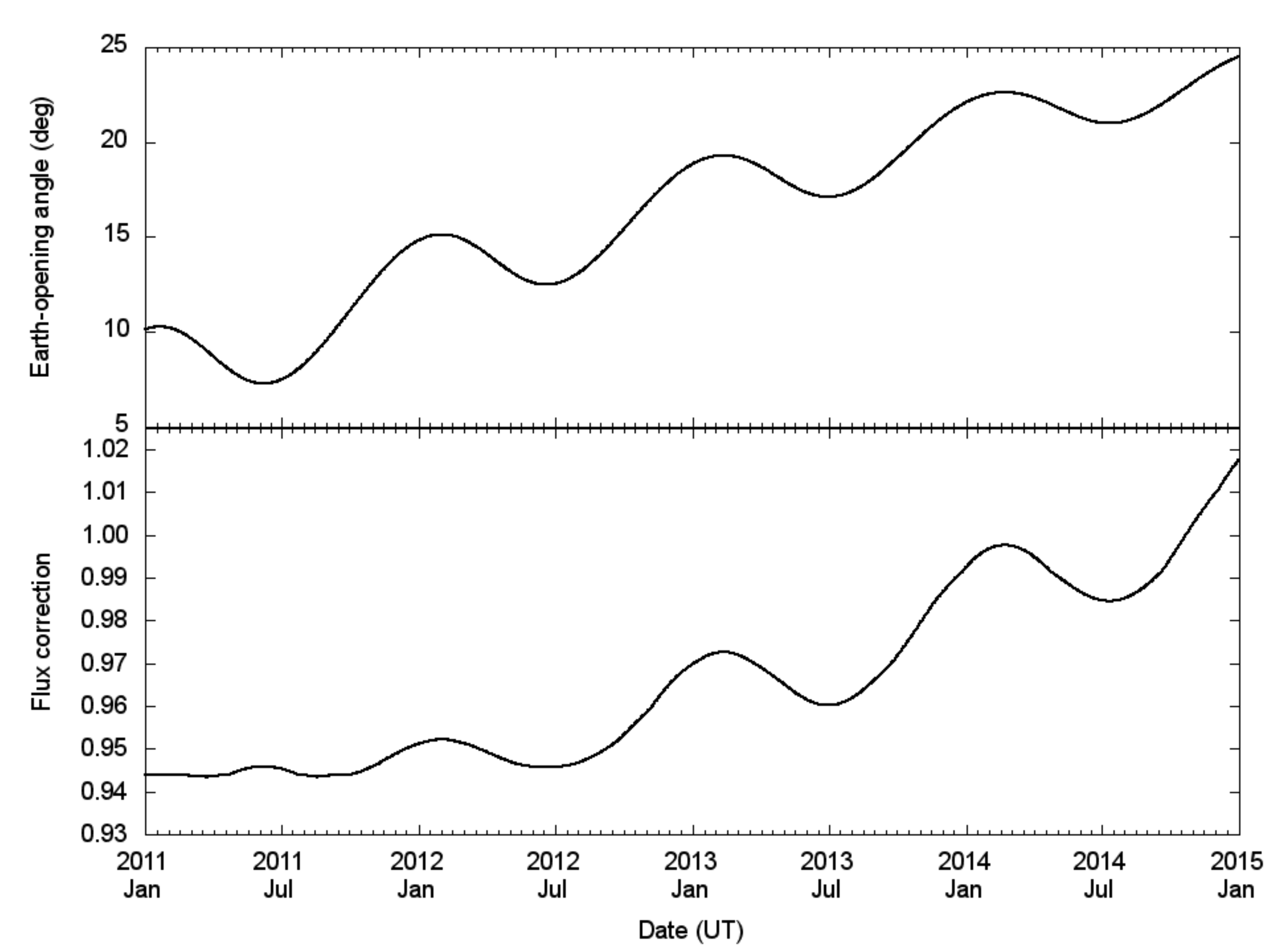}
 \end{center}
 \caption{Top panel: Earth-opening angle of Saturn's rings during the
 observing period of 2011 to 2014. Bottom panel: corresponding
 correction factor for Saturn's flux, relative to a simple disk without rings. \label{ring_correction}}
\end{figure}

To calculate the flux density of Jupiter and Saturn, we modeled the planets as  circular disks with constant brightness temperature \citep{lin2009}. The 3~mm brightness temperatures we adopted are $171.8\pm1.7$~K for Jupiter \citep{page2003, griffin1986} and $149.3\pm4.1$~K for Saturn \citep{ulich1981}. 
Since the synthesized beam of AMiBA-13 is only a few times larger than
the angular size of Jupiter or Saturn, the planets are slightly
resolved. In the extreme case when Jupiter's size is close to
$40$\arcsec, the longest baselines will detect about $15\%$ less flux as
compared to a point-source assumption. In addition, we included the
obscuration effect of Saturn's rings as a function of their
Earth-opening angle following the WMAP-derived model parameters \citep{weiland2011}. Figure~\ref{ring_correction} shows this correction factor compared to a simple disk model of Saturn over the entire observing period from 2011 to 2014. The angular sizes of Jupiter and Saturn and the Earth-opening angle of Saturn's rings are calculated with the Python package 'PyEphem'\footnote{http://rhodesmill.org/pyephem/}.

\subsection{Noise Performance and Observing Efficiency}\label{sec:noise}

Following \citet{lin2009}, we characterize the instrumental efficiency
of AMiBA-13 by examining the net sensitivity as a function of the effective
integration time. Here, the sensitivity is defined as the observed rms
fluctuation of a cleaned map excluding the source region. The 
effective integration time sums up the integration of all valid
(unflagged) visibility channels, accounting for the effects of
relative weighting. Specifically, the effective integration time is
defined as $t_\mathrm{eff} \equiv [(\sum_i w_i)^2 / \sum_i w_i^2] \,t_\mathrm{on\_src}$, 
where $t_\mathrm{on\_src}$ is the physical on-source time, 
$w_i$ is the weight of each visibility, and
the index $i$ runs over all visibility elements. For AMiBA-13, in the ideal case of all instruments working and having identical weighting, the effective integration time and on-source time are related by 
$t_\mathrm{eff}^\mathrm{ideal} / t_\mathrm{on\_src}= (78~\mathrm{baselines}) \times (2~\mathrm{polarizations}) \times  (2~\mathrm{channels}) =  312$.

\begin{figure}[!htb] 
 \begin{center}
  \includegraphics[width=0.45\textwidth,angle=0,clip]{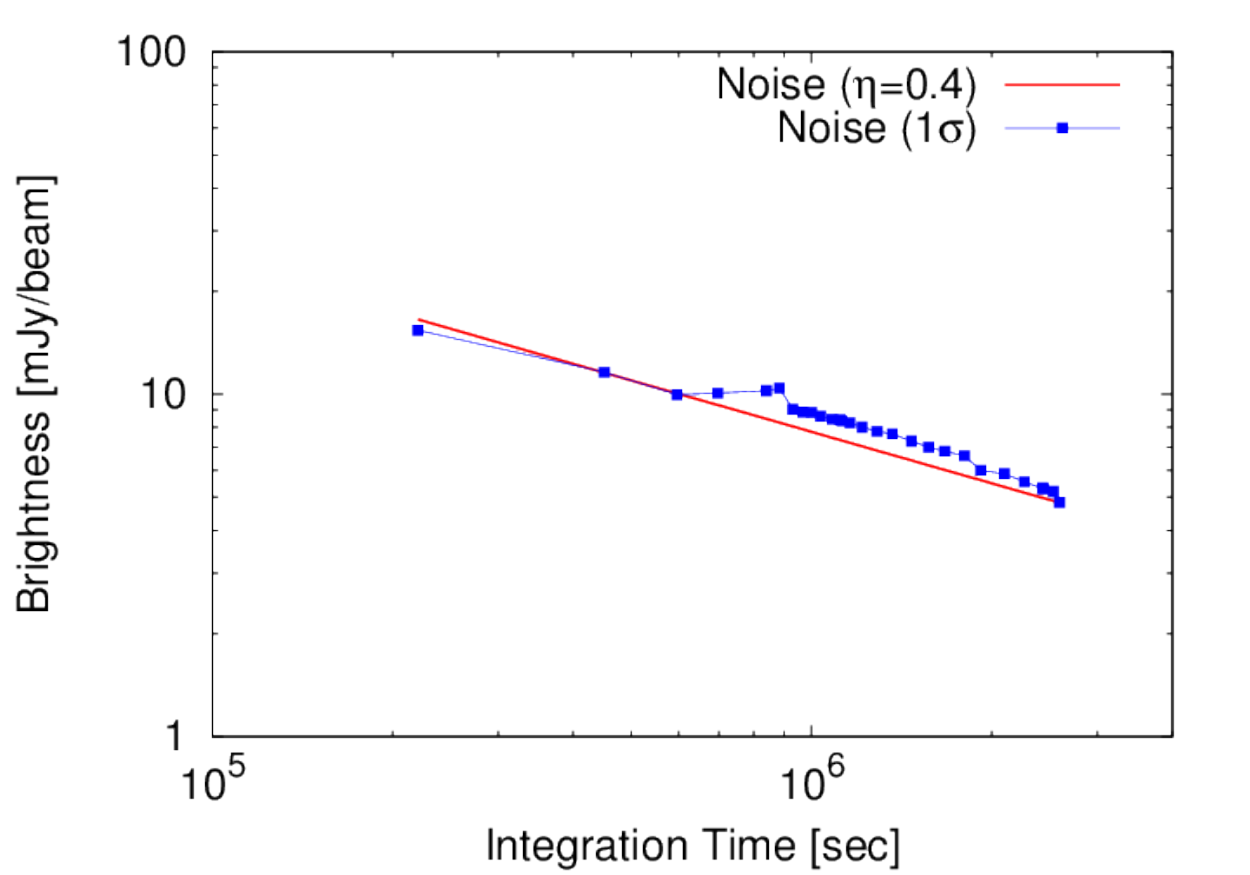}
 \end{center}
 \caption{Sensitivity (rms noise) as a function of effective integration
 time for observations of A1689. \label{snr_a1689}} 
\end{figure}

As an example, Figure~\ref{snr_a1689} shows how the sensitivity depends on the effective integration time for our observations of Abell 1689. In this plot, visibilities are successively multiplied by $(-1)^j$, where $j$ is the index of a data point, so that the signals of the cluster SZE or any other sources are significantly suppressed. 
The figure demonstrates that the variance of noise scales with the inverse of the effective integration time. 
Changing the multiplying factor from $(-1)^j$ to $\exp(i2\pi j/3)$, which cancels the signal for every three integrations, does not change the results. 

The amplitude of the noise scaling curve can further be used to determine the overall efficiency $\eta$ of the array. For AMiBA-13, we obtain $\eta=0.4$, which is comparable to that of the AMiBA-7 system \citep{lin2009}. 
We note, however, that as the efficiency is defined against the effective integration time, which is insensitive to data with a higher noise level, this overall efficiency only reflects particular baselines that have higher signal-to-noise ratios.
Instead, the ratio between the used (un-flagged) integration and the effective integration quantifies the array performance.
This ratio ("Eff. Time" over "Used Time" in Table~\ref{snr_table}) varies from 14\% to 49\% and averages to 28\% for the set of clusters shown in Table~\ref{snr_table}.
This rather small average ratio indicates that a significant fraction of the correlations is much noisier than the rest of the array.

The root cause of the low-efficiency problem is that there are only four delay samples in the correlator, giving two highly coupled output channels. 
As geometric delays vary (due to, e.g., platform deformation), the leakage between the two output channels can change drastically. 
The ratio of the recovered power of the two channels from a flat-spectrum source ideally should be one. In practice, however, it can often reach five or more. 
When this happens in calibration, the suppressed channel can be identified from its magnified noise variance. Moreover, when the power imbalance between the channels is severe, the phase error is also amplified. 
This phase error affects both the suppressed and the enhanced channels. Therefore, it is better to downweight the correlator that is affected by a strong imbalance in order to reduce potential phase errors. 
Hence, we use the sum of the variances of both channels for the inverse-variance weighting (see also Section \ref{sec:phase-error}).

\begin{figure*}[!htb] 
 \begin{center}
 \includegraphics[width=0.45\textwidth,angle=0,clip]{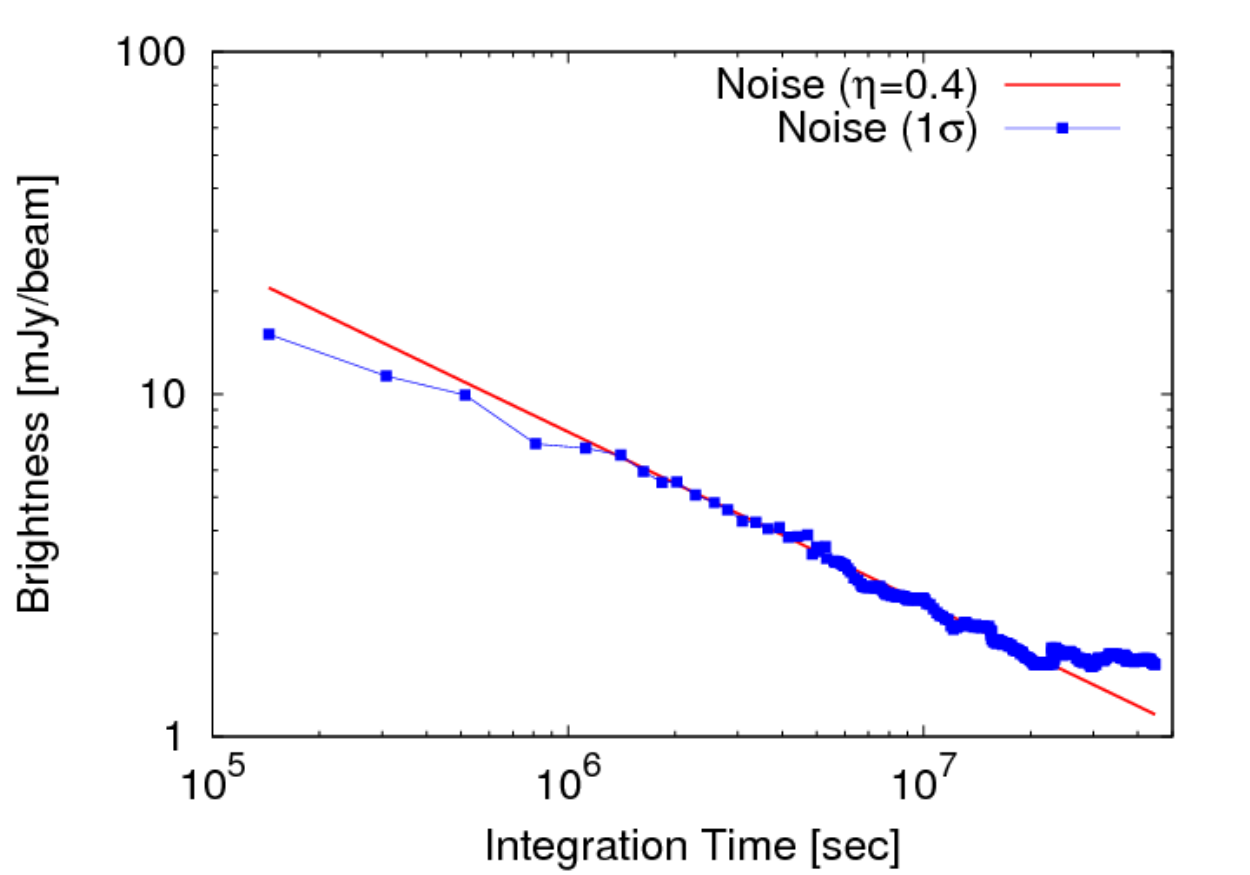}
 \includegraphics[width=0.45\textwidth,angle=0,clip]{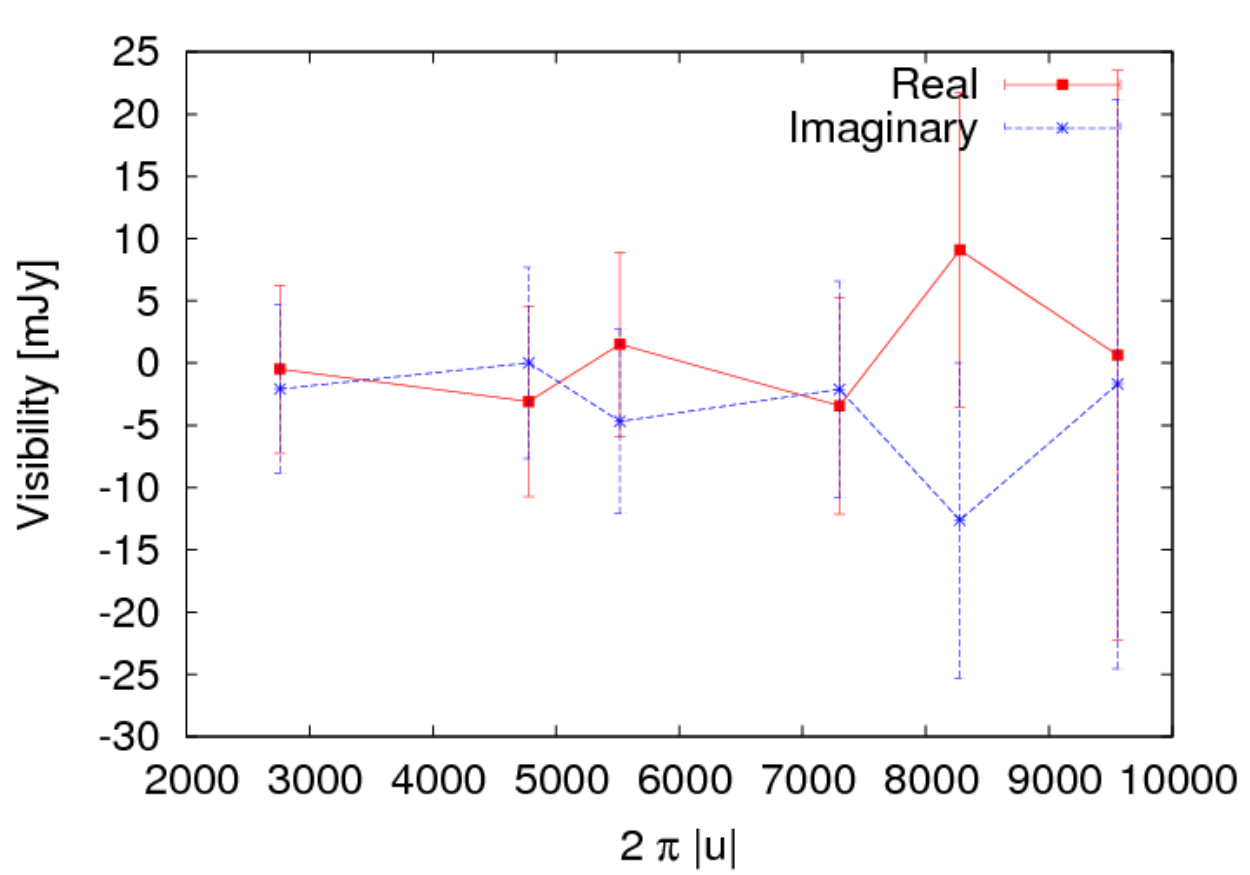}
\end{center}
 \caption{Stacked noise properties of AMiBA-13 observations for twelve selected clusters (Section
 \ref{sec:noise}). 
Left panel: stacked rms noise level as a function of effective
 integration time.
 Right panel: stacked noise visibilities as a function of $uv$-distance. \label{snr_12clusters}}
\end{figure*}

Typically, the sensitivity of our individual cluster observations
reaches a level of 3 to 6\,mJy\,beam$^{-1}$. Here, we further
investigate whether any significant systematics are present below this
limit. We do this by stacking data of many clusters together. 
The left panel in Figure~\ref{snr_12clusters} shows the stacked noise
level as a function of the effective integration time for our sample of
the twelve selected clusters (Table~\ref{coord_table}).
This scaling is obtained by suppressing the cluster signal through phase scrambling. 
This test consistently confirms that the noise scaling holds down to a
level slightly below 2\,mJy\,beam$^{-1}$. 
The right panel in Figure~\ref{snr_12clusters} displays the stacked noise
visibilities as a function of $uv$-distance, shown separately for the
real and imaginary components.
Here, the error bars indicate the
expected level of uncertainty, $\sqrt{\langle\sigma^2\rangle}$, assuming
Gaussian random noise, where  
$\langle\sigma^2\rangle = (\sum_i w_i^2 \sigma_i^2) / (\sum_i
w_i^2)$. The figure displays a noise level that is consistent with zero,
demonstrating that no significant systematics are present.

\subsection{Data Flagging}  \label{sec:data_flagging}
Two integration times are listed for each cluster in Table~\ref{snr_table}. The ``Obs. Time'' shows the on-source integration on each cluster. The actual time spent on the observation is approximately double this value because of the trailing patch observing strategy. The ``Used Time'' shows the remaining integration time after flagging. Table~\ref{flag_table} summarizes the fraction of data flagged by various criteria. 
``Offline'' indicates the fraction of data flagged due to hardware malfunctions (including receivers and correlators). 
``High noise'' sets a limit on the minimum weighting required to be included in the analysis. While including lowly weighed data does not affect the result, we choose to explicitly flag them out to help keep track of the problem. 
``Unstable BL'' and ``U/L band diff.'' are related to the varying delay and band-smearing issue of the broadband analog correlator. 
As mentioned in Section~\ref{sec:delay}, the platform deformation can introduce an additional delay to baselines with already large instrumental delays and cause the visibility to be very sensitive to the combined delay. Its symptom reveals itself as unstable measurements in one-hour planet trackings and also as inconsistent phase/delay measurements between the upper and lower band. Therefore, we set up these criteria to flag them out. 
``Non-Gauss'' catches the occasional glitches in the system that fail a Gaussianity test. 
Finally, ``Cal. flag'' indicates the fraction of data that are flagged as a consequence of flagged calibrator events. 
Overall, 30\% to 50\% of the data in the AMiBA-13 observations are flagged.

\begin{deluxetable*}{lccccccc}
\tablecaption{Summary of Flags for Each Cluster in Percentage \label{flag_table}}
\tablecolumns{8}
\tablehead{
\colhead{Cluster} & \colhead{Offline} & \colhead{High noise} & \colhead{Unstable BL} & \colhead{U/L band diff.} & \colhead{Non-Gauss} & \colhead{Cal. flag} & \colhead{Used data}
}
\startdata
Abell 1689        & 7.6 &12.4 &15.2 & 1.1 & 0.8 & 9.5 &53.6\\
Abell 2163        & 7.7 &10.8 & 3.9 & 0.4 & 0.3 & 4.2 &72.7\\
Abell 209         &31.7 & 4.2 & 6.2 & 1.9 & 0.1 &37.4 &18.5\\
Abell 2261        & 4.3 &14.7 & 9.9 & 0.3 & 0.2 &17.7 &52.8\\
MACS J1115.9+0129 &21.4 & 8.8 & 2.5 & 0.1 & 0.4 & 2.6 &64.3\\
RCS J1447+0828    &22.0 & 5.1 & 5.0 & 2.7 & 0.5 &37.4 &27.4\\
MACS J1206.2-0847 &10.5 & 7.3 & 5.5 & 0.5 & 0.2 & 6.0 &70.1\\ 
MACS J0329.7-0211 &17.4 &11.1 & 4.1 & 1.5 & 0.2 &31.0 &34.8\\
RX J1347.5-1145   &43.4 & 8.1 &10.2 & 0.3 & 0.5 & 0.3 &37.2\\
MACS J0717.5+3745 & 4.5 & 9.7 & 4.8 & 0.1 & 0.5 & 5.1 &75.4\\
MACS J2129.4-0741 &27.7 &13.4 & 2.2 & 0.4 & 0.5 & 4.7 &51.1\\
RCS J2327-02024   &20.3 &13.5 & 6.3 & 1.0 & 0.4 & 1.3 &57.3
\enddata
\tablecomments{Individual flags are discussed in detail in Section \ref{sec:data_flagging}.}
\end{deluxetable*}

\section{Discussion}\label{sec:discussion}

\subsection{Point Source Contamination}\label{sec:pntsrc}
AMiBA aims at measuring the cluster SZE with a single frequency band and only with a compact array. Without outrigger baselines to look for and subtract off point sources in both our target and trailing fields, the AMiBA measurements can potentially be contaminated. 
On the other hand, the choice of 90\,GHz as the observing frequency is to avoid as much as possible the synchrotron sources at lower and the dusty sources at higher frequencies. 
Following \citet{liu2010}, we estimate the contamination from radio sources by extrapolating flux densities from low-frequency catalogs to 94\,GHz, assuming a simple power law spectrum. Potential sources are identified from within an $11\arcmin$ radius (twice the FWHM of the primary beam) of both the target and trailing fields in the NVSS \citep[1.4\,GHz,][]{condon1998}, PMN \citep[4.85\,GHz,][]{griffith1995}, and GB6 \citep[4.85\,GHz,][]{Gregory1996} catalogs. 
If a source is identified in both the 1.4\,GHz and the 4.85\,GHz catalogs, a spectral index is derived and used to estimate its 94\,GHz flux density. If a source is detected in only one of the frequency bands, then we perform a Monte Carlo simulation with the spectral indices drawn from the five-year WMAP point source catalog \citep{wright2009}. 
In particular, if a source is selected from the NVSS catalog but is not detected in the PMN or GB6 catalog, we limit the spectral index so that its flux density does not exceed the detection criteria of the 4.85\,GHz surveys. 
For the twelve clusters shown in this work, no significant radio source with an extrapolated flux of more than 1\,mJy at 90\,GHz is found within our searching radius of both the target and the trailing fields.

\subsection{Interpreting Cluster Results}
One application of AMiBA measurements is to constrain gas pressure distributions in clusters. In this work, we adopt the spherical generalized Navarro--Frenk--White (gNFW) parametric form, first proposed by \citet{nagai2007}, to describe the gas pressure profile:
\begin{equation}
 \mathbb{P}(x) = \frac{P(x)}{P_{500}} 
= \frac{P_0} {(c_{500}x)^{\gamma}\left[1+
				  (c_{500}x)^{\alpha})\right]^{(\beta -\gamma)/\alpha}}, 
\label{eq:gNFW} 
\end{equation}
where $\mathbb{P}(x)$ is the dimensionless form of $P(x)$ that describes the shape of the profile with the scaled radius $x=r/R_{500}$ and $c_{500} = R_{500} / r_{gs}$, the ratio of $R_{500}$\footnote{$R_{500}$ or $R_{500c}$ denotes the radius within which the enclosed mass is 500 times the critical density of the universe at the given redshift. Thus, $M_{500}$ refers to the enclosed mass within $R_{500}$.} to the gas characteristic radius $r_{gs}$;\footnote{The gas characteristic radius $r_{gs}$ is independent of the dark matter characteristic radius $r_s$ conventionally used in the NFW model.} $P_0$ is the deviation from the characteristic pressure $P_{500}$, which is governed by gravity in the self-similar model.
The AMiBA measurement is used to determine $P_0$ and $r_{gs}$ (or equivalently $c_{500}$), while the slope parameters ($\alpha$, $\beta$, $\gamma$) are fixed at the best-fit values found by \citet[][A10, hereafter]{arnaud2010}. 
If a cluster is classified as cool-core or disturbed (Table \ref{coord_table}), the corresponding best-fit values from A10 is used. Otherwise, the best-fit values from the overall sample are used. 
$R_{500}$ is obtained separately from the literature (X-ray or lensing) for each cluster. 

Our data analysis is performed in two steps. The first step is to reconstruct the cluster visibility and remove any residual pointing offsets from the measurement. For this, we assume the cluster to be axisymmetric. 
Therefore, the AMiBA-13 data can be modeled with twelve independent real-valued band powers, $V(|u|)$, coming from six discrete baseline lengths with two spectral channels each, and with two more parameters for the pointing offsets.
These parameters are then determined from the two-dimensional visibility data using a Markov Chain Monte Carlo (MCMC) method.
After obtaining the visibility band powers, the second step is to fit the gNFW profile. A separate MCMC program is used for the profile fitting. Primary beam attenuation is applied to the model in this process before comparing to the band powers. The method is detailed in Wang et~al. ({\it in preparation}). 
In Section~\ref{sec:a1689} we demonstrate and apply this method to Abell 1689.
For optically selected clusters, individual $R_{500}$ estimates may be unavailable or unreliable due to large scatter. In this case, AMiBA SZE data can serve as a mass-proxy. We demonstrate this in Section~\ref{sec:M_1447} with the first targeted SZE detection of the cluster RCS J1447+0828.

\subsection{Combined AMiBA-7 and AMiBA-13 Observations of Abell 1689}\label{sec:a1689}
The rich cluster Abell 1689 at $z=0.183$ is among the most powerful cosmic lenses known to date \citep{Broadhurst2005a,Oguri2005,Limousin2007,UB2008,Coe2010,Diego2015}, exhibiting a high degree of mass concentration in projection of the cluster.
As such, the cluster has been a subject of detailed multiwavelength analyses \citep{lemze2009,Peng2009,Molnar2010HSE,Kawaharada2010,Morandi2011,Sereno2013,Okabe2014,umetsu2015}, and is one of the six clusters observed by AMiBA-7.
A recent Bayesian analysis of the cluster \citep{umetsu2015} shows that combined multiwavelength data favor a triaxial geometry with minor-major axis ratio $0.39 \pm 0.15$ and major axis closely aligned with the line of sight ($22^\circ\pm 10^\circ$). 
This aligned orientation boosts the projected surface mass density of a massive cluster with $M_{200}=(1.7\pm 0.3)\times 10^{15}M_\sun$ \citep{umetsu2015}, and thus explains the exceptionally high lensing efficiency of the cluster.

\begin{figure}[!htb] 
 \begin{center}
 \includegraphics[width=0.45\textwidth,angle=0,clip]{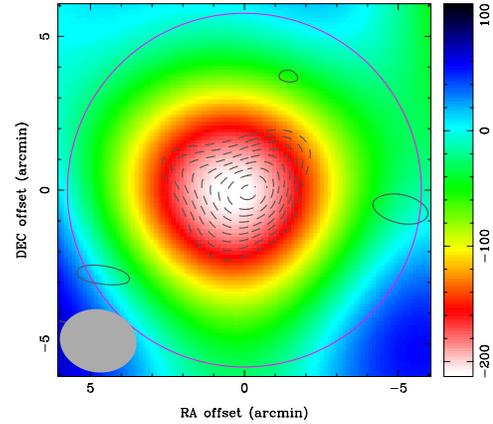}
 \end{center}
 \caption{AMiBA SZE maps of the cluster Abell 1689. The image is $12\arcmin\times 12\arcmin$ in size and centered on the cluster center (Table~\ref{coord_table}). The color image shows the AMiBA-7 observations \citep{wu2009} of the cluster with an rms noise level of $40$\,mJy per AMiBA-7 synthesized beam, that is approximately circular with a FWHM of $6.5\arcmin$. The black contours show the AMiBA-13 observations. The contour levels are shown in units of $\sigma$ and start at $\pm3\sigma$ for positive (solid) and negative (dashed) flux levels, respectively, where $1\sigma$ is about 4\,mJy per beam. The synthesized beam of AMiBA-13 is indicated by a gray-shaded ellipse in the lower-left corner. \label{a1689_clean}}
\end{figure}

Being massive and at a relatively low redshift, the bulk of the SZE signal is beyond the angular scales probed by AMiBA-13 but is largely captured by AMiBA-7. 
Hence, the cluster is well-suited for an examination of how well AMiBA can constrain the cluster pressure profile combining both the AMiBA-7 and AMiBA-13 data. 
Figure~\ref{a1689_clean} shows the two overlaid SZE maps of the cluster observed with the two configurations of AMiBA.

\begin{figure}[!htb] 
 \begin{center}
  \includegraphics[width=0.45\textwidth,angle=0,clip]{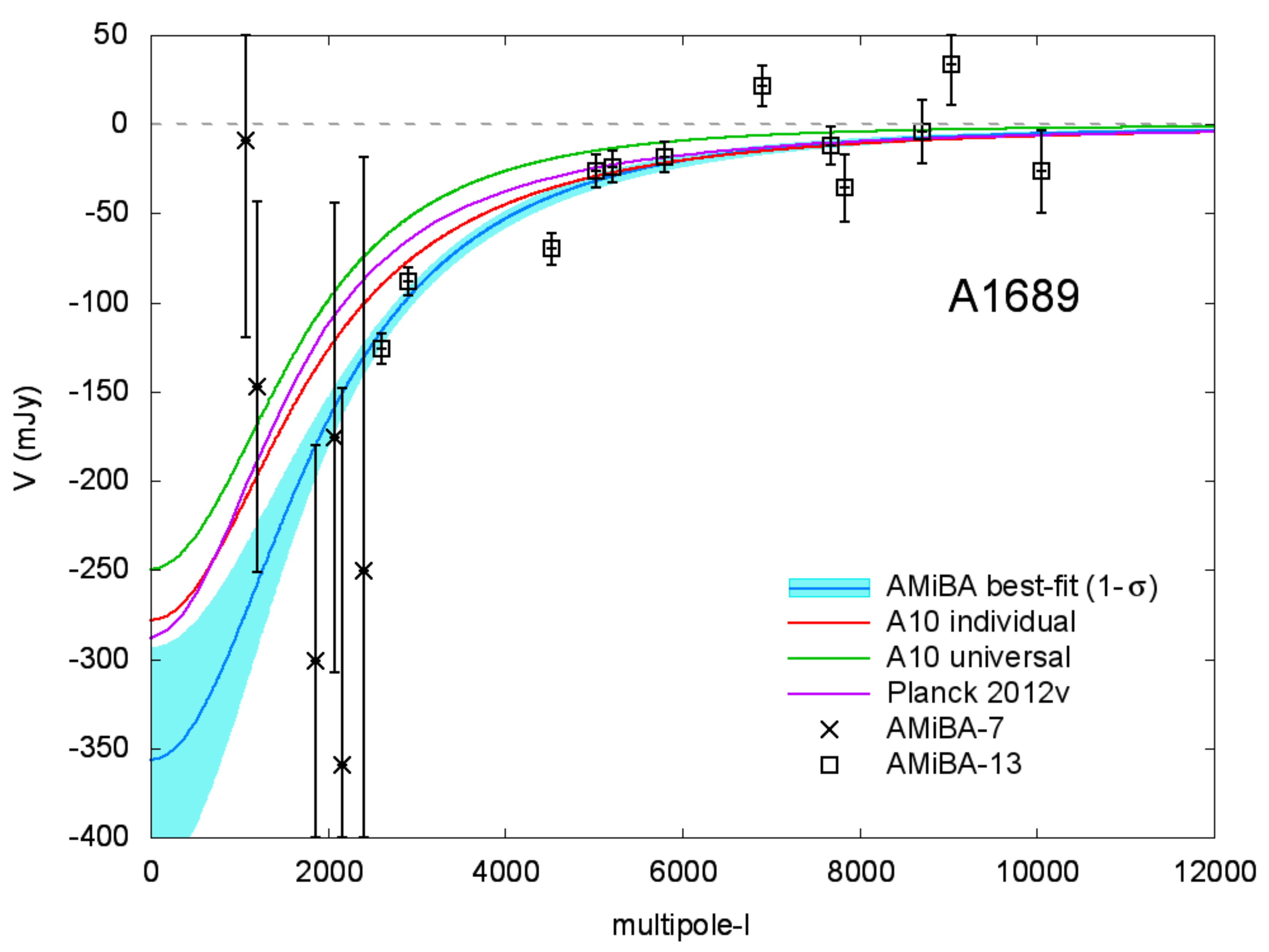}
 \end{center}
 \caption{Visibility flux profile of Abell 1689 as a
 function of angular multipole $l$. The black crosses and squares with
 error bars represent visibility band powers obtained from AMiBA-7
 ($l<2400$) and AMiBA-13 ($l>2400$) observations, respectively. The blue-shaded
 region shows the 68.3\% confidence interval in the marginalized
 posterior distribution of the gNFW pressure profile. 
 Abell 1689 (see Table~C.1 of A10, RXC~J1311.4$-$0120 aka Abell 1689).
 The green line represents the A10 {\em universal} gNFW profile.
 The purple line represents the gNFW profile for the cluster obtained by \citet{planckcollaboration2012v} from {\em Planck} SZ + {\em XMM-Newton} observations. 
 All model predictions are multiplied with a Gaussian with a width of 11\farcm5 in image domain to account for the primary beam attenuation of AMiBA-13.
\label{a1689_combine}} 
\end{figure}

We characterize the gas pressure structure of Abell 1689 with the gNFW profile (Equation~\ref{eq:gNFW}). 
For profile fitting, we have fixed the following structural/shape parameters with the {\em universal} values given in Eq.~(12) of A10: 
\begin{equation}
\label{eq:gNFW_par}
	[\gamma, \alpha, \beta] = [0.3081, 1.0510, 5.4905].
\end{equation}
We adopt $R_{500} = 1.351$\,Mpc, given in Table~C.1 of A10, which is based on an iterative estimation from their {\em XMM-Newton} data using the integrated mass vs integrated Compton-$y$, $M_{500}$--$Y_X$,  scaling relation.   
We apply the flux-loss correction discussed in Section~\ref{sec:phase-error}. 
The resulting AMiBA visibility band powers for Abell 1689 are shown in Figure~\ref{a1689_combine}.
Our best-fit gNFW profile and its $1\sigma$ uncertainty range are also presented in Figure~\ref{a1689_combine}.
For comparison, three additional gNFW pressure profiles determined for Abell 1689 from the literature are reproduced in the same plot.
The first profile (marked A10 individual) is the best fit to the X-ray data of Abell 1689 in A10.
The second profile (marked A10 universal) is the best fit to the X-ray data of the entire sample of 33 clusters in A10.
In both cases, all parameters were free except for the outer slope parameter, $\beta = 5.4905$. 
The third profile is from \citet{planckcollaboration2012v}, where they combined their {\em Planck} SZE data with archival {\em XMM-Newton} X-ray data \citep{planckcollaboration2011xi} and fitted a gNFW profile. In their fitting, $\gamma$ was fixed to 0.31, while the remaining parameters were free. 
In reproducing these profiles, $R_{500} = 1.351$~Mpc and the corresponding $P_{500} = 4.169 \times 10^3$~keV~cm$^{-3}$ were used.
The profiles are multiplied by a Gaussian with a width of $11\farcm5$ to account for the AMiBA-13 primary beam effect and then inverted to $uv$-space for plotting. 
Table~\ref{tab:chi2} summarizes the gNFW parameters of these profiles and their $\chi^2$ values against the AMiBA data.

\begin{deluxetable}{lcccccc}
\tablecaption{gNFW Parameters Determined for Abell 1689 \label{tab:chi2}}
\tablecolumns{7}
\tablehead{
\colhead{Profile Name} & \colhead{$P_0$} & \colhead{$c_{500}$} & \colhead{$\alpha$} & \colhead{$\beta$} & \colhead{$\gamma$} & \colhead{$\chi^2$\tablenotemark{a}}\\
 & 
}
\startdata
AMiBA              & 15.64 & 1.404 & {\bf 1.051}\tablenotemark{b} & {\bf 5.4905}\tablenotemark{b} & {\bf 0.3081}\tablenotemark{b} & 28\\
A10 (individual)   &23.13 & 1.16 & 0.78 & {\bf 5.4905}\tablenotemark{b} & 0.399 & 40\\
A10 (universal)    &8.403 & 1.177 & 1.051 & {\bf 5.4905}\tablenotemark{b} & 0.3081 & 84\\
Planck             &33.95 & 1.76 & 0.77 & 4.49 & {\bf 0.31}\tablenotemark{b} & 57
\enddata
\tablenotetext{a}{The $\chi^2$ are computed against 18 AMiBA reconstructed band powers.}
\tablenotetext{b}{The bold-faced numbers are fixed in their respective fitting.}
\end{deluxetable}

\begin{figure}
[!htb] 
 \begin{center}
  \includegraphics[width=0.45\textwidth,angle=0,clip]{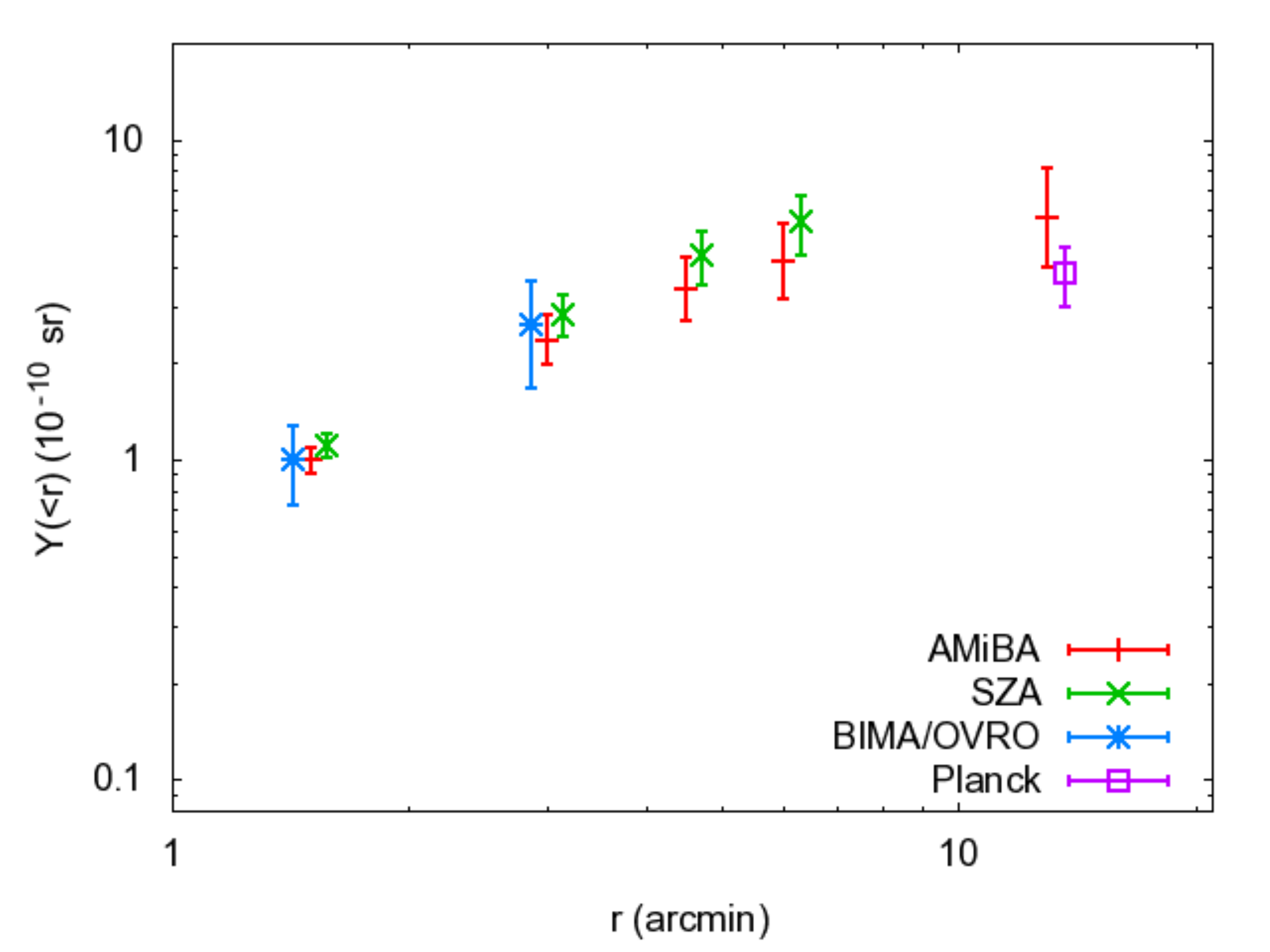}
 \end{center}
 \caption{Multi-scale SZE constraints on the cylindrically integrated
 Compton-$y$, $Y(<r)$, derived for Abell 1689. Our joint AMiBA-7 and AMiBA-13
 constraints (red) are compared with other interferometric (BIMA/OVRO:
 blue; SZA: green) and bolometric ({\em Planck}: purple) SZE observations.
 For visual clarity, results measured at the same enclosing radius are slightly shifted from each other horizontally.
 The AMiBA results, after the
 upward-flux correction (see Section~\ref{sec:phase-error}), are consistent with all
 other SZE observations. The AMiBA uncertainty increases with increasing
 integration radius ($r$), showing that the pressure structure on larger
 angular scales is poorly constrained by AMiBA. \label{fig:cylinder_y}}
\end{figure}

Abell 1689 has also been observed by other interferometers operating at 30\,GHz, namely the Berkeley-Illinois-Maryland Array (BIMA), the Owens Valley Radio Observatory (OVRO), and the Sunyaev-Zel'dovich Array (SZA). The BIMA and OVRO observations are presented in \citet{laroque2006}, while the SZA observations are presented in \citet{gralla2011}.  
\citet{umetsu2015} determined the best-fit gNFW pressure parameters from the BIMA/OVRO and SZA data separately. The gNFW models were then cylindrically integrated to yield $Y(<r)$ at integration radii $r$ probed by the respective instruments. These results are summarized in their Table~6. \citet{umetsu2015} also determined $Y(<13\arcmin)$ from {\em Planck} SZE observations.  
Figure~\ref{fig:cylinder_y} compares the cylindrically integrated $Y(<r)$ measurements from AMiBA and other SZE observations. 
After applying the upward-flux correction, we find that the AMiBA results are in agreement with both the SZA and the BIMA/OVRO results. 
At larger integration radii, the AMiBA constraints are weaker because of
the shallow AMiBA-7 data. The AMiBA results are consistent within a
$1\sigma$ uncertainty with the {\em Planck} results. Further
discussion on the gNFW profile fitting and integrated Compton-y results
for the other clusters in our sample will be presented in a forthcoming
paper (Wang et al., in preparation).

\subsection{SZE Detection of the Cluster RCS J1447+0828}\label{sec:M_1447}

In this section we explore the possibility of using AMiBA SZE observations as a proxy for the total mass of clusters. 
The primary AMiBA SZE observable considered here is the peak flux density $I_\mathrm{SZ}$ in a dirty image, constructed from visiblity data with natural weighting.
This is a direct observable from interferometric AMiBA observations and is closely related to $Y_{2500}=Y(<R_{2500})$, the integrated Compton-$y$ parameter interior to a cylinder of radius $R_{2500}$.
In Appendix~\ref{sec:app.1447} we describe our Monte--Carlo method that simulates the probability distribution $P(I_\mathrm{SZ}|M_{200})$ of the AMiBA SZE observable as a function of halo mass $M_{200}$ (and redshift), given the underlying halo concentration--mass ($c$--$M$) relation and intrinsic distributions of the gNFW pressure profile parameters.

\begin{figure*}[!htb] 
 \begin{center}
  \includegraphics[width=0.45\textwidth,angle=0,clip]{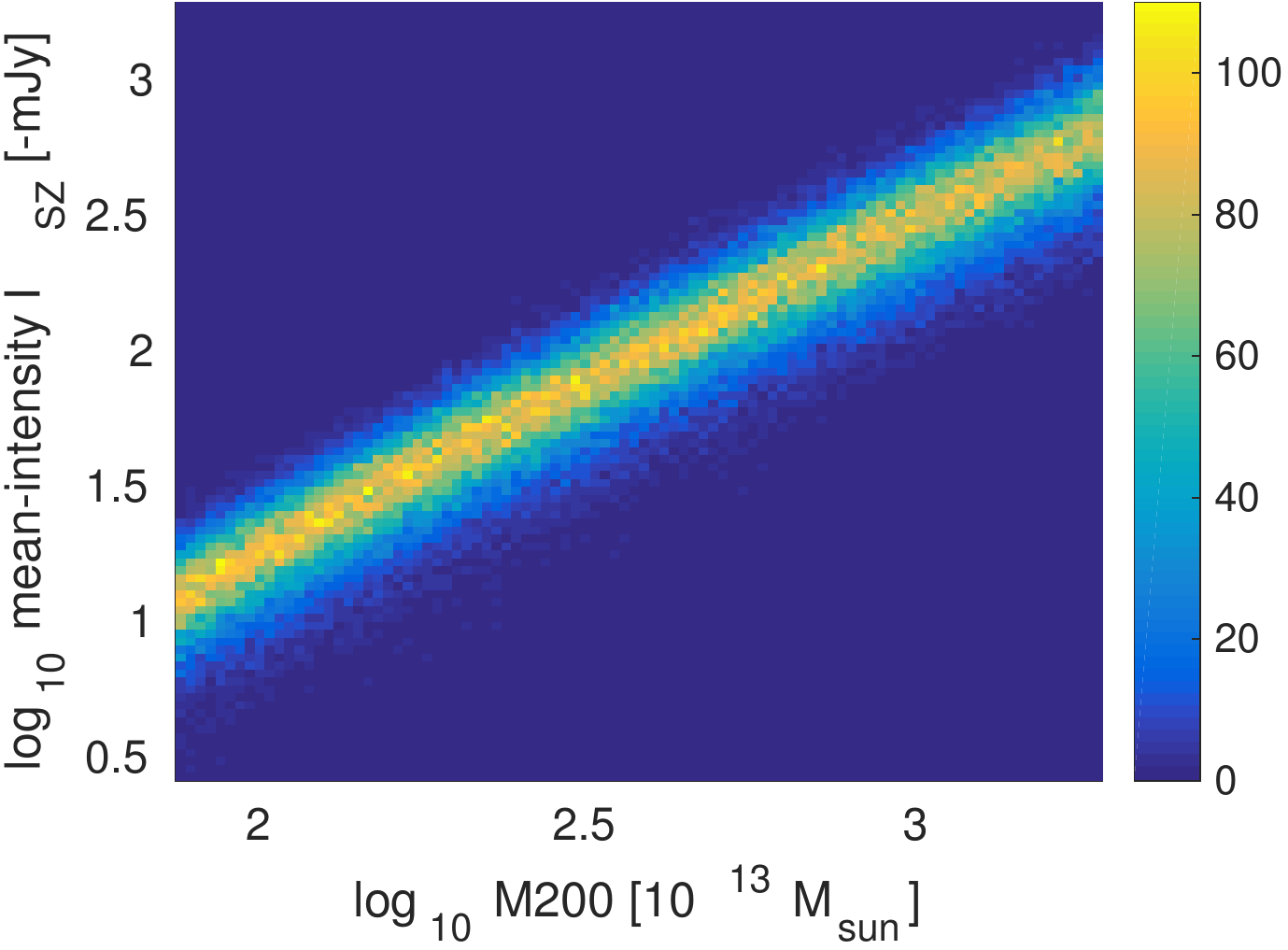}
  \includegraphics[width=0.45\textwidth,angle=0,clip]{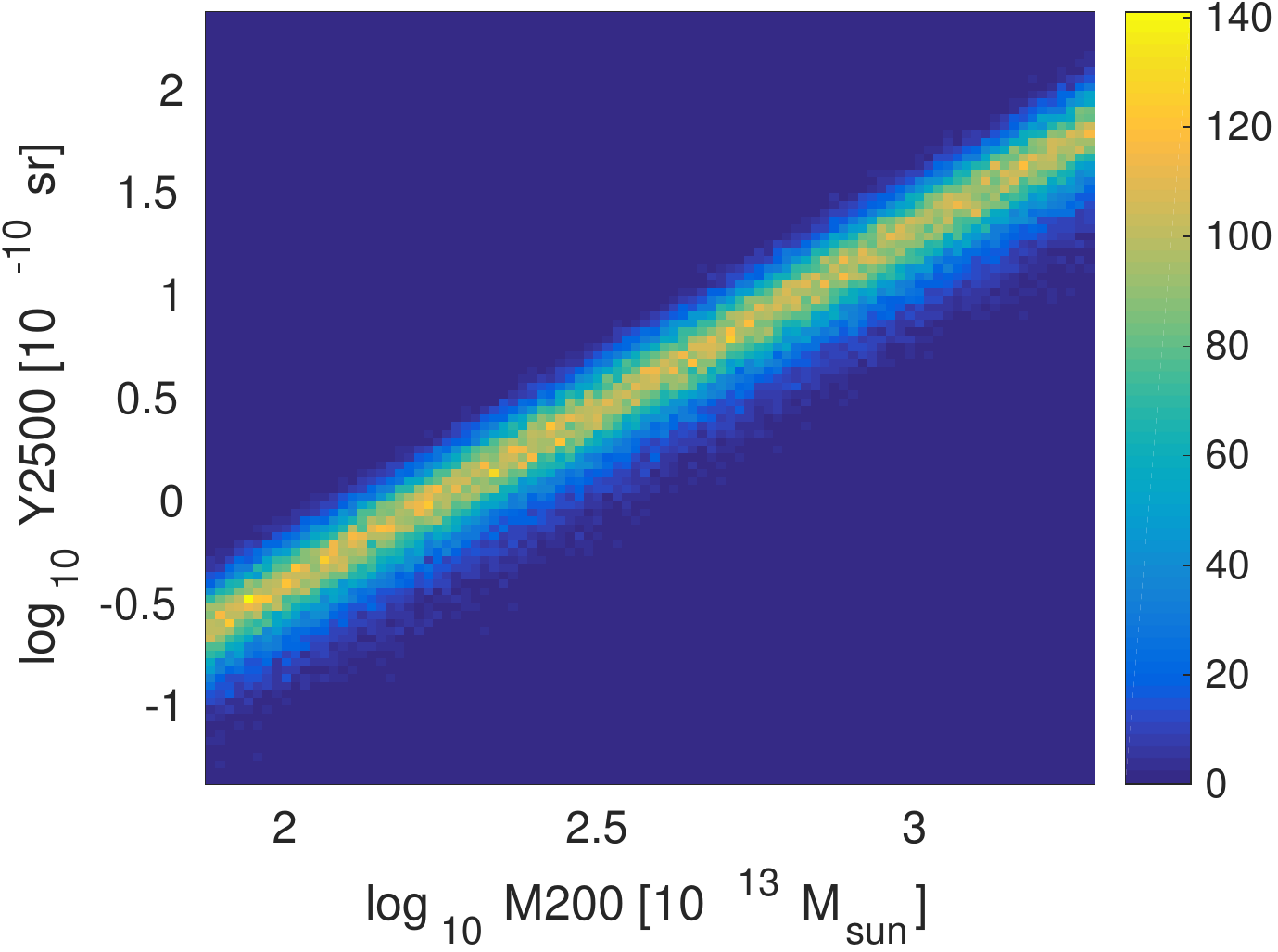}
\end{center}
 \caption{
 Simulated scaling relations for RCS J1447+0828
 assuming that it is a cool-core cluster (see Section~\ref{sec:M_1447}). \textbf{Left:}  $I_\mathrm{SZ}-M_{200}$ relation. \textbf{Right:} $Y_{2500}-M_{200}$ relation. A fixed number of clusters were simulated in each of the logarithmically spaced mass bin. The color scale depicts the number of clusters falling in the logarithmically spaced $I_\mathrm{SZ}$ or $Y_{2500}$ cells at a given mass bin.} \label{fig:scalings}
\end{figure*}

As a demonstration, we choose RCS J1447+0828, an optically selected cluster at $z=0.38$ from the Red-sequence Cluster Survey \citep[][RCS]{gladders2005}. 
Subsequent {\em Chandra} X-ray observations identified it as a strong cool-core cluster \citep{hicks2013}. 
Figure~\ref{fig:scalings} shows the  $I_\mathrm{SZ}$--$M_{200}$ relation predicted for this cluster, with the simulated gNFW parameters drawn from the REXCESS cool-core subsample presented in A10 (see Appendix~\ref{sec:app.1447}). 
Also shown in this figure is the $Y_{2500}$--$M_{200}$ relation obtained
using the same prior for comparison.
The slope of $\log_{10}I_\mathrm{SZ}$ versus $\log_{10}M_{200}$ is 1.21 with a scatter of $\pm 0.13$~dex. The slope of $\log_{10}Y_{2500}$ versus $\log_{10}M_{200}$ is 1.73 with a scatter of $\pm 0.14$~dex.
While $Y_{2500}$ is calculated from the simulated intrinsic cluster profile, the observable $I_\mathrm{SZ}$ is obtained after applying a Gaussian beam with a FWHM of $11\farcm5$ to simulate the primary beam attenuation of AMiBA-13 observations.

\begin{figure}
 \plotone{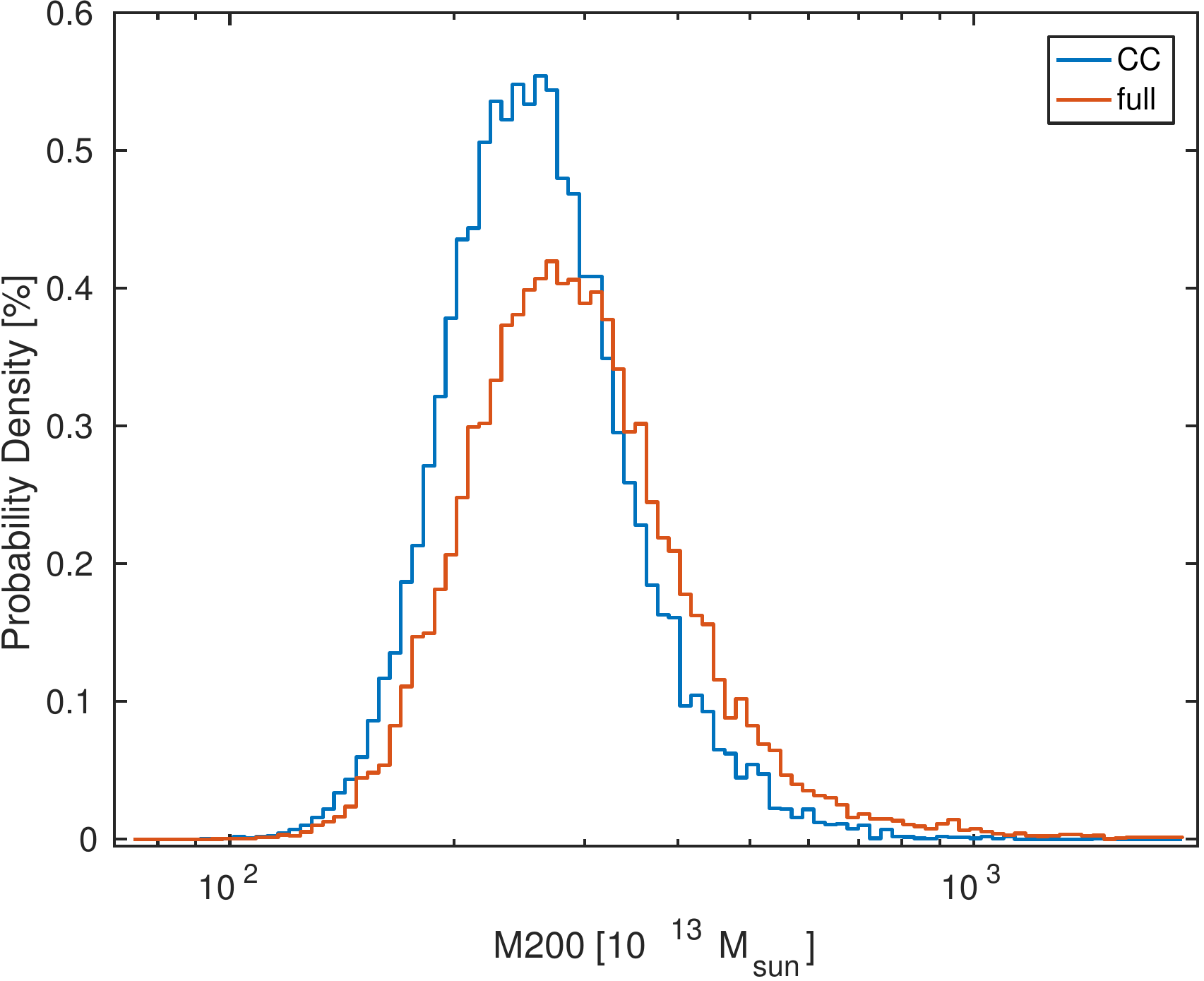}
 \caption{Posterior mass distribution of RCS J1447+0828 derived
 from the $I_\mathrm{SZ}-M_{200}$ scaling relation, given the AMiBA-13 measured peak intensity $I_\mathrm{SZ}= -57.3\pm7.3$\,mJy\,beam$^{-1}$. The blue curve is derived from the cool-core (CC) cluster simulation shown in the left panel of Figure~\ref{fig:scalings}. The red curve shows the posterior mass distribution similarly derived from the full cluster sample simulation (see text). } \label{fig:i-m}
\end{figure}

Multiplying the probability distribution function $P(I_\mathrm{SZ}|M_{200})$ with a Gaussian likelihood of the AMiBA-13 measurement $I_\mathrm{SZ}=-57.3\pm7.3$\,mJy\,beam$^{-1}$ (Table~\ref{snr_table}, after the upward-flux correction) and integrating over $I_\mathrm{SZ}$ yields the posterior distribution of $M_{200}$ for RCS J1447+0828, as shown in Figure~\ref{fig:i-m}.  
Here, we take the biweight estimator of \citet{beers1990} to be the central location ($C_\mathrm{BI}$) and scale ($S_\mathrm{BI}$) of the marginalized posterior mass distribution.
We find $M_{200} = 26.9 \pm 7.4 \times 10^{14} M_{\sun}$.
From the same simulation, we can also construct the mass relation at a higher overdensity, e.g., $I_\mathrm{SZ}$--$M_{2500}$. With the AMiBA-13 measurement, we find $M_{2500} = 6.6 \pm 2.1 \times 10^{14} M_{\sun}$.
The latter result can be compared to the X-ray-based $M_{2500}$ mass estimates of \citet{hicks2013}, who find from {\em Chandra} observations $M_{2500}=4.8^{+0.7}_{-0.5} \times 10^{14} M_{\sun}$ using $Y_\mathrm{X}$ as a mass proxy and $M_{2500}=(6\pm2) \times 10^{14} M_{\sun}$ using $T_\mathrm{X}$.
The AMiBA-13 estimate of $M_{2500}$ is consistent with these measurements within uncertainties.

Alternatively, one can relax the cool-core assumption and simulate the
cluster observable using gNFW parameters drawn from the full REXCESS sample in A10. The posterior distribution of $M_{200}$, also shown in Figure~\ref{fig:i-m}, favors higher masses than the cool-core results: $M_{200} = 31.1 \pm 10.8 \times 10^{14} M_{\sun}$. At a higher overdensity, we obtain $M_{2500} = 7.4 \pm 2.5 \times 10^{14} M_{\sun}$. 
In general, for a given AMiBA-13 SZE measurement, the cool-core and
disturbed cluster priors give lower and higher mass estimates,
respectively, relative to those from the full-sample prior.
The discrepancy is of the same order as the width of posterior mass distributions.

\subsection{Mass Estimates Compared to Lensing Masses}\label{sec:mass-statistics}
In Section~\ref{sec:M_1447}, we showed that, for the cluster RCS J1447+0828,
the AMiBA-13 derived and X-ray based cluster mass estimates are
consistent with each other.
For the rest of the clusters shown in this work, a direct comparison
with gravitational lensing mass measurements can be
made.

\begin{figure*}
 \plottwo{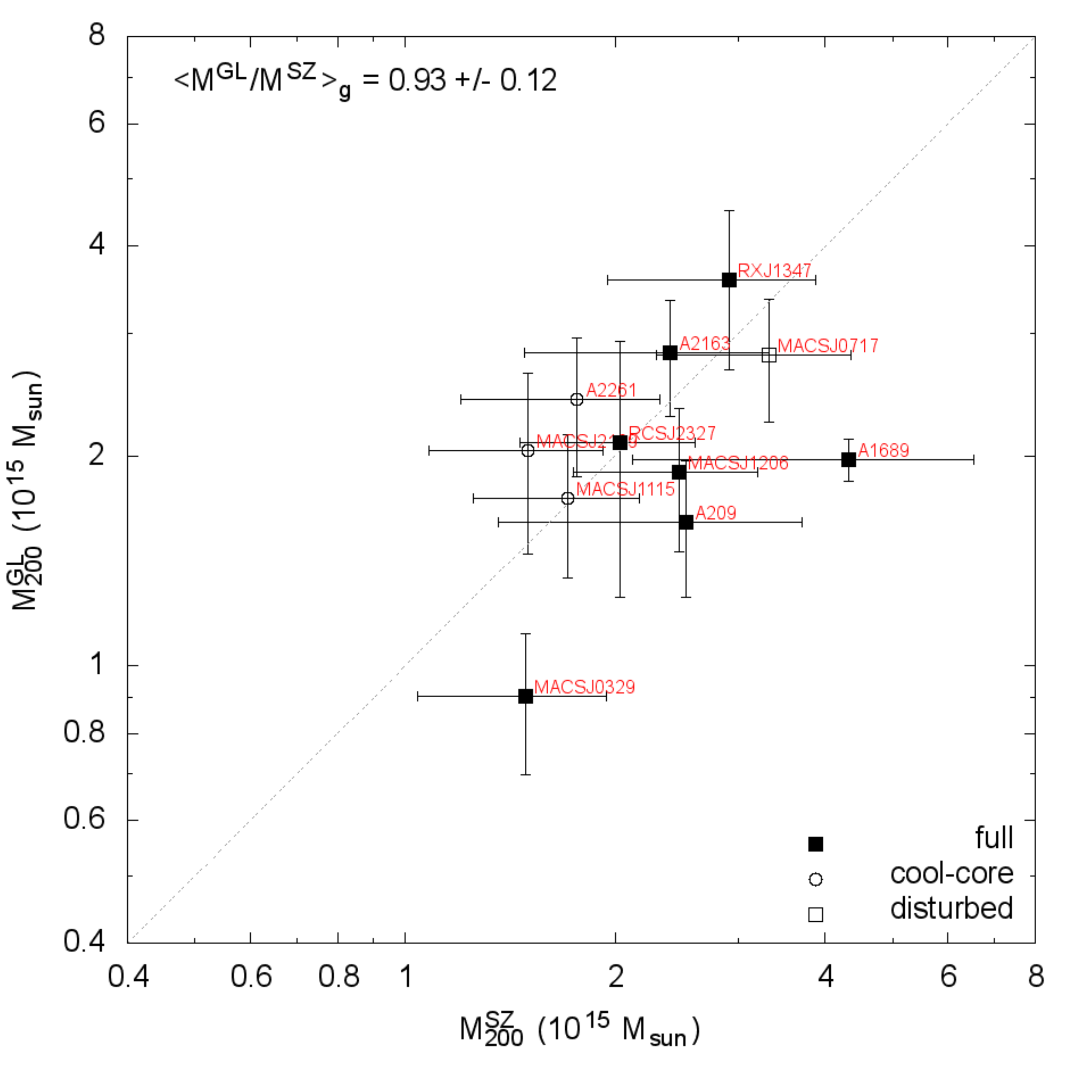}{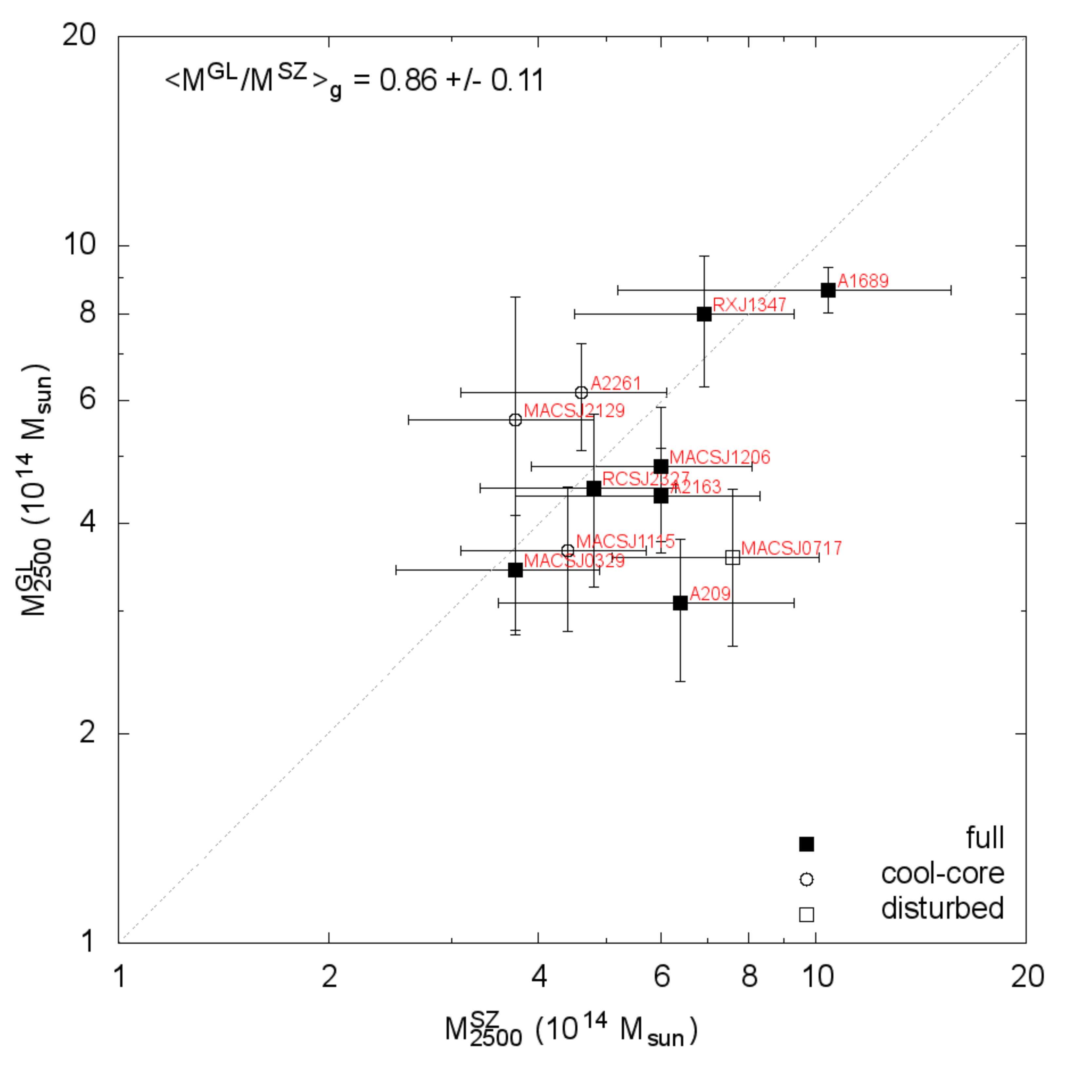}
 \caption{Comparison of gravitational lensing and AMiBA-13 SZ
 derived masses for eleven clusters in our sample. The spherical
 enclosed masses are derived at two overdensities, $M_{200}$ (left) and
 $M_{2500}$ (right). The one-to-one relation is represented by the dotted line. The
 geometric mean of the mass ratios is indicated at the top of each diagram.}\label{fig:mstat} 
\end{figure*}

Table~\ref{mstat_table} and Figure~\ref{fig:mstat} summarize and
compare, for the other eleven clusters,
recent lensing mass measurements taken from the published
literature and our AMiBA-13 results.
The spherical enclosed masses are derived at two over-densities, $M_{200}$ and $M_{2500}$. 
To derive $M^\mathrm{SZ}$ from AMiBA-13 data, we performed Monte-Carlo
simulations for each cluster with gNFW parameters drawn from either the
cool-core or disturbed subsamples, or the full sample of A10, according to the X-ray classification of the cluster \citep[see Table~\ref{coord_table} and] []{sayers2013b}. 
For clusters identified as both cool-core and disturbed (MACS
J0329.7-0211 and RX J1347.5-1145), we make a less-informative assumption
and simulate them with gNFW parameters drawn from the full sample range
(see Section~\ref{sec:M_1447}). 
Figure~\ref{fig:mstat} shows that the three cool-core clusters tend to
have lower $M^\mathrm{SZ}$ estimates with $M^\mathrm{GL}/M^\mathrm{SZ} > 1$.
It hints, albeit with a small sample, that using the less informative
full-sample prior,
which includes the parameter ranges of cool-core and disturbed
subsamples, may be adequate for cluster mass estimation. 
We also quote the geometric mean and uncertainty \citep[e.g.,][]{umetsu2016}
of the mass ratio 
$\langle M^\mathrm{GL} / M^\mathrm{SZ}\rangle_g$, where each cluster is weighted
by its error, and the errors of $M^\mathrm{SZ}$ and
$M^\mathrm{GL}$ are treated as independent. The mean mass ratio
of this sample with eleven clusters shows no significant bias at both
over-densities.

\begin{deluxetable*}{lccccccc}
\tablecaption{Comparison of Mass Estimates with Lensing \label{mstat_table}}
\tablecolumns{7}
\tablehead{
\colhead{Cluster} & \colhead{sim. type\tablenotemark{a}} & \colhead{$M_{200}^\mathrm{SZ}$} & \colhead{$M_{200}^\mathrm{GL}$}\tablenotemark{b} & \colhead{$M_{2500}^\mathrm{SZ}$} & \colhead{$M_{2500}^\mathrm{GL}$}\tablenotemark{b} & \colhead{Lensing ref.}\\
  & & $10^{14} M_{\sun}$ & $10^{14} M_{\sun}$ & $10^{14} M_{\sun}$ & $10^{14} M_{\sun}$ & }
\startdata
Abell 1689         & f & $43.3 \pm 22.1$ & $19.8 \pm 1.4$ & $10.4 \pm 5.2$ & $8.7 \pm 0.65$ & \citet{umetsu2015}\\ 
Abell 2163         & f & $24.0 \pm  9.2$ & $28.2 \pm 5.3$ & $ 6.0 \pm 2.3$ & $4.4 \pm 0.75$ & \citet{okabe2011}\\
Abell 209          & f & $25.3 \pm 11.7$ & $16.1 \pm 3.6$ & $ 6.4 \pm 2.9$ & $3.1 \pm 0.71$ & \citet{umetsu2016}\\ 
Abell 2261         & c & $17.6 \pm  5.6$ & $24.1 \pm 5.4$ & $ 4.6 \pm 1.5$ & $6.2 \pm 1.08$ & \citet{umetsu2016}\\ 
MACS J1115.9+0129  & c & $17.1 \pm  4.6$ & $17.4 \pm 4.0$ & $ 4.4 \pm 1.3$ & $3.7 \pm 0.86$ & \citet{umetsu2016}\\ 
MACS J1206.2-0847  & f & $24.7 \pm  7.3$ & $19.0 \pm 4.4$ & $ 6.0 \pm 2.1$ & $4.8 \pm 1.06$ & \citet{umetsu2016}\\\ 
MACS J0329.7-0211  & f\tablenotemark{c}& $14.9 \pm  4.5$ & $ 9.4 \pm 2.1$ & $ 3.7 \pm 1.2$ & $3.4 \pm 0.67$ & \citet{umetsu2016}\\ 
RX J1347.5-1145    & f\tablenotemark{c}& $29.1 \pm  9.6$ & $35.8 \pm 9.2$ & $ 6.9 \pm 2.4$ & $8.0 \pm 1.70$ & \citet{umetsu2016}\\ 
MACS J0717.5+3745  & d & $33.2 \pm 10.3$ & $28.0 \pm 5.6$ & $ 7.6 \pm 2.5$ & $3.6 \pm 0.91$ & \citet{umetsu2016}\\
MACS J2129.4-0741  & c & $15.0 \pm  4.2$ & $20.4 \pm 5.9$ & $ 3.7 \pm 1.1$ & $5.6 \pm 2.82$ & \citet{applegate2014}\tablenotemark{d}\\
RCS J2327-0204     & f & $20.3 \pm  5.7$ & $20.9 \pm 8.4$ & $ 4.8 \pm 1.5$ & $4.5 \pm 1.25$ & \citet{sharon2015}\\
\hline
$\langle M^\mathrm{GL} / M^\mathrm{SZ}\rangle_g$ &   & \multicolumn{2}{c}{$0.93 \pm 0.12$}  & \multicolumn{2}{c}{$0.86 \pm 0.11$} &
\enddata

\tablenotetext{a}{gNFW parameter range for the Monte-Carlo simulations. f: full sample in A10; c: cool-core clusters subsample; d: disturbed clusters subsample.}
\tablenotetext{b}{Lensing masses have been converted to the cosmology adopted in this work, where $h=0.67$.}
\tablenotetext{c}{For clusters that are classified as both cool-core and disturbed, we choose to use a less-informative prior by drawing simulation parameters from the full sample in A10.}
\tablenotetext{d}{The authors obtained a spherical mass estimate of $M(<1.5\,\mathrm{Mpc})$ for MACS J2129.4-0741 assuming the NFW model with halo concentration $c_{200}=4$. We have converted it to the spherical $M_{200}$ and $M_{2500}$ masses using the same NFW model.}
 
\end{deluxetable*}

\section{Summary and Conclusions}\label{sec:conclusion}

The Yuan-Tseh Lee Array for Microwave Background Anisotropy (AMiBA) is a co-planar interferometer array operating at a wavelength of 3\,mm to measure the Sunyaev-Zel’dovich effect (SZE) of galaxy clusters at arcminute scales.
After an intial phase with seven close-packed 0.6\,m antennas (AMiBA-7), the AMiBA was upgraded to a 13-element array. In the following, we summarize the AMiBA expansion, its commissioning and results from its SZE observing program.

\begin{itemize}

\item 
{\it Array upgrade.}
The expanded AMiBA-13 is comprised of thirteen new lightweight carbon-fiber-reinforced plastic (CFRP) 1.2\,m diameter antennas with a field-of-view (FoV) of $11\arcmin$. 
Despite weighing only 25\,kg, the antennas show excellent structural behaviour with an overall aperture efficiency of about 0.6. 
All antennas are co-mounted on the hexapod-driven CFRP platform in a close-packed configuration, yielding baselines from 1.4\,m to 4.8\,m which sample scales on the sky from $2\arcmin$ to $10\arcmin$ with a synthesized beam of $2\farcm5$.
The shortest baseline of 1.4\,m is chosen to minimize both CMB leakage ($\sim11$\,mJy) and antenna cross-talk ($\sim-135$\,dB).
Additional correlators and six new receivers with noise temperatures between 55-75\,K complete the AMiBA expansion.

\item
{\it Commissioning and new correction schemes.}
For a bandwidth of 20\,GHz, baselines of up to 5\,m and an antenna FoV of $11\arcmin$, instrumental delays need to be within about $\pm22$\,ps for AMiBA-13. Delays are initially measured for every baseline with a movable broadband noise source emitting towards the two receivers, and then further iterated by scanning the Sun with the entire array. Optimized delay solutions for all 78 baselines show an rms scatter of 22\,ps in residual delays. 
Additional pointing-dependent geometrical delays can result from the platform deformation. This repeatable deformation -- entirely negligible for the earlier central close-packed AMiBA-7 -- is modelled both from extensive photogrammetry surveys and the all-sky pointing correlation between two optical telescopes mounted on two different locations on the platform. This yields an all-sky deformation model. Platform deformation-induced antenna misalignments are measured to be within $\pm2\arcmin$, leading to an efficiency loss of a few percent. 
The refined all-sky radio pointing error is about $0\farcm4$ in rms, with a repeatability around $10\arcsec$, which gives an efficiency loss of less than 2\%. 
Overall, platform- and pointing-induced phase errors can reduce the point source flux recovery by $15\%$ to $25\%$. Measured visibilities are, thus, upward corrected by a factor of 1.25 to account for phase decoherence. An additional $\pm5\%$ systematic uncertainty is added in quadrature to the statistical errors. 

\item
{\it Calibration, tests and array performance.}
Flux and gain calibration are done by observing Jupiter and Saturn for at least one hour every night. Obscuration effects due to Saturn's rings are accounted for beyond a simple disk model. Data are flagged and checked against various criteria, such as, high noise, unstable baselines, upper-lower-band differences, non-Gaussianity, and calibration failure. Typically, 20\% to 50\% of the cluster data are flagged.
We demonstrate the absence of further systematics with a noise level consistent with zero in stacked $uv$-visibilities.  Scaling of stacked noise with integration time indicates an overall array efficiency $\eta=0.4$.

\item
{\it Cluster SZE observing program.}
Targeted cluster SZE observations with AMiBA-13 started in mid-2011 and ended in late 2014. Observations are carried out in a lead-trail observing mode.
The AMiBA-13 cluster sample consists of (1) the six clusters observed with AMiBA-7, (2) twenty clusters selected from the Cluster Lensing And Supernova survey with Hubble (CLASH) with complementary strong- and weak-lensing, X-ray, and additional SZE data from Bolocam and MUSTANG, and (3) seven optically selected cluster candidates from the Red-sequence Cluster Survey 2 (RCS2). From these samples, we present maps of a subset of twelve clusters, in a redshift range from 0.183 to 0.700, with AMiBA-13 SZE detections with signal-to-noise ratios ranging from about 5 to 11. Achieved noise levels are between 3 to 7 mJy/beam.
No significant radio point sources, with extrapolated flux levels of more than 1\,mJy at 90\,GHz, are found in the target or trailing fields.

\item
{\it AMiBA SZE detections of A1689 and RCS J1447.}
AMiBA-7 and AMiBA-13 observations of Abell 1689 are combined and jointly fitted to a gNFW model.  Our cylindrically integrated Compton-$y$ values for five radii are consistent with BIMA/OVRO, SZA, and {\em Planck} results.
We report the first targeted SZE detection towards the optically selected cluster RCS J1447$+$0828. We develop a Monte-Carlo approach to predict the AMiBA-13 observed SZE peak flux given an underlying halo concentration-mass relation and intrinsic distributions of gNFW pressure profile parameters. It is then used reversely to constrain halo mass given the AMiBA-13 SZE flux measurement.
Our estimates yield 
$M_{200} = 26.9 \pm 7.4 \times 10^{14} M_{\sun}$
and $M_{2500} = 6.6 \pm 2.1 \times 10^{14} M_{\sun}$.
The AMiBA-13 result for $M_{2500}$ agrees with X-ray-based measurements within error bars.

\item
{\it AMiBA-13 derived total mass estimates versus lensing masses.}
Using the same Monte-Carlo method, we have obtained total mass estimates
     for the other eleven clusters studied in this work and compared
     them with recent lensing mass measurements available in the literature. For
     this small sample, we find that the AMiBA-13 and lensing masses are
     in agreement. The geometric mean of the mass ratios is found to be $\langle M^\mathrm{GL}_{200} / M^\mathrm{SZ}_{200}\rangle_g = 0.93 \pm 0.12$ and $\langle M^\mathrm{GL}_{2500} / M^\mathrm{SZ}_{2500}\rangle_g = 0.86 \pm 0.11$.

\end{itemize}

\acknowledgments
{\bf Acknowledgments}  
Capital and operational funding for AMiBA came from the Ministry of Education and the Ministry of Science and Technology of Taiwan (MoST) as part of the Cosmology and Particle Astrophysics (CosPA) initiative. Additional funding also came in the form of an Academia Sinica Key Project. 
We thank the Smithsonian Astrophysical Observatory for hosting the AMiBA project staff at the SMA Hilo Base Facility. We thank the NOAA for locating the AMiBA project on their site on Mauna Loa. We also thank the Hawaiian people for allowing astronomers to work on their mountains in order to study the universe.
JHPW acknowledges support by the NSC grant NSC 100-2112-M-002-004-MY3 and the MoST grant MoST 103-2628-M-002-006-MY4
Support from the STFC for MB is also acknowledged.
PMK acknowledges support from MoST 103-2119-M-001-009 and from an Academia Sinica Career Development Award.
KU acknowledges support by the MoST grants MOST 103-2112-M-001-030-MY3 and MOST 103-2112-M-001-003-MY3.
We are grateful to Nobuhiro Okabe for useful comments and help with
the lensing mass comparison. We also thank the anonymous referee for their useful comments and suggestions. 

\bibliography{References}

\begin{thebibliography}{79}
\expandafter\ifx\csname natexlab\endcsname\relax\def\natexlab#1{#1}\fi

\bibitem[{{Agudo} {et~al.}(2012){Agudo}, {Thum}, {Wiesemeyer}, {Molina},
  {Casadio}, {G{\'o}mez}, \& {Emmanoulopoulos}}]{agudo2012}
{Agudo}, I., {Thum}, C., {Wiesemeyer}, H., {et~al.} 2012,
  \href{http://dx.doi.org/10.1051/0004-6361/201218801}{\aap, 541, A111}

\bibitem[{{Applegate} {et~al.}(2014){Applegate}, {von der Linden}, {Kelly},
  {Allen}, {Allen}, {Burchat}, {Burke}, {Ebeling}, {Mantz}, \&
  {Morris}}]{applegate2014}
{Applegate}, D.~E., {von der Linden}, A., {Kelly}, P.~L., {et~al.} 2014,
  \href{http://dx.doi.org/10.1093/mnras/stt2129}{\mnras, 439, 48}

\bibitem[{{Arnaud} {et~al.}(2010){Arnaud}, {Pratt}, {Piffaretti},
  {B{\"o}hringer}, {Croston}, \& {Pointecouteau}}]{arnaud2010}
{Arnaud}, M., {Pratt}, G.~W., {Piffaretti}, R., {et~al.} 2010,
  \href{http://dx.doi.org/10.1051/0004-6361/200913416}{\aap, 517, A92}

\bibitem[{{Beers} {et~al.}(1990){Beers}, {Flynn}, \& {Gebhardt}}]{beers1990}
{Beers}, T.~C., {Flynn}, K., \& {Gebhardt}, K. 1990,
  \href{http://dx.doi.org/10.1086/115487}{\aj, 100, 32}

\bibitem[{{Birkinshaw}(1999)}]{birkinshaw1999}
{Birkinshaw}, M. 1999,
  \href{http://dx.doi.org/10.1016/S0370-1573(98)00080-5}{\physrep, 310, 97}

\bibitem[{{Bleem} {et~al.}(2015){Bleem}, {Stalder}, {de Haan}, {Aird}, {Allen},
  {Applegate}, {Ashby}, {Bautz}, {Bayliss}, {Benson}, {Bocquet}, {Brodwin},
  {Carlstrom}, {Chang}, {Chiu}, {Cho}, {Clocchiatti}, {Crawford}, {Crites},
  {Desai}, {Dietrich}, {Dobbs}, {Foley}, {Forman}, {George}, {Gladders},
  {Gonzalez}, {Halverson}, {Hennig}, {Hoekstra}, {Holder}, {Holzapfel},
  {Hrubes}, {Jones}, {Keisler}, {Knox}, {Lee}, {Leitch}, {Liu}, {Lueker},
  {Luong-Van}, {Mantz}, {Marrone}, {McDonald}, {McMahon}, {Meyer}, {Mocanu},
  {Mohr}, {Murray}, {Padin}, {Pryke}, {Reichardt}, {Rest}, {Ruel}, {Ruhl},
  {Saliwanchik}, {Saro}, {Sayre}, {Schaffer}, {Schrabback}, {Shirokoff},
  {Song}, {Spieler}, {Stanford}, {Staniszewski}, {Stark}, {Story}, {Stubbs},
  {Vanderlinde}, {Vieira}, {Vikhlinin}, {Williamson}, {Zahn}, \&
  {Zenteno}}]{bleem2015}
{Bleem}, L.~E., {Stalder}, B., {de Haan}, T., {et~al.} 2015,
  \href{http://dx.doi.org/10.1088/0067-0049/216/2/27}{\apjs, 216, 27}

\bibitem[{{Broadhurst} {et~al.}(2005){Broadhurst}, {Ben{\'{\i}}tez}, {Coe},
  {Sharon}, {Zekser}, {White}, {Ford}, {Bouwens}, {Blakeslee}, {Clampin},
  {Cross}, {Franx}, {Frye}, {Hartig}, {Illingworth}, {Infante}, {Menanteau},
  {Meurer}, {Postman}, {Ardila}, {Bartko}, {Brown}, {Burrows}, {Cheng},
  {Feldman}, {Golimowski}, {Goto}, {Gronwall}, {Herranz}, {Holden}, {Homeier},
  {Krist}, {Lesser}, {Martel}, {Miley}, {Rosati}, {Sirianni}, {Sparks},
  {Steindling}, {Tran}, {Tsvetanov}, \& {Zheng}}]{Broadhurst2005a}
{Broadhurst}, T., {Ben{\'{\i}}tez}, N., {Coe}, D., {et~al.} 2005,
  \href{http://dx.doi.org/10.1086/426494}{\apj, 621, 53}

\bibitem[{{Carlstrom} {et~al.}(2002){Carlstrom}, {Holder}, \&
  {Reese}}]{carlstrom2002}
{Carlstrom}, J.~E., {Holder}, G.~P., \& {Reese}, E.~D. 2002,
  \href{http://dx.doi.org/10.1146/annurev.astro.40.060401.093803}{\araa, 40,
  643}

\bibitem[{{Chen} {et~al.}(2009){Chen}, {Li}, {Hwang}, {Jiang}, {Altamirano},
  {Chang}, {Chang}, {Chang}, {Chiueh}, {Chu}, {Han}, {Huang}, {Kesteven},
  {Kubo}, {Martin-Cocher}, {Oshiro}, {Raffin}, {Wei}, {Wang}, {Wilson}, {Ho},
  {Huang}, {Koch}, {Liao}, {Lin}, {Liu}, {Molnar}, {Nishioka}, {Umetsu},
  {Wang}, \& {Wu}}]{chen2009}
{Chen}, M.-T., {Li}, C.-T., {Hwang}, Y.-J., {et~al.} 2009,
  \href{http://dx.doi.org/10.1088/0004-637X/694/2/1664}{\apj, 694, 1664}

\bibitem[{{Coe} {et~al.}(2010){Coe}, {Ben{\'{\i}}tez}, {Broadhurst}, \&
  {Moustakas}}]{Coe2010}
{Coe}, D., {Ben{\'{\i}}tez}, N., {Broadhurst}, T., \& {Moustakas}, L.~A. 2010,
  \href{http://dx.doi.org/10.1088/0004-637X/723/2/1678}{\apj, 723, 1678}

\bibitem[{{Condon} {et~al.}(1998){Condon}, {Cotton}, {Greisen}, {Yin},
  {Perley}, {Taylor}, \& {Broderick}}]{condon1998}
{Condon}, J.~J., {Cotton}, W.~D., {Greisen}, E.~W., {et~al.} 1998,
  \href{http://dx.doi.org/10.1086/300337}{\aj, 115, 1693}

\bibitem[{{Diego} {et~al.}(2015){Diego}, {Broadhurst}, {Benitez}, {Umetsu},
  {Coe}, {Sendra}, {Sereno}, {Izzo}, \& {Covone}}]{Diego2015}
{Diego}, J.~M., {Broadhurst}, T., {Benitez}, N., {et~al.} 2015,
  \href{http://dx.doi.org/10.1093/mnras/stu2064}{\mnras, 446, 683}

\bibitem[{{Donahue} {et~al.}(2014){Donahue}, {Voit}, {Mahdavi}, {Umetsu},
  {Ettori}, {Merten}, {Postman}, {Hoffer}, {Baldi}, {Coe}, {Czakon},
  {Bartelmann}, {Benitez}, {Bouwens}, {Bradley}, {Broadhurst}, {Ford},
  {Gastaldello}, {Grillo}, {Infante}, {Jouvel}, {Koekemoer}, {Kelson}, {Lahav},
  {Lemze}, {Medezinski}, {Melchior}, {Meneghetti}, {Molino}, {Moustakas},
  {Moustakas}, {Nonino}, {Rosati}, {Sayers}, {Seitz}, {Van der Wel}, {Zheng},
  \& {Zitrin}}]{donahue2014}
{Donahue}, M., {Voit}, G.~M., {Mahdavi}, A., {et~al.} 2014,
  \href{http://dx.doi.org/10.1088/0004-637X/794/2/136}{\apj, 794, 136}

\bibitem[{{Dutton} \& {Macci{\`o}}(2014)}]{Dutton2014}
{Dutton}, A.~A., \& {Macci{\`o}}, A.~V. 2014,
  \href{http://dx.doi.org/10.1093/mnras/stu742}{\mnras, 441, 3359}

\bibitem[{{Gilbank} {et~al.}(2011){Gilbank}, {Gladders}, {Yee}, \&
  {Hsieh}}]{gilbank2011}
{Gilbank}, D.~G., {Gladders}, M.~D., {Yee}, H.~K.~C., \& {Hsieh}, B.~C. 2011,
  \href{http://dx.doi.org/10.1088/0004-6256/141/3/94}{\aj, 141, 94}

\bibitem[{{Gladders} \& {Yee}(2005)}]{gladders2005}
{Gladders}, M.~D., \& {Yee}, H.~K.~C. 2005,
  \href{http://dx.doi.org/10.1086/427327}{\apjs, 157, 1}

\bibitem[{{Gralla} {et~al.}(2011){Gralla}, {Sharon}, {Gladders}, {Marrone},
  {Barrientos}, {Bayliss}, {Bonamente}, {Bulbul}, {Carlstrom}, {Culverhouse},
  {Gilbank}, {Greer}, {Hasler}, {Hawkins}, {Hennessy}, {Joy}, {Koester},
  {Lamb}, {Leitch}, {Miller}, {Mroczkowski}, {Muchovej}, {Oguri}, {Plagge},
  {Pryke}, \& {Woody}}]{gralla2011}
{Gralla}, M.~B., {Sharon}, K., {Gladders}, M.~D., {et~al.} 2011,
  \href{http://dx.doi.org/10.1088/0004-637X/737/2/74}{\apj, 737, 74}

\bibitem[{{Gregory} {et~al.}(1996){Gregory}, {Scott}, {Douglas}, \&
  {Condon}}]{Gregory1996}
{Gregory}, P.~C., {Scott}, W.~K., {Douglas}, K., \& {Condon}, J.~J. 1996,
  \href{http://dx.doi.org/10.1086/192282}{\apjs, 103, 427}

\bibitem[{{Griffin} {et~al.}(1986){Griffin}, {Ade}, {Orton}, {Robson}, {Gear},
  {Nolt}, \& {Radostitz}}]{griffin1986}
{Griffin}, M.~J., {Ade}, P.~A.~R., {Orton}, G.~S., {et~al.} 1986,
  \href{http://dx.doi.org/10.1016/0019-1035(86)90137-5}{\icarus, 65, 244}

\bibitem[{{Griffith} {et~al.}(1995){Griffith}, {Wright}, {Burke}, \&
  {Ekers}}]{griffith1995}
{Griffith}, M.~R., {Wright}, A.~E., {Burke}, B.~F., \& {Ekers}, R.~D. 1995,
  \href{http://dx.doi.org/10.1086/192146}{\apjs, 97, 347}

\bibitem[{{Hasselfield} {et~al.}(2013){Hasselfield}, {Hilton}, {Marriage},
  {Addison}, {Barrientos}, {Battaglia}, {Battistelli}, {Bond}, {Crichton},
  {Das}, {Devlin}, {Dicker}, {Dunkley}, {D{\"u}nner}, {Fowler}, {Gralla},
  {Hajian}, {Halpern}, {Hincks}, {Hlozek}, {Hughes}, {Infante}, {Irwin},
  {Kosowsky}, {Marsden}, {Menanteau}, {Moodley}, {Niemack}, {Nolta}, {Page},
  {Partridge}, {Reese}, {Schmitt}, {Sehgal}, {Sherwin}, {Sievers}, {Sif{\'o}n},
  {Spergel}, {Staggs}, {Swetz}, {Switzer}, {Thornton}, {Trac}, \&
  {Wollack}}]{hasselfield2013}
{Hasselfield}, M., {Hilton}, M., {Marriage}, T.~A., {et~al.} 2013,
  \href{http://dx.doi.org/10.1088/1475-7516/2013/07/008}{\jcap, 7, 008}

\bibitem[{{Hicks} {et~al.}(2013){Hicks}, {Pratt}, {Donahue}, {Ellingson},
  {Gladders}, {B{\"o}hringer}, {Yee}, {Yan}, {Croston}, \&
  {Gilbank}}]{hicks2013}
{Hicks}, A.~K., {Pratt}, G.~W., {Donahue}, M., {et~al.} 2013,
  \href{http://dx.doi.org/10.1093/mnras/stt348}{\mnras, 431, 2542}

\bibitem[{{Hinshaw} {et~al.}(2013){Hinshaw}, {Larson}, {Komatsu}, {Spergel},
  {Bennett}, {Dunkley}, {Nolta}, {Halpern}, {Hill}, {Odegard}, {Page}, {Smith},
  {Weiland}, {Gold}, {Jarosik}, {Kogut}, {Limon}, {Meyer}, {Tucker}, {Wollack},
  \& {Wright}}]{hinshaw2013}
{Hinshaw}, G., {Larson}, D., {Komatsu}, E., {et~al.} 2013,
  \href{http://dx.doi.org/10.1088/0067-0049/208/2/19}{\apjs, 208, 19}

\bibitem[{{Ho} {et~al.}(2009){Ho}, {Altamirano}, {Chang}, {Chang}, {Chang},
  {Chen}, {Chen}, {Chen}, {Han}, {Ho}, {Huang}, {Hwang}, {Iba{\~n}ez-Romano},
  {Jiang}, {Koch}, {Kubo}, {Li}, {Lim}, {Lin}, {Liu}, {Lo}, {Ma}, {Martin},
  {Martin-Cocher}, {Molnar}, {Ng}, {Nishioka}, {O'Connell}, {Oshiro}, {Patt},
  {Raffin}, {Umetsu}, {Wei}, {Wu}, {Chiueh}, {Chiueh}, {Chu}, {Huang}, {Hwang},
  {Liao}, {Lien}, {Wang}, {Wang}, {Wei}, {Yang}, {Kesteven}, {Kingsley},
  {Sinclair}, {Wilson}, {Birkinshaw}, {Liang}, {Lancaster}, {Park}, {Pen}, \&
  {Peterson}}]{ho2009}
{Ho}, P.~T.~P., {Altamirano}, P., {Chang}, C.-H., {et~al.} 2009,
  \href{http://dx.doi.org/10.1088/0004-637X/694/2/1610}{\apj, 694, 1610}

\bibitem[{{Huang} {et~al.}(2010){Huang}, {Wu}, {Ho}, {Koch}, {Liao}, {Lin},
  {Liu}, {Molnar}, {Nishioka}, {Umetsu}, {Wang}, {Altamirano}, {Birkinshaw},
  {Chang}, {Chang}, {Chang}, {Chen}, {Chiueh}, {Han}, {Huang}, {Hwang},
  {Jiang}, {Kesteven}, {Kubo}, {Li}, {Martin-Cocher}, {Oshiro}, {Raffin},
  {Wei}, \& {Wilson}}]{huang2010}
{Huang}, C.-W.~L., {Wu}, J.-H.~P., {Ho}, P.~T.~P., {et~al.} 2010,
  \href{http://dx.doi.org/10.1088/0004-637X/716/1/758}{\apj, 716, 758}

\bibitem[{{Huang} {et~al.}(2011){Huang}, {Raffin}, \& {Chen}}]{huang2011}
{Huang}, Y.-D., {Raffin}, P., \& {Chen}, M.-T. 2011,
  \href{http://dx.doi.org/10.1109/TAP.2011.2122211}{IEEE Transactions on
  Antennas and Propagation, 59, 2022}

\bibitem[{{Huang} {et~al.}(2008){Huang}, {Raffin}, {Chen}, {Altamirano}, \&
  {Oshiro}}]{huang2008}
{Huang}, Y.~D., {Raffin}, P., {Chen}, M.-T., {Altamirano}, P., \& {Oshiro}, P.
  2008, \href{http://dx.doi.org/10.1117/12.786811}{in Society of Photo-Optical
  Instrumentation Engineers (SPIE) Conference Series, Vol. 7012, Society of
  Photo-Optical Instrumentation Engineers (SPIE) Conference Series}

\bibitem[{{Kawaharada} {et~al.}(2010){Kawaharada}, {Okabe}, {Umetsu},
  {Takizawa}, {Matsushita}, {Fukazawa}, {Hamana}, {Miyazaki}, {Nakazawa}, \&
  {Ohashi}}]{Kawaharada2010}
{Kawaharada}, M., {Okabe}, N., {Umetsu}, K., {et~al.} 2010,
  \href{http://dx.doi.org/10.1088/0004-637X/714/1/423}{\apj, 714, 423}

\bibitem[{{Koch} {et~al.}(2008){Koch}, {Kesteven}, {Chang}, {Huang}, {Raffin},
  {Chen}, {Chereau}, {Chen}, {Ho}, {Huang}, {Iba{\~n}ez-Romano}, {Jiang},
  {Liao}, {Lin}, {Liu}, {Molnar}, {Nishioka}, {Umetsu}, {Wang}, {Wu},
  {Altamirano}, {Chang}, {Chang}, {Chang}, {Han}, {Kubo}, {Li},
  {Martin-Cocher}, \& {Oshiro}}]{koch2008}
{Koch}, P., {Kesteven}, M., {Chang}, Y.-Y., {et~al.} 2008,
  \href{http://dx.doi.org/10.1117/12.787852}{in Society of Photo-Optical
  Instrumentation Engineers (SPIE) Conference Series, Vol. 7018, Society of
  Photo-Optical Instrumentation Engineers (SPIE) Conference Series}

\bibitem[{{Koch} {et~al.}(2009){Koch}, {Kesteven}, {Nishioka}, {Jiang}, {Lin},
  {Umetsu}, {Huang}, {Raffin}, {Chen}, {Iba{\~n}ez-Romano}, {Chereau}, {Huang},
  {Chen}, {Ho}, {Pausch}, {Willmeroth}, {Altamirano}, {Chang}, {Chang},
  {Chang}, {Han}, {Kubo}, {Li}, {Liao}, {Liu}, {Martin-Cocher}, {Oshiro},
  {Wang}, {Wei}, {Wu}, {Birkinshaw}, {Chiueh}, {Lancaster}, {Lo}, {Martin},
  {Molnar}, {Patt}, \& {Romeo}}]{koch2009}
{Koch}, P.~M., {Kesteven}, M., {Nishioka}, H., {et~al.} 2009,
  \href{http://dx.doi.org/10.1088/0004-637X/694/2/1670}{\apj, 694, 1670}

\bibitem[{{Koch} {et~al.}(2011){Koch}, {Raffin}, {Huang}, {Chen}, {Han}, {Lin},
  {Altamirano}, {Granet}, {Ho}, {Huang}, {Kesteven}, {Li}, {Liao}, {Liu},
  {Nishioka}, {Ong}, {Oshiro}, {Umetsu}, {Wang}, \& {Wu}}]{koch2011}
{Koch}, P.~M., {Raffin}, P., {Huang}, Y.-D., {et~al.} 2011,
  \href{http://dx.doi.org/10.1086/658675}{\pasp, 123, 198}

\bibitem[{{LaRoque} {et~al.}(2006){LaRoque}, {Bonamente}, {Carlstrom}, {Joy},
  {Nagai}, {Reese}, \& {Dawson}}]{laroque2006}
{LaRoque}, S.~J., {Bonamente}, M., {Carlstrom}, J.~E., {et~al.} 2006,
  \href{http://dx.doi.org/10.1086/508139}{\apj, 652, 917}

\bibitem[{{Lemze} {et~al.}(2009){Lemze}, {Broadhurst}, {Rephaeli}, {Barkana},
  \& {Umetsu}}]{lemze2009}
{Lemze}, D., {Broadhurst}, T., {Rephaeli}, Y., {Barkana}, R., \& {Umetsu}, K.
  2009, \href{http://dx.doi.org/10.1088/0004-637X/701/2/1336}{\apj, 701, 1336}

\bibitem[{{Li} {et~al.}(2010){Li}, {Kubo}, {Wilson}, {Lin}, {Chen}, {Ho},
  {Chen}, {Han}, {Oshiro}, {Martin-Cocher}, {Chang}, {Chang}, {Altamirano},
  {Jiang}, {Chiueh}, {Lien}, {Wang}, {Wei}, {Yang}, {Peterson}, {Chang},
  {Huang}, {Hwang}, {Kesteven}, {Koch}, {Liu}, {Nishioka}, {Umetsu}, {Wei}, \&
  {Proty Wu}}]{li2010}
{Li}, C.-T., {Kubo}, D.~Y., {Wilson}, W., {et~al.} 2010,
  \href{http://dx.doi.org/10.1088/0004-637X/716/1/746}{\apj, 716, 746}

\bibitem[{{Liao} {et~al.}(2010){Liao}, {Proty Wu}, {Ho}, {Locutus Huang},
  {Koch}, {Lin}, {Liu}, {Molnar}, {Nishioka}, {Umetsu}, {Wang}, {Altamirano},
  {Birkinshaw}, {Chang}, {Chang}, {Chang}, {Chen}, {Chiueh}, {Han}, {Huang},
  {Hwang}, {Jiang}, {Kesteven}, {Kubo}, {Li}, {Martin-Cocher}, {Oshiro},
  {Raffin}, {Wei}, \& {Wilson}}]{liao2010}
{Liao}, Y.-W., {Proty Wu}, J.-H., {Ho}, P.~T.~P., {et~al.} 2010,
  \href{http://dx.doi.org/10.1088/0004-637X/713/1/584}{\apj, 713, 584}

\bibitem[{{Liao} {et~al.}(2013){Liao}, {Lin}, {Huang}, {Proty Wu}, {Ho},
  {Chen}, {Locutus Huang}, {Koch}, {Nishioka}, {Cheng}, {Fu}, {Liu}, {Molnar},
  {Umetsu}, {Wang}, {Chang}, {Han}, {Li}, {Martin-Cocher}, \&
  {Oshiro}}]{liao2013}
{Liao}, Y.-W., {Lin}, K.-Y., {Huang}, Y.-D., {et~al.} 2013,
  \href{http://dx.doi.org/10.1088/0004-637X/769/1/71}{\apj, 769, 71}

\bibitem[{{Limousin} {et~al.}(2007){Limousin}, {Richard}, {Jullo}, {Kneib},
  {Fort}, {Soucail}, {El{\'{\i}}asd{\'o}ttir}, {Natarajan}, {Ellis}, {Smail},
  {Czoske}, {Smith}, {Hudelot}, {Bardeau}, {Ebeling}, {Egami}, \&
  {Knudsen}}]{Limousin2007}
{Limousin}, M., {Richard}, J., {Jullo}, E., {et~al.} 2007,
  \href{http://dx.doi.org/10.1086/521293}{\apj, 668, 643}

\bibitem[{{Lin} {et~al.}(2009){Lin}, {Li}, {Ho}, {Huang}, {Liao}, {Liu},
  {Koch}, {Molnar}, {Nishioka}, {Umetsu}, {Wang}, {Wu}, {Kestevan},
  {Birkinshaw}, {Altamirano}, {Chang}, {Chang}, {Chang}, {Chen},
  {Martin-Cocher}, {Han}, {Huang}, {Hwang}, {Iba{\~n}ez-Roman}, {Jiang},
  {Kubo}, {Oshiro}, {Raffin}, {Wei}, {Wilson}, {Chen}, \& {Chiueh}}]{lin2009}
{Lin}, K.-Y., {Li}, C.-T., {Ho}, P.~T.~P., {et~al.} 2009,
  \href{http://dx.doi.org/10.1088/0004-637X/694/2/1629}{\apj, 694, 1629}

\bibitem[{{Liu} {et~al.}(2010){Liu}, {Birkinshaw}, {Proty Wu}, {Ho}, {Locutus
  Huang}, {Liao}, {Lin}, {Molnar}, {Nishioka}, {Koch}, {Umetsu}, {Wang},
  {Altamirano}, {Chang}, {Chang}, {Chang}, {Chen}, {Han}, {Huang}, {Hwang},
  {Jiang}, {Kesteven}, {Kubo}, {Li}, {Martin-Cocher}, {Oshiro}, {Raffin},
  {Wei}, \& {Wilson}}]{liu2010}
{Liu}, G.-C., {Birkinshaw}, M., {Proty Wu}, J.-H., {et~al.} 2010,
  \href{http://dx.doi.org/10.1088/0004-637X/720/1/608}{\apj, 720, 608}

\bibitem[{{Mason} {et~al.}(2010){Mason}, {Dicker}, {Korngut}, {Devlin},
  {Cotton}, {Koch}, {Molnar}, {Sievers}, {Aguirre}, {Benford}, {Staguhn},
  {Moseley}, {Irwin}, \& {Ade}}]{mason2010}
{Mason}, B.~S., {Dicker}, S.~R., {Korngut}, P.~M., {et~al.} 2010,
  \href{http://dx.doi.org/10.1088/0004-637X/716/1/739}{\apj, 716, 739}

\bibitem[{{Merten} {et~al.}(2015){Merten}, {Meneghetti}, {Postman}, {Umetsu},
  {Zitrin}, {Medezinski}, {Nonino}, {Koekemoer}, {Melchior}, {Gruen},
  {Moustakas}, {Bartelmann}, {Host}, {Donahue}, {Coe}, {Molino}, {Jouvel},
  {Monna}, {Seitz}, {Czakon}, {Lemze}, {Sayers}, {Balestra}, {Rosati},
  {Ben{\'{\i}}tez}, {Biviano}, {Bouwens}, {Bradley}, {Broadhurst}, {Carrasco},
  {Ford}, {Grillo}, {Infante}, {Kelson}, {Lahav}, {Massey}, {Moustakas},
  {Rasia}, {Rhodes}, {Vega}, \& {Zheng}}]{merten2015}
{Merten}, J., {Meneghetti}, M., {Postman}, M., {et~al.} 2015,
  \href{http://dx.doi.org/10.1088/0004-637X/806/1/4}{\apj, 806, 4}

\bibitem[{{Molnar} {et~al.}(2010{\natexlab{a}}){Molnar}, {Chiu}, {Umetsu},
  {Chen}, {Hearn}, {Broadhurst}, {Bryan}, \& {Shang}}]{Molnar2010HSE}
{Molnar}, S.~M., {Chiu}, I.-N., {Umetsu}, K., {et~al.} 2010{\natexlab{a}},
  \href{http://dx.doi.org/10.1088/2041-8205/724/1/L1}{\apjl, 724, L1}

\bibitem[{{Molnar} {et~al.}(2010{\natexlab{b}}){Molnar}, {Umetsu},
  {Birkinshaw}, {Bryan}, {Haiman}, {Hearn}, {Shang}, {Ho}, {Locutus Huang},
  {Koch}, {Liao}, {Lin}, {Liu}, {Nishioka}, {Wang}, \& {Proty Wu}}]{molnar2010}
{Molnar}, S.~M., {Umetsu}, K., {Birkinshaw}, M., {et~al.} 2010{\natexlab{b}},
  \href{http://dx.doi.org/10.1088/0004-637X/723/2/1272}{\apj, 723, 1272}

\bibitem[{{Morandi} {et~al.}(2011){Morandi}, {Limousin}, {Rephaeli}, {Umetsu},
  {Barkana}, {Broadhurst}, \& {Dahle}}]{Morandi2011}
{Morandi}, A., {Limousin}, M., {Rephaeli}, Y., {et~al.} 2011,
  \href{http://dx.doi.org/10.1111/j.1365-2966.2011.19175.x}{\mnras, 416, 2567}

\bibitem[{{Mroczkowski} {et~al.}(2012){Mroczkowski}, {Dicker}, {Sayers},
  {Reese}, {Mason}, {Czakon}, {Romero}, {Young}, {Devlin}, {Golwala},
  {Korngut}, {Sarazin}, {Bock}, {Koch}, {Lin}, {Molnar}, {Pierpaoli}, {Umetsu},
  \& {Zemcov}}]{mroczkowski2012}
{Mroczkowski}, T., {Dicker}, S., {Sayers}, J., {et~al.} 2012,
  \href{http://dx.doi.org/10.1088/0004-637X/761/1/47}{\apj, 761, 47}

\bibitem[{{Nagai} {et~al.}(2007){Nagai}, {Kravtsov}, \&
  {Vikhlinin}}]{nagai2007}
{Nagai}, D., {Kravtsov}, A.~V., \& {Vikhlinin}, A. 2007,
  \href{http://dx.doi.org/10.1086/521328}{\apj, 668, 1}

\bibitem[{{Navarro} {et~al.}(1997){Navarro}, {Frenk}, \& {White}}]{navarro1997}
{Navarro}, J.~F., {Frenk}, C.~S., \& {White}, S.~D.~M. 1997,
  \href{http://dx.doi.org/10.1086/304888}{\apj, 490, 493}

\bibitem[{{Nishioka} {et~al.}(2009){Nishioka}, {Wang}, {Wu}, {Ho}, {Huang},
  {Koch}, {Liao}, {Lin}, {Liu}, {Molnar}, {Umetsu}, {Birkinshaw}, {Altamirano},
  {Chang}, {Chang}, {Chang}, {Chen}, {Han}, {Huang}, {Hwang}, {Jiang},
  {Kesteven}, {Kubo}, {Li}, {Martin-Cocher}, {Oshiro}, {Raffin}, {Wei}, \&
  {Wilson}}]{nishioka2009}
{Nishioka}, H., {Wang}, F.-C., {Wu}, J.-H.~P., {et~al.} 2009,
  \href{http://dx.doi.org/10.1088/0004-637X/694/2/1637}{\apj, 694, 1637}

\bibitem[{{Oguri} {et~al.}(2005){Oguri}, {Takada}, {Umetsu}, \&
  {Broadhurst}}]{Oguri2005}
{Oguri}, M., {Takada}, M., {Umetsu}, K., \& {Broadhurst}, T. 2005,
  \href{http://dx.doi.org/10.1086/452629}{\apj, 632, 841}

\bibitem[{{Okabe} {et~al.}(2011){Okabe}, {Bourdin}, {Mazzotta}, \&
  {Maurogordato}}]{okabe2011}
{Okabe}, N., {Bourdin}, H., {Mazzotta}, P., \& {Maurogordato}, S. 2011,
  \href{http://dx.doi.org/10.1088/0004-637X/741/2/116}{\apj, 741, 116}

\bibitem[{{Okabe} {et~al.}(2014){Okabe}, {Umetsu}, {Tamura}, {Fujita},
  {Takizawa}, {Zhang}, {Matsushita}, {Hamana}, {Fukazawa}, {Futamase},
  {Kawaharada}, {Miyazaki}, {Mochizuki}, {Nakazawa}, {Ohashi}, {Ota}, {Sasaki},
  {Sato}, \& {Tam}}]{Okabe2014}
{Okabe}, N., {Umetsu}, K., {Tamura}, T., {et~al.} 2014,
  \href{http://dx.doi.org/10.1093/pasj/psu075}{\pasj, 66, 99}

\bibitem[{{Padin} {et~al.}(2000){Padin}, {Cartwright}, {Joy}, \&
  {Meitzler}}]{padin2000}
{Padin}, S., {Cartwright}, J.~K., {Joy}, M., \& {Meitzler}, J.~C. 2000,
  \href{http://dx.doi.org/10.1109/8.855504}{IEEE Transactions on Antennas and
  Propagation, 48, 836}

\bibitem[{{Page} {et~al.}(2003){Page}, {Barnes}, {Hinshaw}, {Spergel},
  {Weiland}, {Wollack}, {Bennett}, {Halpern}, {Jarosik}, {Kogut}, {Limon},
  {Meyer}, {Tucker}, \& {Wright}}]{page2003}
{Page}, L., {Barnes}, C., {Hinshaw}, G., {et~al.} 2003,
  \href{http://dx.doi.org/10.1086/377223}{\apjs, 148, 39}

\bibitem[{{Peng} {et~al.}(2009){Peng}, {Andersson}, {Bautz}, \&
  {Garmire}}]{Peng2009}
{Peng}, E.-H., {Andersson}, K., {Bautz}, M.~W., \& {Garmire}, G.~P. 2009,
  \href{http://dx.doi.org/10.1088/0004-637X/701/2/1283}{\apj, 701, 1283}

\bibitem[{{Planck Collaboration} {et~al.}(2011{\natexlab{a}}){Planck
  Collaboration}, {Aghanim}, {Arnaud}, {Ashdown}, {Aumont}, {Baccigalupi},
  {Balbi}, {Banday}, {Barreiro}, {Bartelmann}, \&
  et~al.}]{planckcollaboration2011ix}
{Planck Collaboration}, {Aghanim}, N., {Arnaud}, M., {et~al.}
  2011{\natexlab{a}},
  \href{http://dx.doi.org/10.1051/0004-6361/201116460}{\aap, 536, A9}

\bibitem[{{Planck Collaboration} {et~al.}(2011{\natexlab{b}}){Planck
  Collaboration}, {Ade}, {Aghanim}, {Arnaud}, {Ashdown}, {Aumont},
  {Baccigalupi}, {Balbi}, {Banday}, {Barreiro}, \&
  et~al.}]{planckcollaboration2011xi}
{Planck Collaboration}, {Ade}, P.~A.~R., {Aghanim}, N., {et~al.}
  2011{\natexlab{b}},
  \href{http://dx.doi.org/10.1051/0004-6361/201116458}{\aap, 536, A11}

\bibitem[{{Planck Collaboration} {et~al.}(2013){Planck Collaboration}, {Ade},
  {Aghanim}, {Arnaud}, {Ashdown}, {Atrio-Barandela}, {Aumont}, {Baccigalupi},
  {Balbi}, {Banday}, \& et~al.}]{planckcollaboration2012v}
---. 2013, \href{http://dx.doi.org/10.1051/0004-6361/201220040}{\aap, 550,
  A131}

\bibitem[{{Planck Collaboration} {et~al.}(2014){Planck Collaboration}, {Ade},
  {Aghanim}, {Armitage-Caplan}, {Arnaud}, {Ashdown}, {Atrio-Barandela},
  {Aumont}, {Baccigalupi}, {Banday}, \& et~al.}]{PlanckCollaboration2014xvi}
---. 2014, \href{http://dx.doi.org/10.1051/0004-6361/201321591}{\aap, 571, A16}

\bibitem[{{Planck Collaboration} {et~al.}(2015{\natexlab{a}}){Planck
  Collaboration}, {Ade}, {Aghanim}, {Arnaud}, {Ashdown}, {Aumont},
  {Baccigalupi}, {Banday}, {Barreiro}, {Bartlett}, \&
  et~al.}]{planckcollaboration2015xiii}
---. 2015{\natexlab{a}}, ArXiv e-prints,
  \href{http://arxiv.org/abs/1502.01589}{{\sffamily arXiv:1502.01589}}

\bibitem[{{Planck Collaboration} {et~al.}(2015{\natexlab{b}}){Planck
  Collaboration}, {Ade}, {Aghanim}, {Arnaud}, {Ashdown}, {Aumont},
  {Baccigalupi}, {Banday}, {Barreiro}, {Barrena}, \&
  et~al.}]{planckcollaboration2015xxvii}
---. 2015{\natexlab{b}}, ArXiv e-prints,
  \href{http://arxiv.org/abs/1502.01598}{{\sffamily arXiv:1502.01598}}

\bibitem[{{Postman} {et~al.}(2012){Postman}, {Coe}, {Ben{\'{\i}}tez},
  {Bradley}, {Broadhurst}, {Donahue}, {Ford}, {Graur}, {Graves}, {Jouvel},
  {Koekemoer}, {Lemze}, {Medezinski}, {Molino}, {Moustakas}, {Ogaz}, {Riess},
  {Rodney}, {Rosati}, {Umetsu}, {Zheng}, {Zitrin}, {Bartelmann}, {Bouwens},
  {Czakon}, {Golwala}, {Host}, {Infante}, {Jha}, {Jimenez-Teja}, {Kelson},
  {Lahav}, {Lazkoz}, {Maoz}, {McCully}, {Melchior}, {Meneghetti}, {Merten},
  {Moustakas}, {Nonino}, {Patel}, {Reg{\"o}s}, {Sayers}, {Seitz}, \& {Van der
  Wel}}]{postman2012}
{Postman}, M., {Coe}, D., {Ben{\'{\i}}tez}, N., {et~al.} 2012,
  \href{http://dx.doi.org/10.1088/0067-0049/199/2/25}{\apjs, 199, 25}

\bibitem[{{Raffin} {et~al.}(2006){Raffin}, {Koch}, {Huang}, {Chang}, {Chang},
  {Chen}, {Chen}, {Ho}, {Huang}, {Iba{\~n}ez Roman}, {Jiang}, {Kesteven},
  {Lin}, {Liu}, {Nishioka}, \& {Umetsu}}]{raffin2006}
{Raffin}, P., {Koch}, P., {Huang}, Y.-D., {et~al.} 2006,
  \href{http://dx.doi.org/10.1117/12.672020}{in Society of Photo-Optical
  Instrumentation Engineers (SPIE) Conference Series, Vol. 6273, Society of
  Photo-Optical Instrumentation Engineers (SPIE) Conference Series}

\bibitem[{{Raffin} {et~al.}(2004){Raffin}, {Martin}, {Huang}, {Patt}, {Romeo},
  {Chen}, \& {Kingsley}}]{raffin2004}
{Raffin}, P.~A., {Martin}, R.~N., {Huang}, Y.-D., {et~al.} 2004,
  \href{http://dx.doi.org/10.1117/12.554599}{in Society of Photo-Optical
  Instrumentation Engineers (SPIE) Conference Series, Vol. 5495, Society of
  Photo-Optical Instrumentation Engineers (SPIE) Conference Series, ed.
  J.~{Antebi} \& D.~{Lemke}}, 159

\bibitem[{{Ruze}(1966)}]{ruze1966}
{Ruze}, J. 1966, IEEE Proceedings, 54, 633

\bibitem[{{Sayers} {et~al.}(2013){Sayers}, {Czakon}, {Mantz}, {Golwala},
  {Ameglio}, {Downes}, {Koch}, {Lin}, {Maughan}, {Molnar}, {Moustakas},
  {Mroczkowski}, {Pierpaoli}, {Shitanishi}, {Siegel}, {Umetsu}, \& {Van der
  Pyl}}]{sayers2013b}
{Sayers}, J., {Czakon}, N.~G., {Mantz}, A., {et~al.} 2013,
  \href{http://dx.doi.org/10.1088/0004-637X/768/2/177}{\apj, 768, 177}

\bibitem[{{Sereno} {et~al.}(2013){Sereno}, {Ettori}, {Umetsu}, \&
  {Baldi}}]{Sereno2013}
{Sereno}, M., {Ettori}, S., {Umetsu}, K., \& {Baldi}, A. 2013,
  \href{http://dx.doi.org/10.1093/mnras/sts186}{\mnras, 428, 2241}

\bibitem[{{Sharon} {et~al.}(2015){Sharon}, {Gladders}, {Marrone}, {Hoekstra},
  {Rasia}, {Bourdin}, {Gifford}, {Hicks}, {Greer}, {Mroczkowski}, {Barrientos},
  {Bayliss}, {Carlstrom}, {Gilbank}, {Gralla}, {Hlavacek-Larrondo}, {Leitch},
  {Mazzotta}, {Miller}, {Muchovej}, {Schrabback}, {Yee}, \&
  {RCS-Team}}]{sharon2015}
{Sharon}, K., {Gladders}, M.~D., {Marrone}, D.~P., {et~al.} 2015,
  \href{http://dx.doi.org/10.1088/0004-637X/814/1/21}{\apj, 814, 21}

\bibitem[{{Sunyaev} \& {Zeldovich}(1970)}]{Sunyaev1970}
{Sunyaev}, R.~A., \& {Zeldovich}, Y.~B. 1970, Comments on Astrophysics and
  Space Physics, 2, 66

\bibitem[{{Sunyaev} \& {Zeldovich}(1972)}]{Sunyaev1972}
---. 1972, Comments on Astrophysics and Space Physics, 4, 173

\bibitem[{{Ulich}(1981)}]{ulich1981}
{Ulich}, B.~L. 1981, \href{http://dx.doi.org/10.1086/113046}{\aj, 86, 1619}

\bibitem[{{Umetsu} \& {Broadhurst}(2008)}]{UB2008}
{Umetsu}, K., \& {Broadhurst}, T. 2008,
  \href{http://dx.doi.org/10.1086/589683}{\apj, 684, 177}

\bibitem[{{Umetsu} {et~al.}(2016){Umetsu}, {Zitrin}, {Gruen}, {Merten},
  {Donahue}, \& {Postman}}]{umetsu2016}
{Umetsu}, K., {Zitrin}, A., {Gruen}, D., {et~al.} 2016,
  \href{http://dx.doi.org/10.3847/0004-637X/821/2/116}{\apj, 821, 116}

\bibitem[{{Umetsu} {et~al.}(2009){Umetsu}, {Birkinshaw}, {Liu}, {Wu},
  {Medezinski}, {Broadhurst}, {Lemze}, {Zitrin}, {Ho}, {Huang}, {Koch}, {Liao},
  {Lin}, {Molnar}, {Nishioka}, {Wang}, {Altamirano}, {Chang}, {Chang}, {Chang},
  {Chen}, {Han}, {Huang}, {Hwang}, {Jiang}, {Kesteven}, {Kubo}, {Li},
  {Martin-Cocher}, {Oshiro}, {Raffin}, {Wei}, \& {Wilson}}]{umetsu2009}
{Umetsu}, K., {Birkinshaw}, M., {Liu}, G.-C., {et~al.} 2009,
  \href{http://dx.doi.org/10.1088/0004-637X/694/2/1643}{\apj, 694, 1643}

\bibitem[{{Umetsu} {et~al.}(2014){Umetsu}, {Medezinski}, {Nonino}, {Merten},
  {Postman}, {Meneghetti}, {Donahue}, {Czakon}, {Molino}, {Seitz}, {Gruen},
  {Lemze}, {Balestra}, {Ben{\'{\i}}tez}, {Biviano}, {Broadhurst}, {Ford},
  {Grillo}, {Koekemoer}, {Melchior}, {Mercurio}, {Moustakas}, {Rosati}, \&
  {Zitrin}}]{umetsu2014}
{Umetsu}, K., {Medezinski}, E., {Nonino}, M., {et~al.} 2014,
  \href{http://dx.doi.org/10.1088/0004-637X/795/2/163}{\apj, 795, 163}

\bibitem[{{Umetsu} {et~al.}(2015){Umetsu}, {Sereno}, {Medezinski}, {Nonino},
  {Mroczkowski}, {Diego}, {Ettori}, {Okabe}, {Broadhurst}, \&
  {Lemze}}]{umetsu2015}
{Umetsu}, K., {Sereno}, M., {Medezinski}, E., {et~al.} 2015,
  \href{http://dx.doi.org/10.1088/0004-637X/806/2/207}{\apj, 806, 207}

\bibitem[{{Weiland} {et~al.}(2011){Weiland}, {Odegard}, {Hill}, {Wollack},
  {Hinshaw}, {Greason}, {Jarosik}, {Page}, {Bennett}, {Dunkley}, {Gold},
  {Halpern}, {Kogut}, {Komatsu}, {Larson}, {Limon}, {Meyer}, {Nolta}, {Smith},
  {Spergel}, {Tucker}, \& {Wright}}]{weiland2011}
{Weiland}, J.~L., {Odegard}, N., {Hill}, R.~S., {et~al.} 2011,
  \href{http://dx.doi.org/10.1088/0067-0049/192/2/19}{\apjs, 192, 19}

\bibitem[{{Wright} {et~al.}(2009){Wright}, {Chen}, {Odegard}, {Bennett},
  {Hill}, {Hinshaw}, {Jarosik}, {Komatsu}, {Nolta}, {Page}, {Spergel},
  {Weiland}, {Wollack}, {Dunkley}, {Gold}, {Halpern}, {Kogut}, {Larson},
  {Limon}, {Meyer}, \& {Tucker}}]{wright2009}
{Wright}, E.~L., {Chen}, X., {Odegard}, N., {et~al.} 2009,
  \href{http://dx.doi.org/10.1088/0067-0049/180/2/283}{\apjs, 180, 283}

\bibitem[{{Wu} {et~al.}(2009){Wu}, {Ho}, {Huang}, {Koch}, {Liao}, {Lin}, {Liu},
  {Molnar}, {Nishioka}, {Umetsu}, {Wang}, {Altamirano}, {Birkinshaw}, {Chang},
  {Chang}, {Chang}, {Chen}, {Chiueh}, {Han}, {Huang}, {Hwang}, {Jiang},
  {Kesteven}, {Kubo}, {Lancaster}, {Li}, {Martin-Cocher}, {Oshiro}, {Raffin},
  {Wei}, \& {Wilson}}]{wu2009}
{Wu}, J.-H.~P., {Ho}, P.~T.~P., {Huang}, C.-W.~L., {et~al.} 2009,
  \href{http://dx.doi.org/10.1088/0004-637X/694/2/1619}{\apj, 694, 1619}

\bibitem[{{Zitrin} {et~al.}(2015){Zitrin}, {Fabris}, {Merten}, {Melchior},
  {Meneghetti}, {Koekemoer}, {Coe}, {Maturi}, {Bartelmann}, {Postman},
  {Umetsu}, {Seidel}, {Sendra}, {Broadhurst}, {Balestra}, {Biviano}, {Grillo},
  {Mercurio}, {Nonino}, {Rosati}, {Bradley}, {Carrasco}, {Donahue}, {Ford},
  {Frye}, \& {Moustakas}}]{zitrin2015}
{Zitrin}, A., {Fabris}, A., {Merten}, J., {et~al.} 2015,
  \href{http://dx.doi.org/10.1088/0004-637X/801/1/44}{\apj, 801, 44}

\end{thebibliography}

\appendix
\section{Monte-Carlo Simulations for Cluster Mass Estimates}\label{sec:app.1447}

The gNFW pressure profile in Equation~(\ref{eq:gNFW}) is indirectly related to the halo mass through $P_{500}$. With the mass dependence of $P_{500}$ explicitly written out \citep[see e.g.,][]{nagai2007}, the cluster physical pressure profile becomes
\begin{align}\label{eq:gNFW_p}
	P(r)  = & {} 1.65 \times 10^{-3} h(z)^{8/3} \left[\frac{M_{500}}{3 \times 10^{14} h_{70}^{-1} M_{\sun}}\right]^{2/3 + 0.12}\notag\\
	& \times \mathbb{P}(x) h_{70}^2 \mathrm{\, keV\, cm}^{-3}\, ,
\end{align}
where $\cal{P}(x)$ is defined in equation~(\ref{eq:gNFW}).

In the simulation here, we relate the cluster size $R_{500}$ to its halo mass assuming an NFW model \citep{navarro1997} through a concentration -- mass (c$-$M) relation from \citet{Dutton2014},  with redshift evolution:
\begin{align}\label{eq:c-M}
  & \log_{10}c_{200}=a+b \log_{10}\left(M_{200}/\left[10^{12} h_0^{-1} M_{\sun}\right]\right)\, ,\\
  & a=0.520+(0.905-0.520) \exp(-0.617 z^{1.21})\, ,\notag \\
  & b=-0.101+0.026 z\, . \notag
\end{align}

We choose $M_{200}$ to represent the virial mass. For each $M_{200}$, $R_{500}$ and $M_{500}$ were obtained from this c$-$M relation and then substituted in the pressure equation~(\ref{eq:gNFW_p}).

At a given halo mass, Monte--Carlo (MC) simulations draw a set of gNFW parameters\footnote{$\beta=5.4905$ is fixed in this simulation.} ($P_0$, $c_{500}$, $\alpha$, and $\gamma$) from the prior parameter distribution and calculates the desired observable. Here, we assume that (1) the gNFW parameters are independent of halo mass, and (2) the parameters fitted for individual clusters listed in Table~C.1 of A10 are a fair representation of the intrinsic scatter of the parameters. Specifically, three of the parameters are modeled as Gaussian random numbers in logarithmic space ($\log_{10}P_0$, $\log_{10}c_{500}$, $\log_{10}\alpha$), while the remaining parameter ($\gamma$) is modeled as a Gaussian random number in linear space. We make this distinction because $\gamma$ can be zero while the other parameters are all positive definite. 
To account for correlations among the gNFW parameters, we estimate their covariances from the best-fit parameters of the clusters in A10 and reproduce these covariances in our MC simulation. 
The cluster RXC J2319.6-7313 is removed from the A10 sample because its best-fit parameters are quite different from the rest of the sample. 
Additionally, we require that $\alpha > \gamma$ when generating the random parameters to avoid profiles that are uncharacteristic of the A10 sample.
We note that $c_{200}$ is treated as an independent variable, uncorrelated with the gNFW parameters.

\begin{figure*}
  \plotone{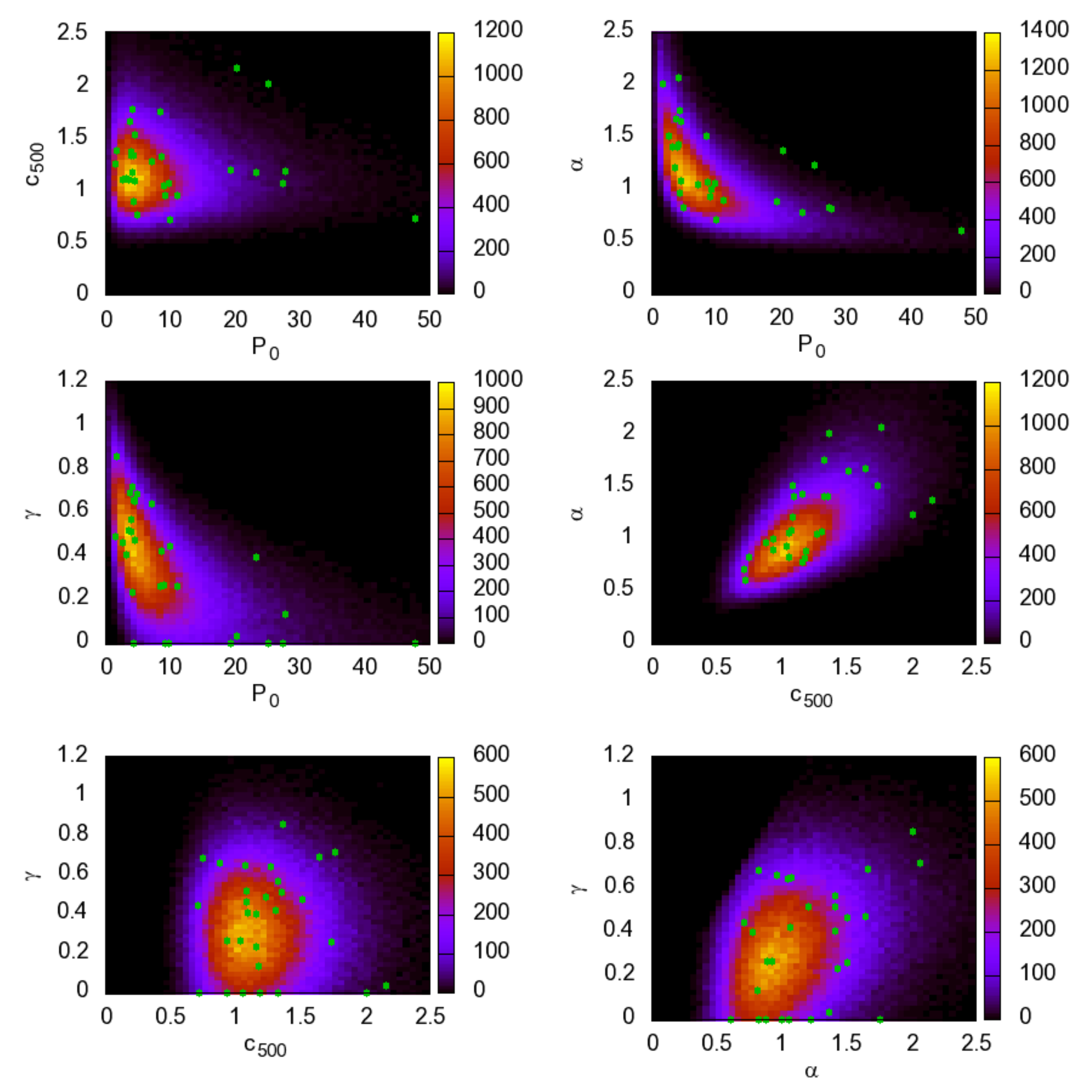}
  \caption{Pair-wise scatters among the gNFW parameters used in our Monte--Carlo simulations. Green points represent the best-fit parameters of the REXCESS clusters from A10. The color map shows the distribution of parameters used in this simulation. The color box indicates the counts in the linearly spaced histogram.}\label{fig:gNFW_scatters}
\end{figure*}

\begin{figure}
 \plotone{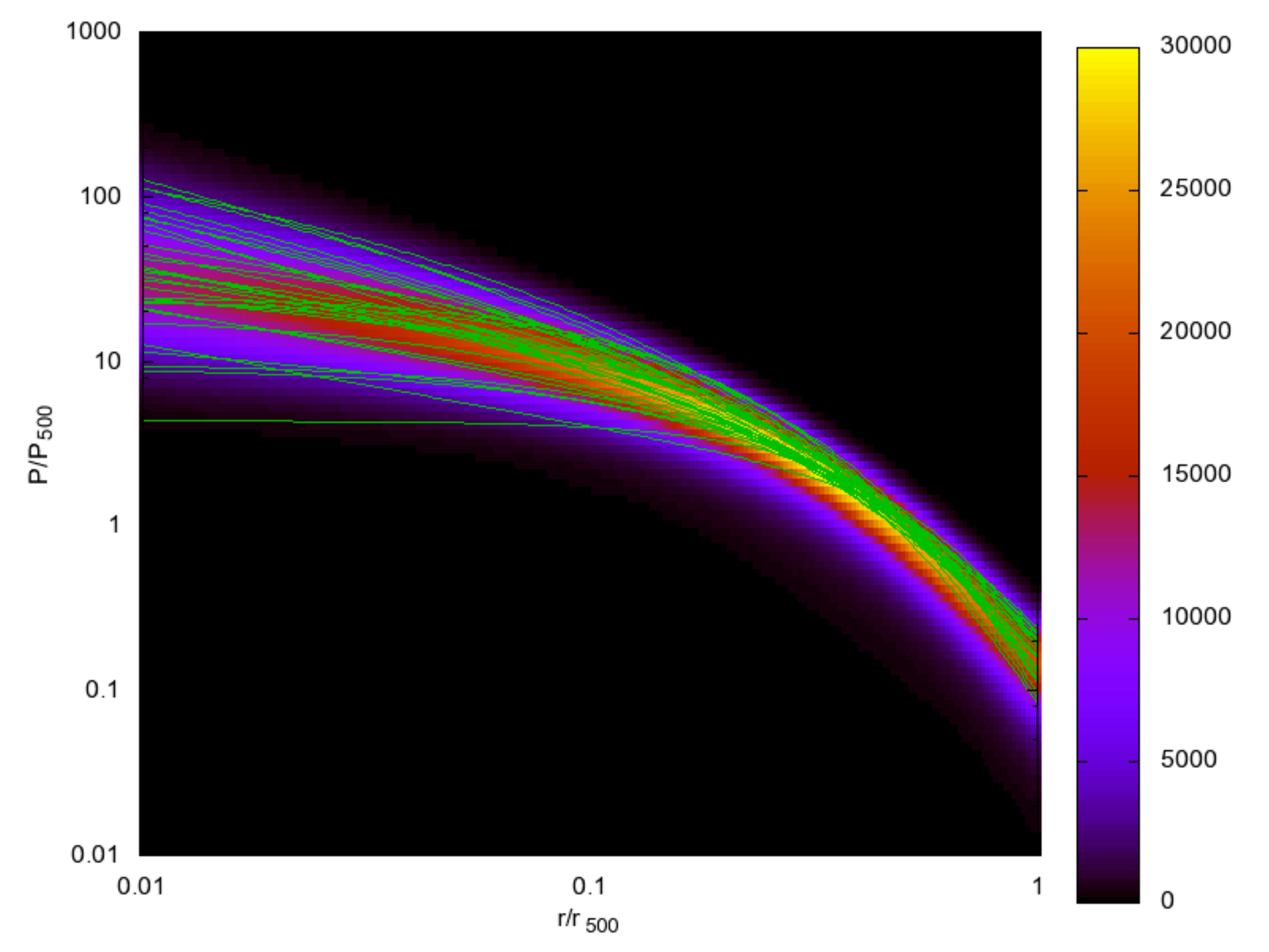}
 \caption{The gNFW profiles produced using parameters from Figure~\ref{fig:gNFW_scatters}. The green lines correspond to the parameters from A10. Color scale indicates the number of the simulated profiles passing through each logarithmically-spaced grid cell.}\label{fig:gNFW_profiles}
\end{figure}

We choose 200 $M_{200}$ values logarithmically sampled from $30$ to $3000$ $\left[10^{13} M_{\sun}\right]$, and for each $M_{200}$ 1000 realizations of SZE profiles are created. Since all AMiBA targets are massive clusters with at least $T_\mathrm{X} > 5$~keV, we include relativistic corrections assuming $T_\mathrm{X} = 10$~keV and note that changing $T_\mathrm{X}$ from 5~keV to 15~keV does not change the results significantly. 
Figure~\ref{fig:gNFW_scatters} shows the pair-wise scatter among the four free parameters for both the A10 clusters and our simulation.
Figure~\ref{fig:gNFW_profiles} shows the gNFW profiles produced using both the A10 results and our simulation parameters.

Although we only show simulations here that model all the clusters together from A10, simulations that restrictively model either the cool-core or the disturbed subset of the cluster sample are done in an analogous way. 
Generally, we adjust our simulations to each cluster in order to connect its halo mass to its AMiBA flux.

The cosmology adopted in these simulations is a flat $\Lambda$CDM model
with $H_0=67.1$\,km\,s$^{-1}$\,Mpc$^{-1}$, $\Omega_\mathrm{m}=0.3175$, and
$T_\mathrm{CMB}=2.725$ K \citep{PlanckCollaboration2014xvi}.

\end{document}